
\font\titlefont = cmr10 scaled\magstep 4
\font\myfont    = cmr10 scaled\magstep 2
\font\sectionfont = cmr10
\font\littlefont = cmr5 
\font\eightrm = cmr8 

\def\ss{\scriptstyle} 
\def\sss{\scriptscriptstyle} 

\newcount\tcflag
\tcflag = 1  

\ifnum\tcflag = 0 \magnification = 1200 \fi  

\global\baselineskip = 1.2\baselineskip 
\global\parskip = 4pt plus 0.3pt 
\global\abovedisplayskip = 18pt plus3pt minus9pt
\global\belowdisplayskip = 18pt plus3pt minus9pt
\global\abovedisplayshortskip = 6pt plus3pt
\global\belowdisplayshortskip = 6pt plus3pt

\def\barsoff{\overfullrule=0pt}


\def\endignore{}
\def\ignore #1\endignore{} 

\newcount\dflag
\dflag = 0


\def\monthname{\ifcase\month 
\or January \or February \or March \or April \or May \or June%
\or July \or August \or September \or October \or November %
\or December 
\fi}

\newcount\dummy
\newcount\minute  
\newcount\hour
\newcount\localtime
\newcount\localday
\localtime = \time
\localday = \day

\def\advanceclock#1#2{ 
\dummy = #1
\multiply\dummy by 60
\advance\dummy by #2
\advance\localtime by \dummy
\ifnum\localtime > 1440 
\advance\localtime by -1440
\advance\localday by 1
\fi}

\def\settime{{\dummy = \localtime %
\divide\dummy by 60%
\hour = \dummy 
\minute = \localtime%
\multiply\dummy by 60%
\advance\minute by -\dummy 
\ifnum\minute < 10 
\xdef\spacer{0} 
\else \xdef\spacer{} 
\fi %
\ifnum\hour < 12 
\xdef\ampm{a.m.} 
\else 
\xdef\ampm{p.m.} 
\advance\hour by -12 %
\fi %
\ifnum\hour = 0 \hour = 12 \fi 
\xdef\timestring{\number\hour : \spacer \number\minute%
\thinspace \ampm}}}



\def\endtitle{}
\def\title#1\endtitle{\vskip.5in\titlefont
\global\baselineskip = 2\baselineskip 
#1\vskip.4in
\baselineskip = 0.5\baselineskip\rm}
 
\def\endauthors{}
\def\authors#1\endauthors{#1}

\def\endabstract{}
\def\abstract#1\endabstract{\vskip .3in%
\centerline{\sectionfont\bf Abstract}%
\vskip .1in
\noindent#1}

\def\nopageonenumber{\footline={\ifnum\pageno<2\hfil\else
\hss\tenrm\folio\hss\fi}}  

\newcount\nsection 
\newcount\nsubsection 
\newcount\nsubsubsection 

\def\section#1{\global\advance\nsection by 1
\nsubsection=0
\nsubsubsection=0
\bigskip\noindent\centerline{\sectionfont \bf \number\nsection.\ #1}
\bigskip\rm\nobreak}

\def\subsection#1{\global\advance\nsubsection by 1
\nsubsubsection=0
\bigskip\noindent\sectionfont \sl \number\nsection.\number\nsubsection)\
#1\bigskip\rm\nobreak}

\def\subsubsection#1{\global\advance\nsubsubsection by 1 
\smallskip\noindent\sectionfont \sl \number\nsection.\number\nsubsection.\number\nsubsubsection)\
#1\smallskip\rm\nobreak}

\def\topic #1{{\medskip\noindent $\bullet$ \it #1:}}

\def\appendix#1#2{\bigskip\noindent%
\centerline{\sectionfont \bf Appendix #1.\ #2} 
\bigskip\rm\nobreak} 


\newcount\nref 
\global\nref = 1 

\def\therefs{} 


\def\ref#1#2{\xdef #1{[\number\nref]} 
\ifnum\nref = 1\global\xdef\therefs{\item{[\number\nref]} #2\ } 
\else
\global\xdef\oldrefs{\therefs}
\global\xdef\therefs{\oldrefs\vskip.1in\item{[\number\nref]} #2\ }%
\fi%
\global\advance\nref by 1
}

\def\listrefs{\vfill\eject\section{References}\therefs}


\newcount\nfoot 
\global\nfoot = 1 

\def\foot#1#2{\xdef #1{(\number\nfoot)} 
\hskip -0.2cm ${}^{\number\nfoot}$ 
\footnote{}{\vbox{\baselineskip=10pt
\eightrm \hskip -1cm ${}^{\number\nfoot}$ #2}}
\global\advance\nfoot by 1
}


\newcount\nfig 
\global\nfig = 1
\def\thefigs{} 

\def\figure#1#2{\xdef #1{(\number\nfig)}
\ifnum\nfig = 1\global\xdef\thefigs{\item{(\number\nfig)} #2\ }
\else
\global\xdef\oldfigs{\thefigs}
\global\xdef\thefigs{\oldfigs\vskip.1in\item{(\number\nfig)} #2\ }%
\fi%
\global\advance\nfig by 1 } 

\def\fig#1{\xdef #1{(\number\nfig)}
\global\advance\nfig by 1 } 


\newcount\ntab
\global\ntab = 1

\def\table#1{\xdef #1{\number\ntab}
\global\advance\ntab by 1 } 


\newcount\cflag
\newcount\nequation
\global\nequation = 1
\def\eqlabel{(1)}

\def\nexteqno{\ifnum\cflag = 0
\global\advance\nequation by 1
\fi
\global\cflag = 0
\xdef\eqlabel{(\number\nequation)}}

\def\lasteqno{\global\advance\nequation by -1
\xdef\eqlabel{(\number\nequation)}}

\def\label#1{\xdef #1{(\number\nequation)}
\ifnum\dflag = 1
{\escapechar = -1
\xdef\draftname{\littlefont\string#1}}
\fi}

\def\clabel#1#2{\xdef\eqlabel{(\number\nequation #2)}
\global\cflag = 1
\xdef #1{\eqlabel}
\ifnum\dflag = 1
{\escapechar = -1
\xdef\draftname{\string#1}}
\fi}

\def\cclabel#1#2{\xdef\eqlabel{#2)}
\global\cflag = 1
\xdef #1{\eqlabel}
\ifnum\dflag = 1
{\escapechar = -1
\xdef\draftname{\string#1}}
\fi}


\def\eeq{}

\def\eqnn #1\eeq{$$ #1 $$}

\def\eq #1\eeq{
\ifnum\dflag = 0
{\xdef\draftname{\ }}
\fi 
$$ #1
\eqno{\eqlabel \rlap{\ \draftname}} $$
\nexteqno}



\def\eol{& \eqlabel \rlap{\ \draftname} \crcr
\nexteqno
\xdef\draftname{\ }}

\def\eeol{& \eqlabel \rlap{\ \draftname}
\nexteqno
\xdef\draftname{\ }}

\def\eolnn{\cr
\global\cflag = 0
\xdef\draftname{\ }}

\def\eeolnn{\xdef\draftname{\ }}

\def\eqa #1\eeq{
\ifnum\dflag = 0
{\xdef\draftname{\ }}
\fi 
$$ \eqalignno{ #1 } $$
\global\cflag = 0}


\def\ie{{\it i.e.\/}}
\def\eg{{\it e.g.\/}}
\def\etc{{\it etc.\/}}
\def\etal{{\it et.al.\/}}
\def\apriori{{\it a priori\/}}

\def\via{{\it via\/}}


\def\anp#1#2#3{{\it Ann.\ Phys. (NY)} {\bf #1} (19#2) #3}

\def\arnps#1#2#3{{\it Ann.\ Rev.\ Nucl.\ Part.\ Sci.} {\bf #1}, (19#2) #3}

\def\ejp#1#2#3{{\it Eur.\ J.\ Phys.} {\bf #1} (19#2) #3}

\def\npb#1#2#3{{\it Nucl.\ Phys.} {\bf B#1} (19#2) #3}
\def\plb#1#2#3{{\it Phys.\ Lett.} {\bf #1B} (19#2) #3}

\def\pra#1#2#3{{\it Phys.\ Rev.} {\bf A#1} (19#2) #3}

\def\prd#1#2#3{{\it Phys.\ Rev.} {\bf D#1} (19#2) #3}

\def\prep#1#2#3{{\it Phys.\ Rep.} {\bf #1} (19#2) #3}
\def\prl#1#2#3{{\it Phys.\ Rev.\ Lett.} {\bf #1} (19#2) #3}

\def\zpc#1#2#3{{\it Zeit.\ Phys.} {\bf C#1} (19#2) #3}


\global\nulldelimiterspace = 0pt



\def\frac#1#2{{{#1} \over {#2}}\,}  
\def\hf{{1\over 2}}
\def\nth#1{{1\over #1}}


\def\Asl{\hbox{/\kern-.7500em\it A}} 
\def\Dsl{\hbox{/\kern-.6700em\it D}} 
\def\dsl{\hbox{/\kern-.5300em$\partial$}}
\def\pxpsl{\hbox{/\kern-.5600em$p$}}
\def\sslsh{\hbox{/\kern-.5300em$s$}}
\def\epssl{\hbox{/\kern-.5100em$\epsilon$}}
\def\delsl{\hbox{/\kern-.6300em$\nabla$}}
\def\lxpsl{\hbox{/\kern-.4300em$l$}}
\def\elxpsl{\hbox{/\kern-.4500em$\ell$}}
\def\kxpsl{\hbox{/\kern-.5100em$k$}}
\def\qxpsl{\hbox{/\kern-.5000em$q$}}
\def\sla#1{\raise.15ex\hbox{$/$}\kern-.57em #1}

\def\pwr#1{\cdot 10^{#1}}



\def\twi{\widetilde}

\def\roughly#1{\mathrel{\raise.3ex\hbox{$#1$\kern-.75em\lower1ex\hbox{$\sim$}}}}
\def\lsim{\roughly<}
\def\gsim{\roughly>}

\def\tw#1{\tilde{#1}}
\def\ol#1{\overline{#1}}





\def\Scc{{\cal C}}

\def\Sch{{\cal H}}

\def\Scl{{\cal L}}

\def\Sct{{\cal T}}

\def\Scv{{\cal V}}

\def\Scy{{\cal Y}}


\def\ssa{{\sss A}}
\def\ssb{{\sss B}}

\def\ssf{{\sss F}}

\def\ssh{{\sss H}}
\def\ssi{{\sss I}}

\def\ssl{{\sss L}}

\def\ssr{{\sss R}}
\def\ssS{{\sss S}}
\def\sst{{\sss T}}

\def\ssv{{\sss V}}
\def\ssw{{\sss W}}

\def\ssy{{\sss Y}}
\def\ssz{{\sss Z}}


\def\pmb#1{\setbox0=\hbox{#1}%
\kern-.025em\copy0\kern-\wd0
\kern.05em\copy0\kern-\wd0
\kern-.025em\raise.0433em\box0}   


\font\jlgtenbrm=cmbx10
\font\jlgtenbit=cmmib10
\font\jlgtenbsy=cmbsy10
\font\jlgsevenbrm=cmbx10 at 7pt
\font\jlgsevenbsy=cmbsy10 at 7pt
\font\jlgsevenbit=cmmib10 at 7pt
\font\jlgfivebrm=cmbx10 at 5pt
\font\jlgfivebsy=cmbsy10 at 5pt
\font\jlgfivebit=cmmib10 at 5pt
\newfam\jlgbrm

\textfont\jlgbrm=\jlgtenbrm
\scriptfont\jlgbrm=\jlgsevenbrm
\scriptscriptfont\jlgbrm=\jlgfivebrm
\newfam\jlgbit

\textfont\jlgbit=\jlgtenbit
\scriptfont\jlgbit=\jlgsevenbit
\scriptscriptfont\jlgbit=\jlgfivebit
\newfam\jlgbsy

\textfont\jlgbsy=\jlgtenbsy
\scriptfont\jlgbsy=\jlgsevenbsy
\scriptscriptfont\jlgbsy=\jlgfivebsy
\newcount\jlgcode
\newcount\jlgfam
\newcount\jlgchar
\newcount\jlgtmp
\def\bolded#1{
        \jlgcode\the#1 \divide\jlgcode by 4096
        \jlgtmp\the\jlgcode \multiply\jlgtmp by 4096
        \jlgfam\the#1 \advance\jlgfam by -\the\jlgtmp
        \divide\jlgfam by 256
        \jlgtmp\the\jlgcode \multiply\jlgtmp by 16
	\advance\jlgtmp by \the\jlgfam
	\multiply\jlgtmp by 256
        \jlgchar\the#1 \advance\jlgchar by -\the\jlgtmp
        \advance\jlgfam by \the\jlgbrm
        \jlgtmp\the\jlgcode
        \multiply\jlgtmp by 16
        \advance\jlgtmp by \the\jlgfam
        \multiply\jlgtmp by 256
        \advance\jlgtmp by \the\jlgchar
        \mathchar\the\jlgtmp
}


\def\Tr{\mathop{\rm Tr}}


\def\bra#1{\langle #1 |}
\def\ket#1{| #1 \rangle}



\def\hc{{\rm h.c.}}
\def\cc{{\rm c.c.}}


\def\GeV{{\rm \ GeV}}
\def\TeV{{\rm \ TeV}}

\input epsf.tex
\nopageonenumber
\baselineskip = 14pt
\barsoff
\def\fp#1#2#3{{\it Fortsch.\ Phys.} {\bf #1} (19#2) #3}

\def\mref#1#2{#1 -- #2}

\def\bk{\item{}}

\def\lrderiv#1#2{#1 {\leftrightarrow
\atop {  }}  {\kern-0.8em \partial_\mu} #2}
\def\pbar{\ol{p}}
\def\sbar{\ol{s}}
\def\fbar{\ol{f}}
\def\mbar{\ol{m}}
\def\mh{m_h}
\def\sw{s_w}
\def\cw{c_w}
\def\gl{g_\ssl}
\def\gr{g_\ssr}
\def\mw{M_\ssw}
\def\mz{M_\ssz}
\def\eff{{\rm eff}}
\def\GF{G_\ssf}
\def\SM{{\sss SM}}

\def\caption#1{\noindent {\baselineskip 10pt \eightrm #1}}


\line{ \hfil CERN-TH,
McGill and TPI-MINN }		

\title
\centerline{A Higgs or Not a Higgs?}
\centerline{{\myfont What to Do if You Discover
a New Scalar Particle}}
\endtitle

\authors
\centerline{C.P. Burgess,${}^a$ J. Matias${}^b$ and M.
Pospelov ${}^c$}
\vskip .07in
\centerline{\it ${}^a$ Physics Department, McGill
University}
\centerline{\it 3600 University St., Montr\'eal,
Qu\'ebec,  Canada, H3A 2T8.}
\vskip .03in
\centerline{\it ${}^b$ Theory Division, CERN, CH-1211
Gen\`eve 23, Switzerland.}
\vskip .03in
\centerline{\it ${}^c$ Theoretical Physics Institute,
University of Minnesota}
\centerline{\it 431 Tate Laboratory of Physics,
Minneapolis, MN, USA 55455.}
\endauthors

\vskip .25in
\centerline{\sectionfont August 2001}

\vskip .75in

\abstract
\vskip .15in
We show how to systematically analyze what may be inferred
should a new scalar particle be discovered in collider
experiments.  Our approach is systematic in the sense that
we perform the analysis in a manner which minimizes
\apriori\ theoretical assumptions as to the nature of the
scalar particle. For instance, we do {\it not} immediately
make the common assumption that a new scalar  particle is a
Higgs boson, and so must interact with a strength
proportional  to the mass of the particles with which it
couples.  We show how to compare  different observables,
and so to develop a decision tree from which the nature  of
the new particle may be discerned. We define several categories
of models, which summarize the kinds of distinctions which
the first experiments can make.
\endabstract


\vfill\eject
\section{Introduction}

Suppose that you have just learned that a new scalar
particle has been discovered. After your immediate euphoria
there are a number of things which you should do (\eg: call
friends, pay off bets, feverishly write papers, book
flights to  Stockholm, \etc). After this, your next wish
will probably be to know what the new discovery means. Is
it the Standard Model (SM) Higgs?  Is this the first sign
of supersymmetry? If so, if the scalar is neutral is it a
Higgs or is it a sneutrino? Is it a  technipion?

\ref\outa{
M. Carena, P. Zerwas and the Higgs Physics Working Group,
{\it Physics at LEP-2}, published in CERN yellow
report, CERN-96-01 edited by G. Altarelli, T. Sjostrand and F. Zwirner,
(hep-ph/9602250).}

\ref\outaa{M. Carena, J. Conway, H. Haber and J. Hobbs et al.,
Report of the Higgs Working Group of the RunII Workshop,
Fermilab, 1999, see
http://fnth37.fnal.gov/higgs.html.
}

\ref\outb{J. Gunion, H. Haber, G. Kane and S. Dawson, {\it The Higgs
Hunter's Guide}, Addison-Wesley, 1990 and references therein.}

\ref\outbb{
F. de Campos, M.C. Gonzalez-Garcia and S.F. Novaes, Phys. Rev. Lett. 
{\bf 79} 
(1997) 5210, M.C. Gonzalez-Garcia S.M. Lietti and S.F. Novaes, 
Phys. Rev. {\bf D57} (1998) 7045,
M.C. Gonzalez-Garcia, Int. J. Mod. Phys. {\bf A14} (1999) 3121;
J.A. Oller, Phys. Lett. B477 (2000) 187.
}

We hope that once these reactions have passed --- or even
before --- you  will remember and reread this paper, since
our goal here is to show how to  answer these, and other,
issues. We intend here to show how to combine  current
experimental results with the new information about the new
particle, and infer what its properties are in as
unprejudiced a manner as is now possible. We have no
particular axe to grind, and so wish to make this inference
in a  manner which does not build in from the beginning lots
of theoretical prejudices  as to what the new scalar means.
We use the language of effective theories to efficiently organize
the extant experimental information in a way which allows a
relatively objective comparison of the evidence in favour
of the various theoretical possibilities.  We mean to
complement in this way the very many detailed studies of
the implications for Higgs searches of various specific
models \mref\outa\outbb, and to provide a general language within which
such models may be efficiently compared.

Our main assumption in our analysis is that, at least for a
short time, only one (or a few) scalar particles are
initially discovered, and that any others of the zoo of
undiscovered particles are reasonably heavy compared to the
scales presently being scoured for Higgses. Here `reasonably
heavy' might mean as heavy as, say, several hundred GeV, which puts
such particles out of reach of experiments at LEP, HERA and
the current generation of hadron colliders.  This assumption
has two key advantages: $(i)$ it is broad enough to include
most of the models which are of current interest, and
$(ii)$ it is quite predictive since it permits a systematic
parameterization of the scalar particle couplings in terms
of  the effective theory which is obtained when all of the
heavier particles are integrated out.
Any `non-decoupling' and slowly decoupling effects of these heavier
particles will be automatically encoded amongst
the effective couplings of this lagrangian.

The other main assumption we make is that the Yukawa
couplings of the newly observed scalar are dominantly
flavour-diagonal. Although we make this assumption mainly on the
grounds of simplicity, we do not believe it to be a major
limitation on the applicability of our analysis because
of the impressive limits which exist on many types of
flavour-changing processes. These typically require the
couplings to light fermions of the lightest scalar state
in most models to be approximately flavour diagonal.
We nevertheless regard a model-independent study of
the bounds on flavour-changing scalar couplings to be
worthwhile to pursue, but defer such an analysis to
future work.

Our presentation proceeds in the following way. First, the
next section (\S2) presents the most general
low-dimension interactions which are possible between new
scalars and the other well-known elementary particles. So
long as experiments cannot reach energies high enough to
probe the next threshhold for new physics the effective
couplings which appear in this effective action encode all
of the information that can be learnt, even
in principle, about the new scalars. Then, \S3 and \S4
relate these effective interactions to observables in order
to see in a general way which kinds of experiments are
sensitive to which kinds of scalar-particle properties. \S3
concentrates on scalar production and decays, while \S4
specializes to the contribution of virtual scalar exchange
for processes having no scalars in the initial or final
state. Contact with specific models is made in \S5, where
the effective couplings of \S2 are computed as functions of
the underlying couplings for various choices for the models
which might describe this underlying, higher-energy
physics.

The many studies of scalar-particle phenomenology in particular
models shows that tree-level perturbation theory can be insufficiently
accurate in some situations, due to the importance of
next-to-leading-order (NLO) corrections or to genuine nonperturbative
effects. It is therefore important to be clear how these contributions
arise within the effective-lagrangian approach used here. If the
important nonleading contributions involve high-energy particles,
then they must be included in \S5, where these degrees of freedom
are integrated out to generate the effective couplings. If, on the
other hand, the important nonleading terms involve only light
particles -- typically QCD effects, in practice -- they must be
included in \S3\ and \S4\ where the low-energy theory is used
to compute expressions for observables. We emphasize that the
use of tree level, say, in \S3 and \S4\ is {\it not} inconsistent with
obtaining the higher-order (or nonperturbative) contributions to
observables, so long as these arise at high energies and
have been included when computing the effective couplings in \S5.

All of our results are finally pulled together in \S6, in which
we show which observables best differentiate amongst the
various kinds of possible models for scalar-particle
physics.  We discuss in this section a `decision tree'
which may be used to decide whether the new particle is an
element of an `elementary' electroweak multiplet or  is a
composite boson; or whether it is an electroweak doublet or
a member of another multiplet; if it is a supersymmetric
scalar or the familiar Higgs from the SM, \etc. In particular,
we use the possible low-energy couplings to divide models
into 16 categories. We present these categories as being the
proper expression of the information which experimenters are
likely to be able to obtain shortly after the discovery of any
new scalar, in that they can fairly quickly differentiate
from which category of model a new scalar originates. It is also
possible to differentiate amongst models within any particular 
category, but this is likely to take longer as it requires more
detailed information.

\section{General Effective Interactions for New Spinless
Particles}

As stated in the introduction, the central assumption which
organizes our analysis is that only one (or a few) new
scalar particles are initially found, with all other new
particles being sufficiently heavy to  continue evading
detection, at least initially. It is important to emphasize
that, given the current state of the experimental art, these
undiscovered  heavy particles need not actually be
excessively heavy.

For example,  it might happen that new scalars are
discovered with  masses near 100 GeV, but that all other
undiscovered new particles in the  underlying theory have
masses which are at least 200 GeV. This kind of mass
hierarchy is already sufficient to ensure the validity of
the considerations presented here. Of course, the heavier
any other undiscovered particles may be, the better
approximation it is to use only the lowest-dimension
interactions of the effective theory.

Under these assumptions the interactions of the observed
particles are described in terms of the lowest-dimension
interactions within the effective theory obtained by
integrating out all of the heavier undiscovered particles.
These effective interactions must be constructed from local
operators involving only fields which correspond to the
observed particle spectrum. They must also respect all of
the symmetries which are believed to hold exactly, and
which act only among the particles which appear in the
low-energy theory. That is, they must be invariant with
respect to Lorentz and  electromagnetic and $SU_c(3)$
(colour) gauge invariance.

\ref\wbgb{T. Appelquist and C. Bernard, \prd{22}{80}{200};
A.C. Longhitano, \prd{22}{80}{1166}; \npb{188}{81}{118};
M.S. Chanowitz and M.K. Gaillard, \npb{261}{85}{379}; 
M.S. Chanowitz, M. Golden and H. Georgi, \prd{36}{87}{1490};
M.S. Chanowitz, \arnps{38}{88}{323};
R.D. Peccei and X. Zhang, \npb{337}{90}{269};
B. Holdom and J. Terning, \plb{247}{90}{88};
J. Bagger, S. Dawson and G. Valencia, report BNL-45782;
M. Golden and L. Randall, \npb{361}{91}{3}; 
B. Holdom, \plb{259}{91}{329};
A. Dobado, M.J. Herrero and D. Espriu, \plb{255}{91}{405};
R.D. Peccei and S. Peris, \prd{44}{91}{809};
A. Dobado and J.M. Herrero, report CERN-TH-6272/91.}

\ref\usesandabuses{
C.P. Burgess and D. London,
\prl{69}{92}{3428}; \prd{48}{93}{4337}.}

\ref\sscalars{
F. Boudjema, E. Chopin, \zpc{73}{96}{85};
V.A. Ilin, A.E.Pukhov, Y. Kurihara, T. Shimizu, T. Kaneko,
\prd{54}{96}{6717}; A. Djouadi, W. Kilian and M. Muhlleitner and
P.M. Zerwas, \ejp{10}{99}{27}, 
hep-ph/0001169; 
D.A.Demir, hep-ph/9902468;
W. Hollik and
S: Penaranda, hep-ph/0108245;
P. Osland, \prd{59}{99}{055013};
M. Battaglia, E.Boos, W.-M. Yao, hep-ph/0111276
}

Notice that (linearly-realized) invariance with respect to
the SM electroweak gauge group, $SU_\ssl(2) \times
U_\ssy(1)$, should {\it not} be imposed \apriori, unless it
becomes established that the observed degrees of freedom
actually do fill out electroweak multiplets. Indeed,
determining the evidence in favour or against this
possibility is part of the main motivation for the analysis
we here present. (It is important to realise in this regard
that there is no physical difference between completely
ignoring $SU_\ssl(2)\times U_\ssy(1)$ gauge invariance
and nonlinearly realizing it by introducing a collection
of would-be Goldstone bosons \wbgb. This choice is
purely a matter of convenience, and is similar to the choice
between using a unitary or renormalizable gauge in a
renormalizable gauge theory: the nonlinear realization
brings ease of loop calculations and power-counting;
while ignoring the gauge symmetries makes the physical
particle content and interactions easier to see \usesandabuses.)

If one were to also know that the energetics which makes the 
scalar choose to take a symmetry breaking v.e.v. is the same as 
in the minimal SM, Higgs self-interactions must also be studied.
We will not address this issue in this work and refer the reader
to the  literature  \outb and \sscalars.

\ref\unitarityviolation{
J.M. Cornwall, D.N. Levin and G. Tiktopoulos,
\prd{10}{74}{1145};
M.S. Chanowitz, M. Golden and H. Georgi,
\prd{36}{87}{1490}.}

Of course, in the event that the effective theory does not
couple in a gauge-invariant way to the known massive 
spin-one particles, $Z^0$ and $W^\pm$ the effective theory
must violate unitarity at sufficiently high energies
\unitarityviolation, \usesandabuses. 
Far from invalidating the use of such
effective lagrangians at low energies, high-energy
pathologies such as this are invaluable because they
indicate the energies at which the  effective description
fails. As such they provide an upper bound on the masses of
other degrees of freedom, whose interactions cure the
high-energy unitarity problems of the low-energy effective
theory.

\subsection{Effective Interactions Having Dimension $\le$ 4}

\ref\bigfit{I. Maksymyk, C.P. Burgess and D. London,
\prd{50}{94}{529};
C.P Burgess, S. Godfrey, H. K\"onig, D. London and I.
Maksymyk,
\plb{326}{94}{276};  \prd{49}{94}{6115}.}

We now turn to the enumeration of the possible interactions
which can arise in the effective theory. We take its particle content
to consist of  the usual garden-variety fermions and gauge bosons,
plus a recently-discovered collection of $N$ neutral scalar
bosons: $h_i$, $i = 1,\dots,N$. Only interactions which explicitly
involve the hypothetical newly-discovered scalar are listed
here, although a list of corresponding effective
interactions amongst the presently-known particles are
presented and analyzed in a similar spirit in
ref.~\bigfit.

With these particles, and assuming electromagnetic and
colour gauge invariance (and Lorentz invariance), the most
general possible  lowest-dimension interactions are:
\eq
\label\scaopssum
\Scl_\eff = \Scl^{(2)} + \Scl^{(3)} +
\Scl^{(4)} + \Scl^{(5)} + \cdots
\eeq
with the dimension-two and -three operators given by
\eq
\label\dimtwoops
\eqalign{
\Scl^{(2)} &= - \, \hf  m^2_i \; h_i^2 \cr
\Scl^{(3)} &= - \, { \nu_{ijk} \over 3!} \; h_i h_j h_k
- \, {a_\ssz^i \over 2} \; Z_\mu Z^\mu \;
h_i - a^i_\ssw \; W^*_\mu W^\mu \; h_i ,\cr }
\eeq
where the Einstein summation convention applies to all
repeated indices, and the reality of $\Scl_{\rm eff}$
implies the reality of all of the coupling constants.

More interactions arise at dimension four.:
\eq
\label\dfrprlm
\Scl^{(4)} = \Scl_{\rm kin} + \Scl^{(4)}_{\rm scalar} +
 \Scl^{(4)}_{\rm fermion} + \Scl^{(4)}_{\rm vector},
\eeq
with $\Scl_{\rm kin} =  - \, \hf  \partial^\mu h_i \;
\partial_\mu h_i$~\foot\metric{In our conventions the metric
is $\eta_{\mu \nu}={\rm diag}(-,+,+,+)$} 
denoting the usual kinetic terms for
canonically-normalized scalar fields,
\eq
\label\dimfourops
\eqalign{
\Scl^{(4)}_{\rm scalar} &= - \, {\lambda_{ijkl} \over 4!}
\; h_i h_j h_k h_l   \cr
\Scl^{(4)}_{\rm fermion} &= -\sum_{Q(f) = Q(f')}
\ol{f} \Bigl( y^i_{ff'} + i  \gamma_5 z^i_{ff'}
\Bigr) \, f' \; h_i , \cr  }
\eeq
and
\eq
\label\dimfouropsv
\Scl^{(4)}_{\rm vector} =
- \left( {b_\ssz^{ij} \over 4} \; Z_\mu Z^\mu +
{b_\ssw^{ij}\over 2} \; W^*_\mu W^\mu \right)
\; h_i h_j   -  \; {g_\ssz^{ij} \over 2} \;
\lrderiv{h_i}{h_j}
 \; Z^\mu  .
\eeq
The various coefficients in this last expression satisfy
numerous reality and symmetry conditions. For example, all
of the bosonic effective couplings are real, and are
symmetric under the interchange of their subscripts $i$ and
$j$ --- \eg\ $b_\ssz^{ij} = b_\ssz^{ji}$ --- except for
$g_\ssz^{ij}$, which is antisymmetric.  Notice that this
antisymmetry of $g_\ssz^{ij}$ implies that the
corresponding interaction does not arise if there is only
one neutral scalar.

\topic{Unphysical Scalars}

When computing loops using this effective lagrangian it is
usually convenient to work in a manifestly renormalizable
gauge.\foot\stillcanhere{It is always possible to choose such a
gauge, even when $\ss \Scl_\eff$ is not explicitly $\ss SU_\ssl(2)
\times U_\ssy(1)$ gauge invariant, by rewriting the lagrangian
using a nonlinear realization of the gauge group \usesandabuses.}
In these gauges there are two unphysical scalars, $z$ and $w$,
which become the longitudinal spin states of the $Z$ and $W$
in unitary gauge. In these gauges the neutral scalar $z$ must
be included as one of the scalars participating in the effective
interactions just described, as well as including the related
couplings involving the charged scalar $w$.

\subsection{Some Dimension-5 Operators}

\ref\barbieri{R. Barbieri and A. Strumia, \plb{462}{99}{144}.}

With very few exceptions, interactions having dimension
five or higher are not  required for our purposes, since
their effects are negligible compared with those just
listed. This is guaranteed so long as all particles which
are integrated out in producing this effective lagrangian
are sufficiently heavy. This is fortunate because
higher-dimensional interactions can be as numerous as the
proverbial grains of sand on the beach.
An analysis of some higher dimensional operators can be found
in ~\barbieri.

Among the exceptions mentioned in the previous paragraph
are interactions which couple the scalars to photons and
gluons, since these interactions have no lower-dimension
counterparts with which to compete. There are four
interactions of this type which arise at lowest dimension:
\eq
\label\dimfiveops
\Scl^{(5)}_{g,\gamma} = - c_g^i \; G^\alpha_{\mu\nu}
G^{\mu\nu}_\alpha \; h_i - \tilde{c}_g^i \;
G^\alpha_{\mu\nu}
\twi{G}^{\mu\nu}_\alpha \; h_i - c_\gamma^i \;
F_{\mu\nu} F^{\mu\nu} \; h_i - \tilde{c}_\gamma^i \;
F_{\mu\nu} \twi{F}^{\mu\nu} \; h_i ,
\eeq
where $F_{\mu\nu}$ and $G^\alpha_{\mu\nu}$ are,
respectively, the electromagnetic and gluon field
strengths, and a tilde over a field strength denotes the
usual Hodge dual: $\twi{F}_{\mu\nu} = \hf \;
\epsilon_{\mu\nu\lambda\rho} \; F^{\lambda \rho}$.
Some dimension-five interactions involving the photon
and the $Z$ boson are also useful to include for the same
reasons:
\eq
\label\dimfiveZgamma
\Scl^{(5)}_{\ssz\gamma}=-c_{\ssz\gamma}^i
Z_{\mu\nu} F^{\mu\nu} \; h_i
-\tilde c_{\ssz\gamma}^iZ_{\mu\nu} 
\twi{F}^{\mu\nu} \; h_i ,
\eeq
where $Z_{\mu\nu}=\partial_\mu Z_\nu -\partial_\nu Z_\mu$.

\subsection{Special Case 1: Only One New Neutral Scalar}

An important special case is the (second-most) pessimistic
scenario in which only a single neutral scalar is found ---
\ie\ $N=1$.  (This is the case considered in detail in many
of the later sections.) Denoting the sole new scalar field in this case
by $h$, the above effective interactions simplify
considerably, to become:
\eq
\label\dimtwoopsa
\Scl^{(2)} + \Scl^{(3)} = - \, { \mh^2 \over 2} \;
h^2  - \, { \nu \over 3!} \; h^3 - \, {a_\ssz \over 2} \;
Z_\mu Z^\mu \; h -  a_\ssw \; W^*_\mu W^\mu \; h  ,
\eeq
and
\eq
\label\dimfouropsa
\Scl^{(4)}_{\rm int} = -  \sum_{Q(f) = Q(f')}
\ol{f} \Bigl( y_{ff'} + i \gamma_5  z_{ff'} \Bigr) \,
f' \; h - \, {\lambda \over 4!} \; h^4
 - \left( {b_\ssz \over 4} \; Z_\mu Z^\mu +
{b_\ssw\over 2} \; W^*_\mu W^\mu \right) \; h^2 .
\eeq

The dimension-five interactions of eqs.~\dimfiveops\ and
\dimfiveZgamma\ are also possible in this case:
\eq
\label\dimfiveopsa
\eqalign{
\Scl^{(5)}_{g,\gamma} &= - c_g \; G^\alpha_{\mu\nu}
G^{\mu\nu}_\alpha \; h - \tilde{c}_g \; G^\alpha_{\mu\nu}
\twi{G}^{\mu\nu}_\alpha \; h - c_\gamma \;
F_{\mu\nu} F^{\mu\nu} \; h - \tilde{c}_\gamma \;
F_{\mu\nu} \twi{F}^{\mu\nu} \; h \cr
&\qquad\qquad\qquad
- c_{\ssz\gamma}
Z_{\mu\nu} F^{\mu\nu} \; h
-\tilde c_{\ssz\gamma} Z_{\mu\nu} 
\twi{F}^{\mu\nu} \; h . \cr}
\eeq

\subsection{Special Case 2: Two Scalars Subject to a
Conservation Law}

Another important special case arises when there is more
than one scalar but conservation laws exist which forbid
many of the terms in $\Scl_{\rm eff}$. This would happen,
for example, if the new scalar carried a conserved quantum
number such as lepton number. (The sneutrino would be this
kind of scalar, for example, in supersymmetric models if
lepton number should not be broken.) In this section we
identify the lowest-dimension couplings which can survive
in this case, assuming there to be only one new (complex)
scalar, $\Sch =  (h_1 + i h_2)/\sqrt2$.

Assuming the $W$ and $Z$ do not also carry this quantum
number, this kind of scalar can have only the following
low-dimension interactions with bosons:
\eq
\label\dimtwoopsb
\eqalign{
\Scl_{\rm bose}^{(2)} + \Scl_{\rm bose}^{(3)}
+ \Scl^{(4)}_{\rm bose} &= - \, \mh^2 \;
\Sch^* \, \Sch   - \, {\lambda \over 4} \; (\Sch^* \,
\Sch)^2  \cr
& \qquad - \left( {b_\ssz \over 2} \; Z_\mu Z^\mu +
b_\ssw \; W^*_\mu W^\mu \right)
\; \Sch^* \, \Sch   -   i g_\ssz  \; \lrderiv{\Sch^*}{\Sch}
\; Z^\mu . \cr}
\eeq

Most of the dimension-four fermion-scalar couplings
considered above must also vanish for this kind of scalar.
This is because there are not many potentially conserved
global quantum numbers which it is possible for the fermions
to carry, given only the
known particles and low-dimension interactions. The only
candidates are the accidental symmetries of the Standard
Model itself: baryon number, $B$, and the flavour of each
generation of lepton, $L_e$, $L_\mu$ and $L_\tau$.
For instance, if $B(\Sch)
\ne 0$, then there are no dimension-four $B$-conserving
fermion interactions possible at all because the requirement
of colour neutrality automatically implies the $B$ neutrality
of all Lorentz-scalar fermion bilinears. The same is also true
if total lepton number, $L = L_e + L_\mu + L_\tau$, is
conserved and is carried by $\Sch$.

At dimension four the only nontrivial possibilities arise if $\Sch$ carries
only some of $L_e$, $L_\mu$ or $L_\tau$ and either
total lepton number is not conserved, or it is conserved but
is not carried by $\Sch$ (as might happen if $L_e(\Sch) =
- L_\mu(\Sch)$, say). In this case nontrivial
dimension-four couplings are possible between $\Sch$ and
neutrinos, $\nu$, and/or between $\Sch$ and charged leptons,
$\ell$:
\eq
\label\dimfouropsb
\Scl^{(4)}_{\rm fermi} =
- \Bigl[\ol{\ell}_a \Bigl( y^\ell_{ab} + i \gamma_5
z^\ell_{ab} \Bigr) \, \ell_b + \ol{\nu}_a
\Bigl( y^\nu_{ab} + i \gamma_5
z^\nu_{ab} \Bigr) \, \nu_b \Bigr] \; \Sch + \cc .
\eeq

The precluding of so many low-dimension interactions by the
assumed conservation law makes some of the higher-dimension
operators more important than they would be otherwise. In
particular, only the operators of 
eqs.~\dimfouropsb\ mediate $\Sch$ decay, 
and these operators are also
forbidden if either $B$ or $L$ are conserved and carried by
$\Sch$. Unless
$\Sch$ decays are themselves forbidden by $B$ or $L$
conservation, decays must in this case be mediated by
operators of even higher dimension. For example if $B(\Sch)
= -L(\Sch) = 1$ and both of these symmetries are unbroken,
then the decay $\Sch \to n \; \ol\nu$ is allowed, but the
lowest-dimension effective operators which can mediate
scalar decay first arise at dimension seven, such as:
\eq
\label\sampHdecay
\Scl_{\rm decay}^{(7)} = \kappa \,
\epsilon^{\alpha\beta\gamma}
\; \Sch \; (\ol{q}_\alpha q^c_\beta) \; (\ol{q}_\gamma \nu).
\eeq
In any case, for the present purposes we must keep in mind
the possibility that the new scalar might be very
long-lived in this scenario.

\subsection{Special Case 3: The Standard Model}

The Standard Model itself furnishes what is probably the
most important special case to consider. It is a particular instance
of the single-scalar scenario described earlier. Because, true
to its name, the SM really does provide the standard against
which other models are compared, we treat this example in
more detail than the special cases just considered. (This
more detailed discussion is duplicated for other models
of interest in section 5, below.) We consider first the
tree level contributions to the effective lagrangian, and
then discuss the nature of the radiative corrections to these
tree-level results.

\subsubsection{Tree-Level Predictions}

The dominant SM contributions to the effective
interactions of dimension four or less arise at tree level,
and are given in terms of ratios of the relevant particle
masses to the fundamental expectation value, $v = (\sqrt2 \;
\GF)^{-1/2} = 246$ GeV.

The explicit expressions at
dimension three are:
\eq
\label\SMdthree
\nu = {6 \, \mh^2 \over v} ,\qquad
a_\ssz = {2 \, \mz^2 \over v} = {e \mz \over \sw \cw},\qquad
a_\ssw = {2 \, \mw^2 \over v} = {e \mw \over \sw} , \qquad
\hbox{SM(tree)}
\eeq
where $e$ is the electromagnetic coupling, while $\sw$ $\cw$
are the sine and cosine of the weak mixing angle, $\theta_\ssw$.
The dimension four interactions are:
\eq
\label\SMdfourb
\lambda = {6 \, \mh^2 \over v^2} ,\qquad
b_\ssz = {2 \, \mz^2 \over v^2} = {e^2 \over 2\, \sw^2
\cw^2} , \qquad
b_\ssw = {2 \, \mw^2 \over v^2} = {e^2 \over 2\, \sw^2},
\qquad
\hbox{SM(tree)}
\eeq
and
\eq
\label\SMdfourf
y_{ff'} = {m_f \over v} \; \delta_{ff'}, \qquad
z_{ff'} = 0 \qquad
\hbox{SM(tree)}.
\eeq
The purpose of the rest of this paper is to find to what
extent predicted relationships, such as these, amongst the
effective couplings can be experimentally established using
current (and future) data.

\subsubsection{SM Radiative Corrections}

Of course the SM predictions of eqs.~\SMdthree, \SMdfourb\
and \SMdfourf\ are modified at one-loop level and beyond.
These corrections are not required for most of the present
purposes because upcoming experiments will not be sensitive
to small corrections to these relations. The same point also
holds in most -- but not all (see below) -- of the models considered
in what follows, so for many purposes it suffices to restrict
our calculations to tree level.

\ref\cgradcorn{J. Ellis, M.K. Gaillard and D.V. Nanopoulos,
\npb{106}{76}{292}}

\ref\cgradcors{T. Inami, T. Kubota and Y. Okada, \zpc{18}{83}{69};
A. Djouadi, M. Spira and P.M. Zerwas, \plb{264}{91}{440}}

\ref\spira{M. Spira, A. Djouadi, D. Graudenz and P.M. Zerwas,
\npb{453}{95}{17}}

\ref\spiran{M. Spira, \fp{46}{98}{203}}

\ref\tildec{M. Pospelov, hep-ph/9511368 (unpublished).}

The only exceptions to this statement arise when the
leading-order prediction is zero --- or very small because
of suppressions by small factors, such as light particle
masses --- and if the same is not true for the radiative
corrections. It is therefore important to examine carefully
the predictions for unusually small effective couplings. One
might worry that the vanishing of $z_{ff'}$ in
eq.~\SMdfourf, and the extremely small predictions there
for $y_{ff'}$, might be suspicious on these grounds.
Although this worry can be justified for
some other models, the
small size of these couplings is not changed when higher
SM loops are considered. For the $y_{ff'}$ this is
because vanishing Yukawa coupling imply the existence of
new chiral symmetries, and these symmetries ensure that the
higher-loop corrections are themselves also proportional to
the tree-level values, $\delta y_{ff'} \propto y_{ff'}$.
Similar considerations apply to the $z_{ff'}$ since this
coupling breaks the discrete symmetry, $CP$. Loop
corrections to these are therefore also suppressed by the
very small size of SM $CP$-violation.
(See refs.
\mref\cgradcorn\tildec\ for one and two-loop
calculations of effective couplings in
the SM.)

\fig\trianglegraph

\midinsert
\centerline{\epsfxsize=4.5cm\epsfbox{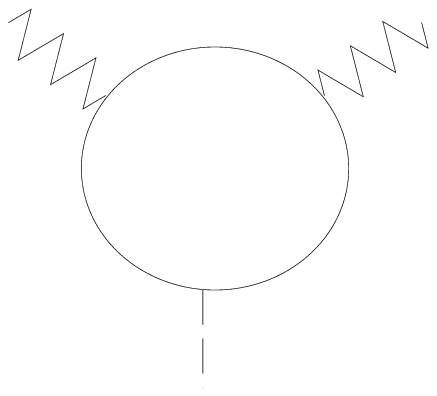}}

\centerline{{\rm Figure \trianglegraph:}}
\medskip
\caption{The Feynman graph
which contributes the leading contribution to the effective
dimension-five vertices of eqs.~\dimfiveops\ and
\dimfiveZgamma\ at one loop.\medskip}
\endinsert

An example of an important SM radiative correction arises
if $\Scl_\eff$ is applied at scales below the top-quark
mass. At these scales the $t$ quark has been integrated
out, and this integration induces the dimension-five
operators of eq.~\dimfiveops, through the fermion loop
of Fig.~\trianglegraph. Using the flavour-diagonal
Yukawa couplings, $y = y_{ff}$ and $z = z_{ff}$, of
eq.~\dimfourops, and working to leading order in the
inverse fermion mass, $1/m$, the effective couplings
which result are:
\eq
\label\genfdimfv
c_k = {y \; \alpha_k \; \Scc_k \over 6 \pi \, m}, \qquad
\tw c_k = {z \; \alpha_k \; \Scc_k \over 6 \pi \, m}
\qquad \hbox{(heavy-fermion loop)} .
\eeq
Here $k = \gamma,g$ denotes either the photon or gluon
and $\alpha_\gamma = \alpha$ and $\alpha_g = \alpha_s$
are their respective fine structure constants.
$\Scc_k$ denotes the quadratic Casimir of the corresponding
gauge generators, $t_a$, as represented on the fermions:
$\Tr(t_a \, t_b) = \Scc \; \delta_{ab}$. Explicitly,
for photons: $\Scc_\gamma = Q_f^2 \; N_c(f)$ where
$Q_f$ is the fermion charge in units of $e$,
and $N_c(f) = 1(3)$ if $f$ is a lepton
(quark); while for gluons: $\Scc_g = \hf$ for quarks (and
$\Scc_g = 0$ for leptons).

The generalization of eq.~\genfdimfv\ to the $h$-$Z$-$\gamma$
effective interaction is straightforward. Starting from $c_\gamma$
or $\tw c_\gamma$
one simply replaces one factor of the fermion charge with the vector part of its
coupling to the $Z$: $e \, Q_f \to e   g_\ssv/ (\sw\cw)$, and multiplies
by 2 to compensate for the fact that the two spin-one particles
are no longer identical. The result is
\eq
\label\Zgammagenf
c_{\ssz\gamma}= 
\left( {y \; \alpha  \over 3 \pi \, m } \right) \; {N_c \, Q_f \, g_\ssv
\over  \sw \cw}, \qquad
\tw c_{\ssz\gamma} = \left( {z \; \alpha  \over 3 \pi \, m } \right) 
\; {N_c \, Q_f \, g_\ssv \over  \sw \cw} \; ,
\eeq
where $g_\ssv$ is normalized such that $g_\ssv = \hf \; T_{3f} - Q_f \sw^2$
for a SM fermion. Here $T_{3f}$ is the third component of the fermion's 
weak isospin. 

Specializing these expressions to the SM tree-level Yukawa
couplings, and including the next-to-leading QCD
corrections \mref\cgradcorn\spiran, finally gives the
$t$-quark contributions $\tw c_g = \tw
c_\gamma = 0$ and:
\eq
\label\tqdimfv
c_g = {\alpha_s \over 12 \pi \, v}
\;\left(1 + {11 \, \alpha_s \over 4 \pi} \right), \quad
c_\gamma = {2 \alpha \over 9 \pi \, v}
\;\left(1 - { \alpha_s \over \pi} \right) \quad \hbox{and} \quad
c_{\ssz\gamma}= {\alpha (1- 8 \sw^2/3) \over  6\pi v \sw \cw}
\left( 1 - {\alpha_s\over\pi} \right).  
\eeq
We quote subleading $\alpha_s$ corrections to these
couplings because these corrections can be numerically
significant. We do so for both the photon and gluon
couplings even though there are other corrections to
the same order in $\alpha_s$ to the gluon coupling
which cannot be absorbed into an overall coefficient
of the effective coupling, $c_g$ \spira. We postpone
our more detailed discussion of these other corrections to
our later applications to $h$ production in hadron colliders.
As we shall see, although it is important that all such
contributions be considered in order to have the complete
QCD corrections, these do not depend on the heavy
degrees of freedom, permitting the heavy physics to
be usefully summarized by the QCD-corrected effective
couplings $c_g$ and $\tw{c}_g$.

\subsubsection{Other SM Contributions to 
$hgg$ and $h\gamma\gamma$ Interactions}

As might be expected given the insensitivity of result
\tqdimfv\ to the heavy-particle masses (in this case
$m_t$), the couplings $c_i$ and $\tw{c}_i$ are potentially
useful quantities for experimentally differentiating
amongst various theoretical models. In order to identify
those contributions which depend on new degrees of
freedom, it is useful to summarize here the other
SM contributions to the processes $h \to gg$ and
$h \to \gamma\gamma$,

For general processes, such as the reaction $e^+e^-
\to h \gamma$ considered in a later section, box graph
and other contributions make it impossible
to summarize all one-loop SM results
as corrections to the $h\gamma\gamma$ vertex.
Furthermore since many of our potential applications
are to energies $\sqrt{s} \gsim \mw,\mz$,
the effects of virtual $W$ and $Z$ particles
do not lend themselves to an analysis in terms
of local effective interactions within some sort of
low-energy effective lagrangian.

\fig\graptres
\topinsert
\centerline{\epsfxsize=9.5cm\epsfbox{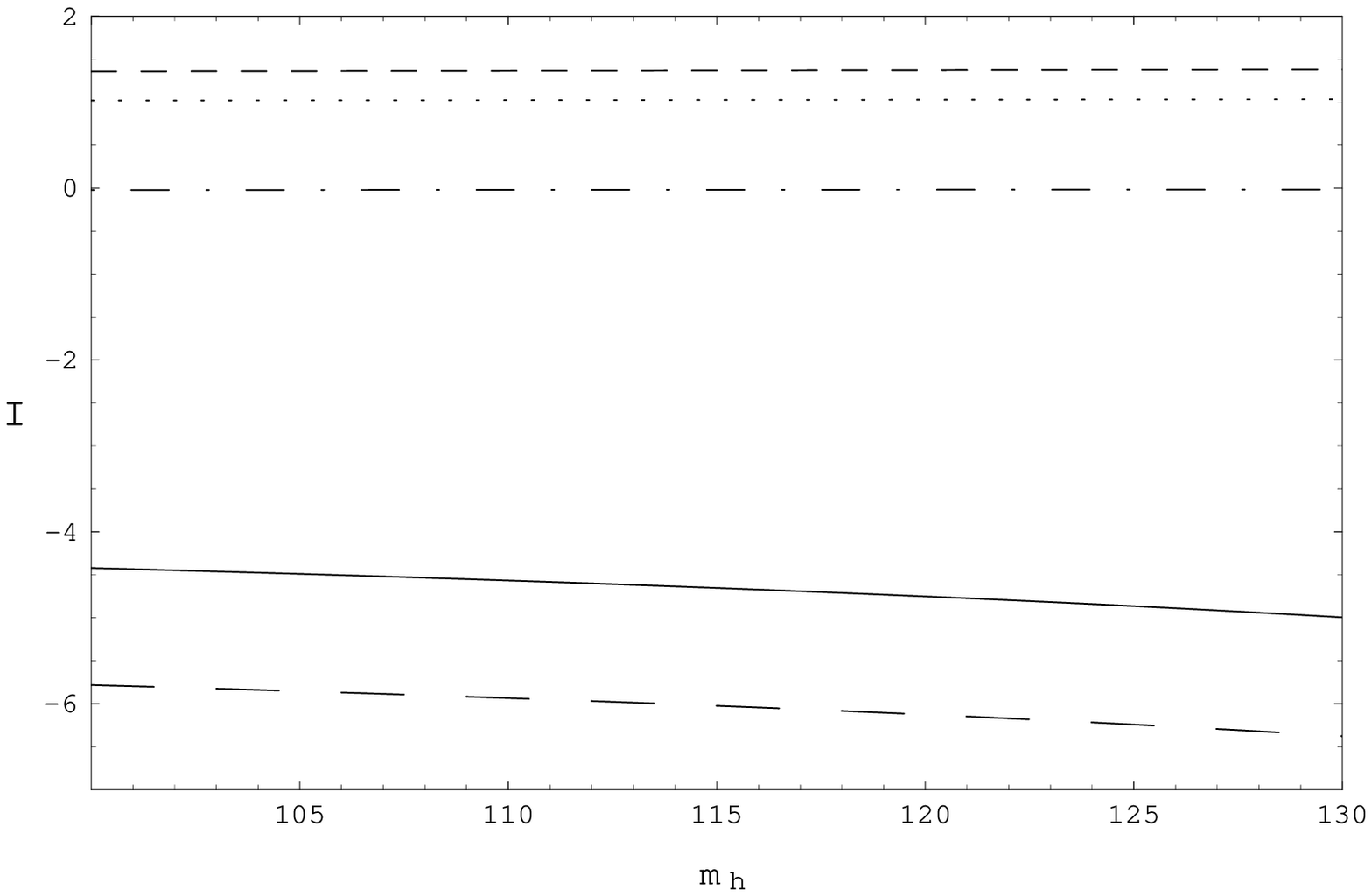}}

\centerline{{\rm Figure \graptres:}}
\medskip
\caption{A comparison of the different contributions to
$\ss c_\gamma$ and $\ss c_g$ as a function of the Higgs mass.
Concerning $\ss c_\gamma$, the long dashed line stands for the
$\ss W$ contribution, $\ss -I_1 \left( {\mw^2 \over \mh^2} 
\right)$, the short-dashed line represents the top 
contribution, $\ss 3 \left( {2\over 3}\right)^2 \; I_\hf\left(
{m^2_t \over \mh^2} \right)$, the dashed-dotted lines 
are the real and imaginary part of the$\ss  b$ quark contribution 
(which practically overlap on this scale). The dotted line is 
the top quark contribution to $\ss c_g$: $\ss I_\hf\left(
{m^2_t \over \mh^2} \right)$.\medskip}
\endinsert

There is an important class of reactions for which
SM loop contributions can be expressed quite generally
in terms of the effective couplings $c_i$ and $\tw{c}_i$,
however. These consist of processes for which both gluons
(or photons) and the scalar $h$ are on shell, such as
the decays $h \to gg$ or $h \to \gamma\gamma$, or
parton-level processes like gluon fusion or
photon-photon collisions. The contribution of light
SM particles, like electrons or light quarks, to
these interactions can be regarded as contributions to
the effective couplings $c_i$ and $\tw{c}_i$
even though an effective lagrangian treatment of the
particles in these processes is not strictly justified.
This is possible because the gauge invariance of the $hgg$ (or
$h\gamma\gamma$) vertex forces it to have the same
tensor structure as have the operators of eq.~\dimfiveops,
up to invariant $q^2$-dependent functions which become constants
when evaluated on shell.

\ref\notetome{
A.I. Vainshtein, M.B. Voloshin, V.I. Sakharov and M. Shifman, {\it Sov. J.
Nucl. Phys.} {\bf 30} (1979) 711.
}

We now record the contributions to $c_i$ and $\tw{c}_i$
which are obtained in this way for the contributions
of SM particles. Evaluating the contribution of spins
0, $\hf$ and 1 to the one-loop graph, Fig.~\trianglegraph,
or differentiating the vacuum polarization with
respect to the Higgs {\it v.e.v.} gives $\tw c_k = 0$ and:
\eq
\label\SMcforms
\eqalign{
c_g^{\sss SM} & =  {\alpha_s \over 12 \pi \, v}
\sum_q  I_\hf\left[{m^2_q \over \mh^2} \right] ,\cr
c_\gamma^{\sss SM} & =
c_\gamma^{\sss SM}(\hbox{up-type fermions})+
c_\gamma^{\sss SM}(\hbox{down-type fermions})
+ c_\gamma^{\sss SM}(W) \cr
&=
{\alpha \over 6 \pi \, v}\; \left[
\sum_{q} 3 Q_q^2 \; I_\hf\left(
{m^2_q \over \mh^2} \right)
+ \sum_{\ell}  \; I_\hf\left(
{m^2_\ell \over \mh^2} \right) -
I_1 \left( {\mw^2 \over \mh^2} \right)
\right],\cr}
\eeq
where the spin-dependent functions, $I_s(r)$, are given by
ref.~\cgradcorn,\spira,\notetome,:
\eq
\label\gcfnresults
\eqalign{
I_0(r)  &= 3r \, \Bigl[1 + 2r \, f(r) \Bigr] \to -\,
\nth4 + O\left({1\over r}\right),\cr
I_\hf(r) &= 3\, \Bigl[2r + r(4r -1) \, f(r) \Bigr]
\to 1 + O\left({1\over r}\right),\cr
I_1(r)  &= 3\, \left[3r + \hf - 3r(1-2r)
\, f(r) \right] \to {21\over 4} +
O\left({1\over r}\right) . \cr}
\eeq
The large-mass limit is displayed explicitly in these
expressions, and the function $f(r)$ is given by
\eq
\label\defoff
f(r) = \left\{ \matrix{-2 \left[ \arcsin \left(
{1 \over 2 \sqrt{r}} \right) \right]^2 &
\hbox{if}\quad r > \nth4 ; \cr
\hf \, \left[ \ln \left( {\eta_+ \over
\eta_-} \right) \right]^2 - {\pi^2 \over 2}
+ i \pi \ln \left( {\eta_+ \over \eta_-}
\right) & \hbox{if}\quad r < \nth4 ; \cr}
\right. \eeq
with $\eta_\pm = \hf \pm \sqrt{\nth4 - r}$.
Similar expressions for the $c_{\ssz\gamma}$ couplings can be found in 
   ref. [2].

Notice that eqs.~\SMcforms\ reduce to expressions \tqdimfv\
when specialized to only the $t$-quark, in the large-$m_t$ limit.
Notice also that, in contrast with the cancellation of the $m_t$
dependence in the ratio $y_t/m_t = 1/v$
in eq.~\tqdimfv, the contribution of light-particle loops to
scalar-photon and scalar-gluon couplings are
suppressed by a power of the light mass over $\mh$.
As a result, the total
scalar-photon coupling tends to be dominated by loops
containing heavy particles, for which the effective
lagrangian description is quite good.

Some remarks are in order about the numerical size of the
various contributions (see Fig.~\graptres), for a light Higgs
 between 100 to 130 GeV. The
$W$ contribution always dominates, with
$-I_1 \left( {\mw^2 \over \mh^2} \right)$
ranging between
$-5.78$ and $-6.39$ between these Higgs masses, while the
lowest-order top quark contribution $3 {\left({2
\over 3}\right)}^2 I_\hf \left(
m^2_t \over \mh^2 \right)$ ranges between $1.36$ and $1.38$.
QCD corrections and  the $b$ quark contribution are also
numerically significant in what follows, since even though they
are small -- less than $10\%$ of the top-quark contribution -- new physics
contributions are typically of the same order.

\section{Connecting to Observables: Production and Decay}

Given the assumption that all new particles (apart from
the  hypothetical newly-discovered scalar) are heavy, {\it
all} underlying models must reduce, in their low-energy
implications, to the effective  lagrangian of the previous
section. It follows that if  empirical access is limited to
this low-energy regime, then measurements of the effective
couplings provide the only possible information
available  with which to experimentally distinguish the
various underlying possibilities.

There are therefore two key questions.
\item{Q1.}
How are the effective
couplings best measured? That is, which
experimental results depend on which of the effective
interactions?
\item{Q2.}
What do measurements of the effective
interactions teach us about the underlying physics? (That
is, how do the effective couplings depend on the more
fundamental couplings of the various possible underlying
models?

It is the purpose of the next two sections to make the connection
between $\Scl_{\rm eff}$ and observables, and connections
to underlying models are made in \S5. Since any
unambiguous observation of the new scalar particle(s)
involves the detection of their production and decay, this
section starts with these two kinds of processes.
The discussion of  the indirect influence of virtual scalars
on interactions involving other particles is
the topic of the following section, \S4. Since these
virtual effects provide important constraints on the nature
of the scalar particles, the bounds they imply are also
included in \S4.

\subsection{Scalar Decays}

The dominant scalar decays are described by those
interactions in $\Scl_{\rm eff}$ which are linear in the
new scalar field(s).  There are three kinds of such terms,
giving couplings to fermions, massive gauge bosons and
massless gauge bosons. Which of these gives the dominant
scalar decay mode depends on the relative size of the
corresponding effective couplings, making an experimental
study of the branching  ratios for the various kinds of
decays a first priority once such a scalar is discovered.
\bigskip

\subsubsection{Decays to Fermions}

Consider first scalar decay into a fermion antifermion
pair. The scalar rest-frame differential decay rate into
polarized fermion pairs, $h \to f(p,s)\ol{f'}
(\pbar ,\sbar)$, depends on the effective Yukawa couplings
in the following way:
\eq
\label\polfdecay
\eqalign{
{d\Gamma_{\rm pol} \over d^3p } &=
{N_c \over 32 \pi^2  \mh  E \ol{E}  } \;
\Bigl[ (|y|^2 + |z|^2) (-p\cdot \pbar - m \mbar \;
s \cdot \sbar) + (|y|^2 - |z|^2) (p \cdot \sbar  \;
\pbar \cdot s  - p \cdot \pbar  \; s \cdot \sbar) \cr
&\qquad  +i(y z^* - y^* z) (m  \; s \cdot \pbar +
\mbar  \; \sbar \cdot p)   \Bigr] .\cr
}\eeq
Here $E$ ($\ol{E}$) is the energy of the fermion
(antifermion) in  the rest frame of the decaying scalar,
while $m$ and $\mbar$ are their masses, and
$s^\mu$ and $\sbar^\mu$ are their spin four-vector.  $N_c$
is a colour factor, given by $N_c =3$ if the daughter
fermions are quarks,  and by $N_c=1$ otherwise.

$y = y_{ff'}$ and $z = z_{ff'}$ denote the relevant
effective Yukawa couplings of $\Scl^{(4)}_{\rm fermion}$.
Notice that it is in principle possible to measure
separately the modulus of both $y$ and $z$, as well as
their relative phase, so long as both the polarizations and
decay distributions  of the daughter fermion are measured.

Unfortunately, in the most likely scenario it will not be
possible to measure the polarizations of the daughter
fermions. In this case rotation invariance ensures the
decay is isotropic in the decaying scalar's rest frame,
leaving only the total unpolarized partial rate as an
observable for each decay channel. In this case the
rest-frame partial decay rate becomes:
\eq
\label\fdecayrate
\Gamma_{f f'} = {N_c \mh \over 8 \pi}
\left[ (|y|^2 + |z|^2) (1 - 2r_+) - \,
{2 m \mbar \over \mh^2} \;  (|y|^2 - |z|^2) \right]
(1 - 4r_+ + 4 r_-^2)^\hf  ,
\eeq
where $r_\pm = (m^2 \pm \mbar^2)/(2 \mh^2)$. Clearly a
measurement of $\Gamma_{ff'}$ only is insufficient in itself to measure
{\it both} $|y|$ and $|z|$  separately.

\subsubsection{Decays to $W$'s and $Z$'s}

If $\mh > 160$ GeV, then decays into pairs of electroweak
gauge bosons are possible. In this case the rest-frame rate
for decays into polarized  bosons,
$h \to W^-(p,s)W^+(\pbar,\sbar)$, is:
\eq
\label\polgbdecay
{d \Gamma_{\rm pol} \over d^3 p } =
{|a_\ssw|^2 \over 32 \pi^2 \mh E^2} \; \Bigl( s\cdot
\sbar  \Bigr) .
\eeq
Here $s^\mu$ and $\sbar^\mu$ are the polarization vectors
for the daughter gauge bosons.

Since eq.~\polgbdecay\ shows that the measurement of
the $W$ polarizations in this decay gives no additional information
about the values of the effective couplings, we specialize
to the unpolarized partial rate for this decay, which is:
\eq
\label\gbdecay
\Gamma_{WW} = {|a_\ssw|^2 \over 64 \pi}
\left( {\mh^3 \over M_\ssw^4} \right)
\left( 1 - 4 \;{ M_\ssw^2 \over \mh^2} +
12 \; { M_\ssw^4 \over \mh^4} \right)
\left( 1 - 4 \; {M_\ssw^2 \over \mh^2} \right)^\hf.
\eeq

\ref\kmar{T.G. Rizzo, \prd{22}{80}{389}; W.-Y.Keung and W.J. Marciano,
\prd{30}{84}{248}.}
\ref\cahn{R.N.Cahn, {\it Rept. Prog. Phys.} {\bf 52} (1989) 389}

Similarly, if $\mh > 180$ GeV then the decay $h \to ZZ$ is
allowed.  The expression for the partial rate for this decay
is given by making the replacements $\mw \to \mz$ and
$|a_\ssw|^2 \to \hf \; |a_\ssz|^2$ in expression \gbdecay.

$h$ decay into massive gauge bosons with the $W$'s or $Z$'s off-shell
can also be important, especially for a light Higgs. Under certain circumstances
this may be obtained straightforwardly from the SM result 
\mref\spiran\cahn.
For example, if only the trilinear scalar/gauge-boson couplings are important,
then the decay rates may be obtained simply by multiplying the SM
expressions given in \spiran\ or \cahn\ by the overall factor
$|a_\ssw/a^{SM}_\ssw|^2$ (or $|a_\ssz/a^{SM}_\ssz|^2$ as appropriate),
where the SM couplings, $a^{SM}_\ssw$ and $a^{SM}_\ssz$, are given
explicitly by eq.~\SMdthree. An example of where this procedure could fail
would be the final state $W/Z + f \ol{f'}$, say, if the effective Yukawa
couplings ($y$ and $z$) are important, since this requires the inclusion of
diagrams directly coupling the scalar to fermions which are usually
neglected in the SM.

\subsubsection{Decays to Photons}

Decays of neutral scalars
may be described in terms of the dimension-five
operators of eq.~\dimfiveops. The decay
rate into polarized photons, $h \to
\gamma(p,\lambda) \gamma(\tw p,\tw\lambda)$, is most simply
computed in a gauge for which $p\cdot s = p \cdot \tw s =
\tw p \cdot s =
\tw p \cdot \tw s = 0$, where
$s^\mu(p,\lambda)$ and $\tw s^\mu(\tw p,\tw\lambda)$ are
the photon polarization vectors, with $\lambda$ and $\tw\lambda$
their helicities. The result is isotropic in
the scalar rest frame, with rate:
\eq
\label\twphpoldc
\Gamma_{\rm pol}(h \to \gamma \gamma) =
{\mh^3 \over 8\pi} \; \left[\left|c_\gamma \right|^2 \;
\left( 1 - | s\cdot \tw s |^2 \right) +
\left|\tilde{c}_\gamma \right|^2 \;  | s\cdot \tw s |^2
\right].
\eeq
Measurement of the photon polarization therefore permits,
in principle, a
disentangling of the two relevant couplings, $
|c_\gamma|^2$ and $|\tilde{c}_\gamma |^2$.

The unpolarized decay rate may be computed
straightforwardly, giving:
\eq
\label\twophdecay
\Gamma(h \to \gamma \gamma) = {\mh^3 \over 4\pi} \;
\left( \left|c_\gamma \right|^2 + \left|\tilde{c}_\gamma
\right|^2  \right).
\eeq

\subsubsection{$Z\gamma$ decays}

The decay of the neutral scalar into a photon and $Z$-boson may occur
if $\mh>m_Z$. The unpolarized decay rate, calculated with the use of
the effective interaction \dimfiveZgamma, is given by
\eq
\label\Zgammadecay
\Gamma(h \to Z \gamma)= {\mh^3 \over 8\pi}
\left(1-{\mz^2 \over \mh^2} \right )^3 
\left( \left| c_{\ssz\gamma} \right|^2 + 
\left|\tilde c_{\ssz\gamma} \right|^2 \right).
\eeq

\subsubsection{Inclusive Decays into Hadrons}

\ref\inclusive{G. Sterman and S. Weinberg, \prl{39}{77}{1436}}

The dimension-five operators of eq.~\dimfiveops\ also
include $h$-gluon interactions. These are more difficult to
relate to exclusive decay rates because of the extra
complication of performing the hadronic matrix element of
the gluon operators. Such matrix-element complications do
not arise for inclusive decays, however \inclusive, which
we therefore describe here.

For states like our hypothetical scalars,  which are much
more massive than the QCD scale, the total hadronic decay
rates are well approximated by the  perturbative sum over
the partial rates for decays into all possible quarks and
gluons.  In the present instance this implies:
\eq
\label\haddecay
\Gamma(h \to {\rm hadrons}) = \Gamma_{qq'}+
\Gamma_{qt}+ \Gamma_{tt}+\Gamma_{\rm gluons} ,
\eeq
where we divide the sum over quarks into those involving
two, one or no top quarks, since the top-quark
contributions can arise only for sufficiently massive
scalars.

Neglecting light quark masses,
the quark decays are given by eq.~\fdecayrate:
\eq
\label\qdecays
\eqalign{
\Gamma_{qq} &= {3 \mh \over 8 \pi} \sum_{qq'}
\left(\left|y_{qq'}\right|^2 + \left|z_{qq'}\right|^2
\right) , \cr
\Gamma_{qt} &= \Gamma(h\to q\ol{t}) +
\Gamma(h \to t \ol{q}) = {3 \mh \over 4 \pi} \sum_{q}
\left( \left|y_{qt}\right|^2 +  \left|z_{qt}\right|^2
\right)
\left( 1 - {m_t^2 \over \mh^2} \right)^2 , \cr
\Gamma_{tt} &= {3 \mh \over 8 \pi}
\left[ \left|y_{tt}\right|^2  \left( 1 - {4 m_t^2 \over
\mh^2}
\right) + \left|z_{tt}\right|^2 \right]
\left( 1 - {4 m_t^2 \over \mh^2} \right)^\hf . \cr }
\eeq
where the expression for $\Gamma_{qt}$ uses the Hermiticity
property $y_{qt} = y^*_{tq}$, which follows from the reality
of $\Scl_{\rm eff}$.

Keeping in mind the gluon colour factor, $N = 8$, the
decay to gluons is  given by the analogue of eq.~\twophdecay:
\eq
\label\gluondecay
\Gamma_{\rm gluons} = {2 \mh^3 \over \pi} \;
\left( \left|c_g \right|^2 + \left|\tilde{c}_g \right|^2
\right).
\eeq

Clearly, depending on the size of the various effective
couplings, decay processes such as these can be used to
determine the magnitudes of the couplings to gauge bosons
and leptons, as well as some information about the Yukawa
couplings to quarks. Disentangling the  couplings to
different quark flavours requires a separation of the
hadronic decays into specific exclusive decay modes.
Although this can be cleanly done for heavy quarks ---
$c,b,t$, say --- it will  inevitably be complicated by
hadronic matrix-element uncertainties  for light quarks and
gluons.

\subsection{Scalar Production (Electron Colliders)}

Scalar particle detection will provide information about
the effective couplings which contribute to the scalar
production, in addition to the information which may be
extracted by studying the scalar decays. This section is
devoted to summarizing the production rates which arise if
the scalars are produced in electron (or muon) colliders,
like  SLC.

\fig\prdngraphs

\midinsert
\centerline{\epsfxsize=8.5cm\epsfbox{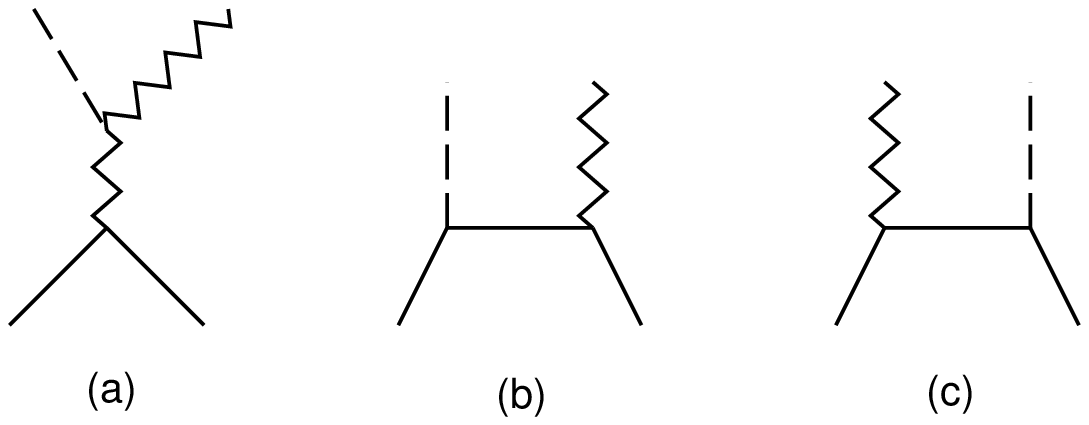}}

\centerline{{\rm Figure \prdngraphs:}}
\medskip
\caption{The Feynman graphs
which contribute the leading contribution to the reaction
$\ss f \ol{f} \to h V$, for $\ss V = Z,\gamma$. For $\ss V=Z$ the
$\ss hZZ$ vertex is as given by eq.~\dimtwoops, while
for $\ss V=\gamma$ the $\ss h\gamma\gamma$ vertex comes
from eq.~\dimfiveops.\medskip}
\endinsert

In electron machines neutral scalars can be emitted by
any of the participants in the basic SM reaction. Since we
imagine the scalars to be too heavy to be themselves
directly produced  in $Z^0$ decays, or to be produced in
association with two gauge bosons, the main mode of
single-scalar production is then due to the reactions
$e^+e^- \to h \, Z$ or $e^+e^- \to h \, \gamma$, with
the subsequent decay of the final $h$ (and $Z)$. The
lowest order contributions to these processes arising
within the effective theory correspond to the
Feynman graphs of Fig.~\prdngraphs.

Since the reactions differ in their detailed features depending on
whether it is a $\gamma$ or $Z$ which accompanies the scalar,
we now consider each case separately. In order to use these
results in later applications, we do not immediately specialize
to electrons in the initial state, quoting instead our expressions
for the more general process $f\ol{f} \to Vh$ (with $V = Z$
or $\gamma$), with an arbitrary
initial fermion.

\subsubsection{The Reaction $f \ol{f} \to Zh$}

\fig\zhprodvs
 
\topinsert
\centerline{\epsfxsize=6.8cm\epsfbox{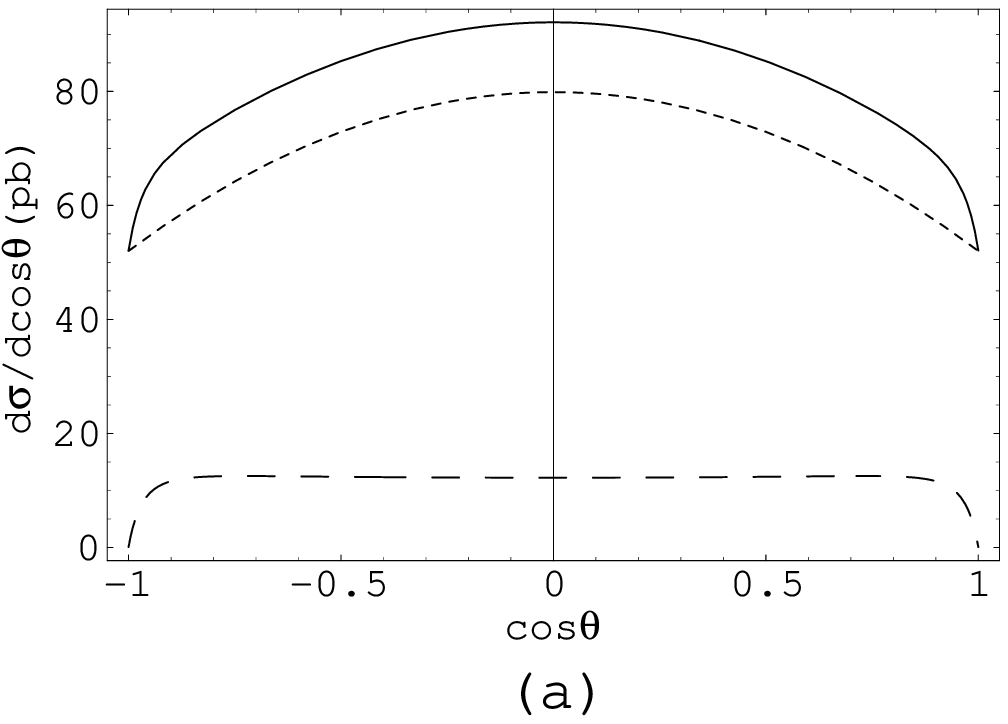}
\epsfxsize=6.8cm\epsfbox{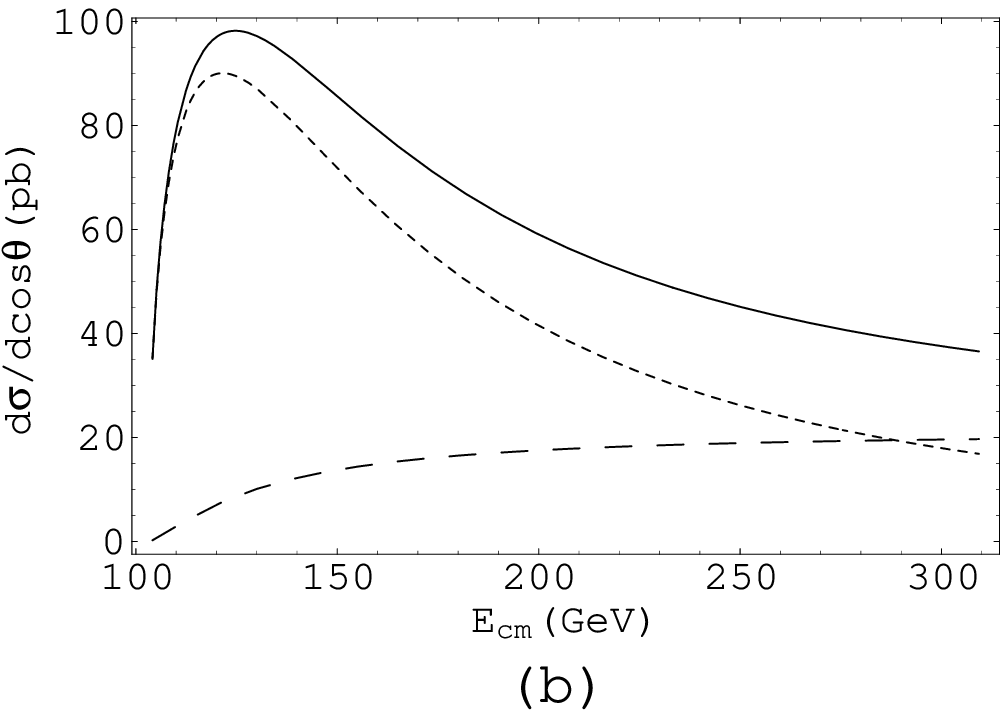}}
 
\centerline{{\rm Figure \zhprodvs:}}
\medskip
\caption{
Reaction  $\ss e^+e^- \to Zh$: (a) Differential production
cross section as a function of $\ss \theta$, the CM scattering angle. 
The
figure assumes the electron
has 140 GeV in the CM frame, as well as
SM fermion-$\ss Z$ couplings,
a scalar mass $\ss \mh = 115$ GeV, and
the effective couplings $\ss a_\ssz = e \mz/\sw\cw$,
and $\ss |\Scy|^2 = |y|^2 + |z|^2 = 0.01 \; e^2$.
The short-dashed line shows the contribution where the
$\ss h$ is emitted from the $\ss Z$ line, and the long-dashed line
gives the same with emission from either of the initial fermions.
Their sum is represented by the solid line.
(b) Differential production cross section
evaluated at $\ss \cos\theta=0$ as a function of the energy. Same
couplings 
and significance
of the lines as in (a).\medskip}
\endinsert

We give the results from evaluating the graphs of
Fig.~\prdngraphs\ using the dimension-three effective coupling,
of eqs.~\dimtwoops, for the $ZZh$ vertex.\foot\nophots{We 
do not include graph $\ss (a)$ with an intermediate photon,
using interaction \dimfiveZgamma\ because this interaction has higher
dimension than the one used. For many models it is also suppressed
by loop factors. 
This neglect should be borne in mind when handling
models for which the couplings $\ss a_\ssz^i$ are suppressed
to be of the same order of $\ss c^i_{\ssz \gamma}$.}
We also work in the limit of vanishing mass for the 
initial-state fermion, and use unitary gauge for the internal $Z$
boson. (Notice that in the present context vanishing
fermion mass is {\it not} equivalent to vanishing scalar Yukawa
couplings.)
In the approximation that we neglect the fermion masses,
graph ($a$) does not interfere with graphs
($b$) and ($c$) due to their differing helicity structure. We
find therefore:
\eq
\label\sigmasum
{d\sigma \over du dt} \; (f \fbar \to h Z) =
{d\sigma_a \over du dt} + {d\sigma_{bc} \over du dt}  ,
\eeq
with
\eq
\label\dsgmforma
{d\sigma_a \over du dt} =  { \alpha
\left| a_\ssz \right|^2 (\gl^2 + \gr^2)  \over 16
\sw^2 \cw^2 \; s^2} \; \left[  { s + (t - \mz^2)
(u - \mz^2) /\mz^2 \over \left| s - \mz^2 +
i \Gamma_\ssz \, \mz \right|^2 } \right]
\delta(s + t + u - \mh^2 - \mz^2) ,
\eeq
and
\eq
\label\dsgmformbc
\eqalign{
{d\sigma_{bc} \over du dt}  &=
{ \alpha \; |\Scy|^2  \over 8 \sw^2 \cw^2
} \;
\left\{  { (\gl - \gr)^2 \over \mz^2 \, s}  +
(\gl^2 + \gr^2)  \left( {1 \over t^2 } +
 {1 \over u^2} \right)  \left( { ut - \mh^2
\mz^2 \over s^2 } \right)   \right. \cr
&\qquad \qquad \left. + { 4 \gl \gr \over ut s^2}
\Bigl[ ut + \mh^2 ( s - \mz^2) \Bigr] \right\}
\delta(s + t + u - \mh^2 - \mz^2) . \cr}
\eeq
Here $s, t$ and $u$ are the usual Mandelstam variables,
with $t = - (p - k)^2$ where $p^\mu$ and $k^\mu$ are
the 4-momenta of the incoming electron and outgoing $Z$
boson. The constants
$\gl$ and $\gr$ are the effective couplings of the fermion
to the $Z$, normalized so that their SM values would be:
$\gl^\SM = T_{3f} - Q_f \, \sw^2$ and
$\gr^\SM = - Q_f \, \sw^2$. $a_\ssz$
and $\Scy = y_{ff} + i z_{ff}$ similarly denote
the relevant effective couplings of $h$ to the $Z$
and the fermion.

These imply the following expressions for the integrated
cross section, $\sigma = \sigma_a + \sigma_{bc}$:
\eq
\label\sgmforma
\sigma_a =  { \alpha
\left| a_\ssz \right|^2 (\gl^2 + \gr^2)  \over 96
\sw^2 \cw^2 \; \mz^2} \; \left[  { \lambda^{1/2} \;
(\lambda + 12 \, s \, \mz^2)  \over
s^2 \; (s - \mz^2)^2} \right]  ,
\eeq
and
\eqa
\label\sgmformbc
\sigma_{bc}  & =
{ \alpha \; |\Scy|^2  \over 8 \sw^2
\cw^2 } \;
\left\{ \left[ {(\gl - \gr)^2 \; s\over \mz^2}
- 4 (\gl^2 + \gr^2 - \gl \; \gr) \right] {\lambda^{1/2}
\over s^2}
\right.\eol
&\qquad \left.
- \; {2 \over s^2} \; \left[ (\gl^2 + \gr^2) (s - \mh^2 -
\mz^2)
+ {4 \gl \, \gr \, \mh^2 (s - \mz^2) \over
s - \mh^2 - \mz^2 } \right] \ln \left|
{\mz^2 + \mh^2 - s + \lambda^{1/2} \over
\mz^2 + \mh^2 - s - \lambda^{1/2} } \right| \right\}
,\eeolnn
\eeq
where $\lambda = (s - \mh^2 - \mz^2)^2 - 4 \mz^2 \mh^2$.

Some of the implications of these expressions are illustrated by
Fig.\zhprodvs-a and  Fig.\zhprodvs-b, which plot the dependence
of the cross section on the electron's centre-of-mass (CM) energy
and on the CM scattering angle between the outgoing $Z$ and the
incoming electron. Inspection of these plots reveals
the following noteworthy features:

\item{1.} {\sl In general all three graphs, ($a$), ($b$) and ($c$), are
required.}

It is common practice to only consider graph ($a$) when
computing the $Zh$ production rate within the Standard
Model and many of its popular extensions.
This is because the electron-scalar coupling in these models is
proportional to the electron mass, and so is negligibly
small. Indeed, expressions \dsgmforma\ and \sgmforma\
reproduce the SM results once the replacement for $a_\ssz$
from eq.~\SMdthree\ is made.
The neglect of diagrams ($b$) and ($c$), which have
scalar emission  occuring from the electron lines, in
comparison with graph ($a$) is not always
\apriori\ justified, however, since models exist for which
the electron Yukawa couplings are not so small. Since one of the central
issues requiring addressing should a new scalar be found
is precisely the question of whether its Yukawa couplings
are related to masses, we do not prejudge the result here,
and so keep all of graphs ($a$), ($b$) and ($c$).

\item{2.} {\sl Graphs ($a$) and ($b$), ($c$) differ in the $\cos\theta$
dependence they predict.}

According to Fig.~\zhprodvs-a, graphs ($a$) and graphs
($b$) and ($c$) differ in the dependence on CM scattering
angle they predict for the $Zh$ production cross section.
In principle, given sufficient accuracy, this difference could
be used to distinguish the two kinds of contributions from
one another experimentally. The nature of this difference depends on
the value of $\mh$, with scalar-{\it strahlung} from the initial
fermions peaking more strongly about $\cos\theta =
\pm 1$ for smaller scalar masses.

\item{3.} {\sl Graphs ($b$) and ($c$) predict strongly rising
dependence on energy.}

It can happen that energy dependence furnishes a
more useful discriminator
between the two kinds of production processes, as is illustrated
by Fig.~\zhprodvs-b. The high-energy limit of the $Zh$ production
cross section depends sensitively on the form of the Yukawa
couplings, as may be seen from the growth of the cross section
which $\sigma_{bc}$ predicts for $s \gg \mz^2 , \mh^2$.

This strongly-rising cross section is typical of theories
which involve massive spin-one particles which are
not gauge bosons for  linearly-realized gauge symmetries
\unitarityviolation, \usesandabuses. Notice, for instance, that it would not
arise in $\gamma h$ production (as we shall shortly see
explicitly) because the singular term at
high energies is proportional to $(\gl - \gr)^2$, which vanishes
for photons. The singular behaviour does not arise in the
SM because of a cancellation between the contribution of
graphs ($b$) and ($c$) with the fermion-mass
dependence --- which is neglected here --- of graph $(a)$.
Such a cancellation is possible within the SM
because the linearly-realized $SU_\ssl(2) \times U_\ssy(1)$
gauge invariance relates the Higgs yukawa coupling, $y_f$,
to the fermion masses.

Since we do not assume \apriori\ that our hypothetical new
scalar falls into a simple $SU_\ssl(2) \times U_\ssy(1)$
multiplet with the other known particles, we cannot assume
that similar cancellations occur between eq.~\sgmforma\ and
\sgmformbc\ in our effective theory when $s \gg \mz^2$.
Indeed, the failure of these cancellations, if seen, would be good news.
The unitarity violations which follow from this failure at
sufficiently high energies mean that the low-energy
approximation used to make sense of the effective
lagrangian is breaking down. And this means that the
threshholds for the production of more new particles must
be encountered before this occurs. If we should find
ourselves lucky enough to experimentally see such strongly
rising cross sections, we could confidently expect the
discovery of further new particles to follow the new scalar
particle under discussion here.
\bigskip

\subsubsection{The Reaction $f \ol{f} \to \gamma h$}

Under certain circumstances the contributions to
$\gamma h$ production may be computed by
evaluating Figs.~\prdngraphs\ using couplings taken
from eqs.~\dimfourops\ and \dimfiveops. In this
section we state the necessary circumstances
for these equations to apply, and give simple expressions for the result
which follows in many cases of interest.

Generally, use of effective couplings is
justified provided the momentum flowing into
the effective vertex is sufficiently small
compared with the scale of the physics which
was integrated out to produce the effective
theory. For example, if the effective operators
of eq.~\dimfiveops\ are obtained by integrating
out a loop involving a particle of mass $m_f$,
then use of the effective coupling in low energy
processes amounts to the neglect of corrections
of order (external momenta)/$m_f$.

If all of the external scalar and photons
(or gluons) are on shell, then the only invariant
external mass scale is set by $\mh$, permitting
an effective calculation so long as relative contributions
of order $\mh^2/m_f^2$ are negligible. This is the case
for the $h$ decays considered earlier, for example,
as well as for gluon-gluon or photon-photon fusion
within hadron colliders in some regimes of energy
and scattering angle.
For $h\gamma$ production in $e^+e^-$ machines,
however, the virtual boson can be strongly off-shell
and so a calculation in terms
of an effective operator is only justified
up to corrections of order $Q^2/m_f^2$,
where $Q^2$ is the invariant momentum transfer
carried by the virtual particle.
\fig\phhprodvs
\topinsert
\centerline{\epsfxsize=6.8cm\epsfbox{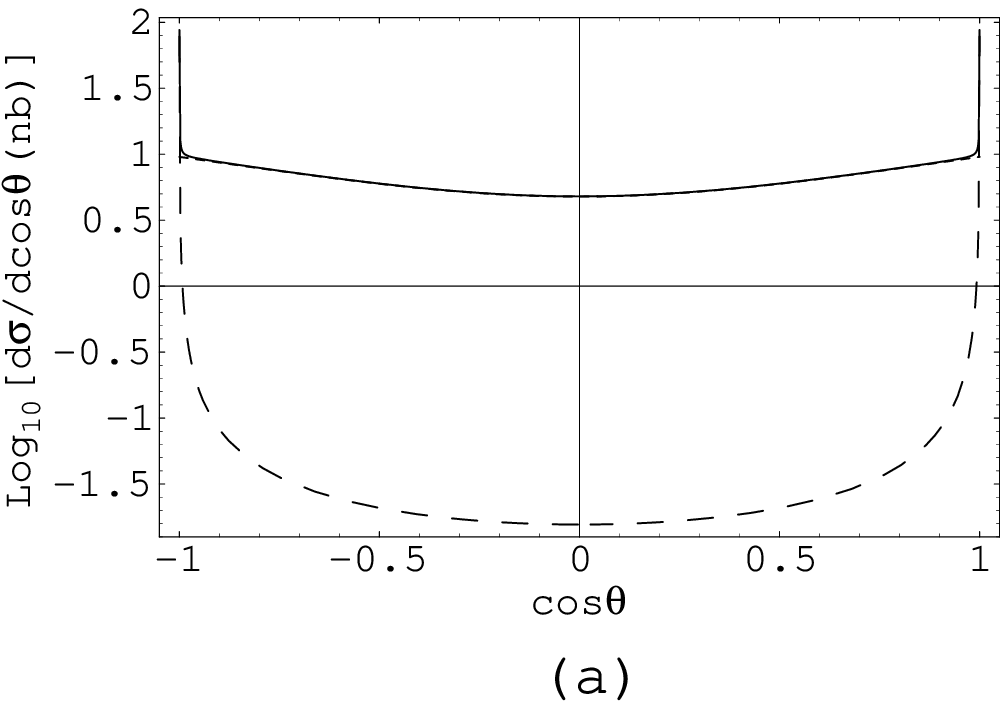}
\epsfxsize=6.8cm\epsfbox{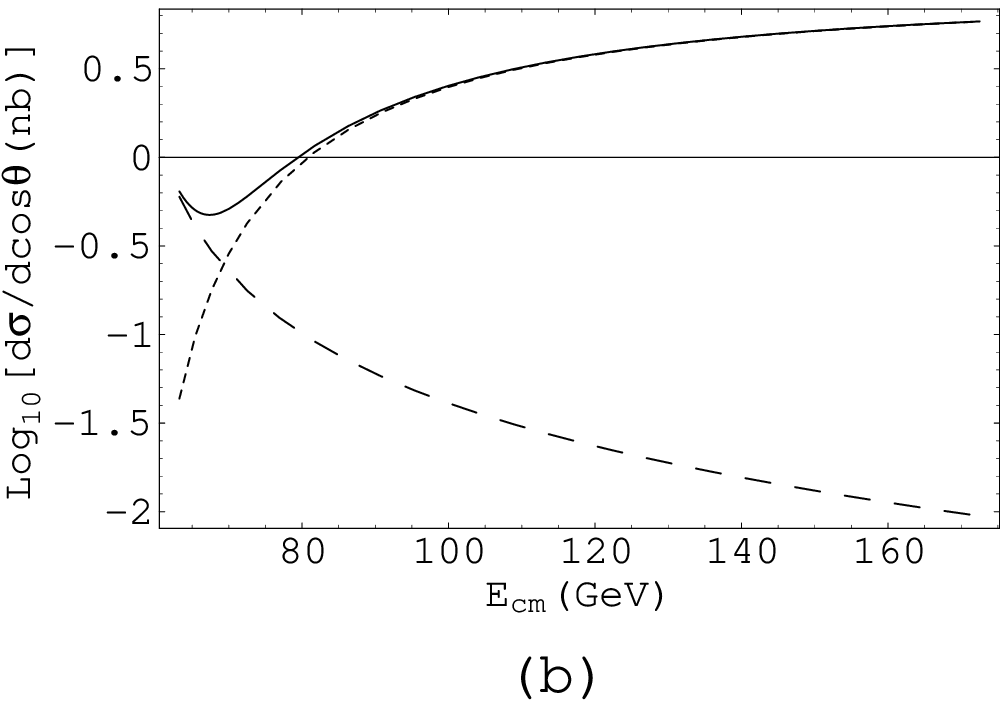}} 
%
\centerline{{\rm Figure \phhprodvs:}}
\medskip
\caption{The new-physics part of the differential production
cross section ($\ss \tw{c}_\gamma$ and yukawa) for the
reaction $\ss e^+e^- \to \gamma h$: (a)
as a function of CM scattering angle $\ss \cos\theta$
evaluated at a CM energy of 140 GeV. The
figure assumes QED fermion-photon couplings,
a scalar mass $\ss \mh = 115$ GeV, effective
 couplings $\ss \tw{c}_\gamma = 1/(246 \GeV)$,
$\ss \tw{c}_{Z\gamma} = 0$
and $\ss |\Scy|^2 = |y|^2 + |z|^2 = 0.01 \; e^2$.
The short-dashed line corresponds to
the $\ss h-\gamma$ vertex
contribution, long-dashed is the bremstrahlung contribution and the solid
line stands for the total. And (b) as a function of the CM Energy
and evaluated at  $\ss \cos\theta=0$ with same couplings and
input parameters.\medskip}
\endinsert 
\ref\hgamSMa{A. Barroso, J. Pulido and J.C. Rom\~ao,
\npb{267}{93}{1}.}
\ref\hgamSMb{A. Abbasabadi, D. Bowser-Chao,
D.A. Dicus and W.W. Repko,
\prd{52}{95}{3919}.}
The contributions to $f \ol f \to h \gamma$ 
of effective operators like $c^i_\gamma$,
$\tw c^i_{\gamma}$, $c^i_{\ssz\gamma}$ and
$\tw c^i_{\ssz\gamma}$ are more difficult to 
compute in a model-independent way 
if it happens that they can interfere with other graphs. Indeed,
experience with specific models shows that this often happens,
since the couplings of these dimension-five effective 
interactions are usually suppressed
by loop factors, and so embedding them into tree graphs gives
results which can interfere with other one-loop graphs. For 
instance, even though the contribution to $f\ol f \to h \gamma$
of a heavy top quark in the SM is well described by inserting
the effective coupling $c_\gamma^h$ of eq.~\tqdimfv\ into graph
($a$) of fig.~\prdngraphs, the result interferes with other
amplitudes, such as loop graphs involving $\gamma h$ emission
from a virtual $W$ boson \hgamSMa, \hgamSMb. This same
interference can happen more generally, such as with a loop graph
involving $h \gamma$ emission from a virtual $W$ (using
the effective coupling $a_W$, say). 

There are important cases for which this kind of interference
does not occur, and so where a simpler statement of the 
$c^i_{g,\gamma}$ and/or $\tw c^i_{g,\gamma}$ contributions 
to $f \ol f \to h \gamma$ can be made. Interference might be
forbidden, for example, by approximate symmetries like
CP invariance. As we shall see, CP invariance generally requires
the vanishing of the couplings $y^h_{fg}$, $a^h_\ssw$,
$a^h_\ssz$ and $c^h_{g,\gamma}$
for a CP-odd pseudoscalar, $h$, but permits nonzero couplings 
of the type $z^h_{fg}$ and $\tw c^h_{g,\gamma}$. In this
case only the graphs of fig.~\prdngraphs\ play any role
in $f \ol f \to h \gamma$, and so the cross section 
may be directly evaluated from these with the result
$d\sigma(f\fbar\to \gamma h) = d\sigma_\ssv
+ d\sigma_{\rm yuk}$, where
\eq
\label\dhphdta
\eqalign{
{d \sigma_\ssv \over dt \, du} &=
{\alpha (t^2+u^2) \over s^3} \;
\left[ \left| Q_f \; \tw{c}_\gamma
+\left( {g_\ssv \tw{c}_{\ssz\gamma} \over \sw\cw}\right)~
{s\over s- \mz^2+i\Gamma_\ssz \mz}\right|^2 \right. \cr
&\qquad\qquad \left. 
+ \left( {g_\ssa^2 \tw{c}_{\ssz\gamma}^2 \over \sw^2 \cw^2 }\right)
\; \left| {s \over  s  - \mz^2+i\Gamma_\ssz \mz }\right|^2
\right] \; \delta(s + t + u - \mh^2) ,\cr  }
\eeq
and we have included the contributions of both virtual
photon and virtual $Z$ exchange. Here the effective $Z$-fermion
couplings are $g_\ssv = \hf (g_\ssl + g_\ssr)$ and $g_\ssa
= \hf (g_\ssl - g_\ssr)$, normalized so that their SM values
would be $g^\SM_\ssv = \hf \; T_{3f} - Q_f \sw^2$ and
$g^\SM_\ssa = \hf \; T_{3f}$. 

Evaluating graphs ($b$) 
and ($c$) of the same figure -- which do not interfere with
graph ($a$) for massless initial fermions -- gives the 
following result for $d\sigma_{\rm yuk}$:
\eq
\label\dhphdtb
{d \sigma_{\rm yuk} \over dt \, du} =
{\alpha \, |\Scy|^2 \, Q_f^2 \over 4 \, s^2} \; \left[
{u \over t} + {t \over u} + 2 \left( 1 + {\mh^2 \, s
\over u \, t} \right) \right] \; \delta(s + t + u - \mh^2).
\eeq

%
%

We remind the reader that this last result, like the previous
ones, assumes fermion masses are negligible in comparison with 
the
Mandelstam variables $s, t$ and $u$. Unlike the earlier 
expressions
this neglect can cause trouble in eq.~\dhphdtb , since the 
quantities
t or u approach zero when the outgoing photon is collinear with
the incoming fermion or antifermion, and near threshhold when
$s \sim m^2$. The breakdown of eq.~\dhphdtb\  in these 
situations
reflects the usual infrared problems which are associated with
the multiple emission of soft and/or collinear photons. As such
this equation should be replaced in these regimes by the
result which does not neglect fermion masses, and by the usual
Bloch-Nordsieck summation over soft-photon emission.

Although interference makes the analogous results for the production
of a CP-even scalar more difficult to compute it may still be done, such
as by judiciously modifying the analytic SM contributions 
of ref.~\hgamSMb. We do not pursue this observation further here, 
however, concentrating instead on the properties of
expressions \dhphdta\ and \dhphdtb. These results 
are plotted for illustrative choices for
the parameters in Fig.~\phhprodvs-a
and Fig.~\phhprodvs-b. We make the following observations:

\item{1.} {\sl Graph ($a$) differs strongly from graphs
($b$) and ($c$) on the dependence on CM scattering angle
it predicts.}

Fig.~\phhprodvs-a shows that graphs ($b$) and ($c$)
imply the well-known strong forward-peaking of the
bremstrahlung cross section. This contrasts with the
flatter dependence on scattering angle which follows
if the $h$ is emitted from the virtual boson line. These
two properties make the differential cross section near
$\theta = {\pi \over 2}$ a good probe of the effective
couplings, $c_\gamma$, $\tw{c}_\gamma$, 
$c_{\ssz\gamma}$ and $\tw{c}_{\ssz\gamma}$.

\item{2.} {\sl For $\gamma h$ production graphs
($b$) and ($c$) do not have a rising high-energy
limit.}

Because the photon is massless, its gauge symmetry
must be linearly realized (on peril of violations of
unitarity and/or Lorentz invariance), and so the cross
section, $\sigma_{bc}$, for $\gamma h$ production
does not share the rising high-energy limit found for
$Zh$ production. For photons it is instead the cross section
due to $h$ emissions from the $\gamma$ line, $\sigma_{a}$,
which rises at high energies, eventually implying a breakdown
of the low-energy approximation.  In either case it is clear that
the high-energy behaviour of both the
$Zh$ and $\gamma h$ production cross sections depends
sensitively on whether the new scalars assemble into
a linear realization of the electroweak $SU_\ssl(2)
\times U_\ssy(1)$ gauge symmetry.

\subsection{Scalar Production (Hadron Colliders)}

The
production cross section of a Higgs at an hadron collider is 
 more involved to compute
than in electron colliders. This is because the possibly
large size of the scalar Yukawa couplings makes more graphs
important at the parton level than is the case, say, for a
SM Higgs. Unfortunately, a complete discussion of all
parton processes using the couplings of  the general
effective lagrangian goes beyond the scope of this paper.

We instead content ourselves to recording expressions for
the production processes in the case that these Yukawa
couplings can be neglected. This is sufficiently general a
situation to still include a great many of the most popular
models. In this case production is dominated by one or two
parton processes, depending on the CM energy at which the
collision occurs. We next consider the most important of these.

\subsubsection{Gluon Fusion}

\ref\ggmn{H. Georgi, S. Glashow, M.Machacek and D. Nanopoulos,
\prl{40}{78}{692}.}

\ref\whpreferred{
A. Stange, W. Marciano and S. Willenbrock, \prd{50}{94}{4491};
Future Electroweak Physics at the Fermilab
Tevatron, D. Amidei and R. Brock, in ref.~\outa;
M. Spira, in ref.\outaa\ (hep-ph/9810289).}

\ref\cmw{M. Carena, S. Mrenna and C.E.M. Wagner,
 \prd{60}{99}{075010} (hep-ph/9808312).}

Gluon-gluon fusion \ggmn\ is by far the dominant production
mechanism for scalar bosons at the LHC (with $\sqrt s=14 \TeV$)
throughout the scalar mass range of current interest, and in
particular for very low scalar masses. For the lower energies
of the next Tevatron run ($\sqrt s=2 \TeV$) scalar-emission
processes like $q {\bar q}^{\prime} \rightarrow h W$ and
$q {\bar q} \rightarrow h Z$ are also important, and indeed
may be preferred \whpreferred, \cmw\ due to the large
QCD background which can swamp the dominant gluon-fusion
production mechanism, to the extent that the produced scalar
dominantly decays through the $b \bar b$ channel (which need
not be true in a generic model). We present results for the parton-level
cross section for $Wh$ and $Zh$ production below, after first
discussing gluon fusion.

The parton-level cross section of the gluonic process
$gg \to h$, mediated by the effective interactions of
eq.~\dimfiveops\ is $\hat\sigma = \sigma_0 \;
\delta(\hat s - \mh^2)$, with
\eq \label\sgma
\sigma_0={\pi \over 4} \left( |c_g|^2+ |{\tilde c}_g|^2 \right) ,
\eeq
where $\hat s$ is the parton-level Mandelstam invariant.
The lowest-order contribution to $c_g$ by a heavy fermion
is given by eq.~\genfdimfv\ (or, as specialized to the SM
top quark contribution -- which is dominant -- by
eq.~\tqdimfv).

Eq.~\sgma\ implies the following lowest-order cross section for
scalar particle production by gluon fusion in $pp$ collisions:
\eq \label\sg
\sigma_{\sss LO}(pp \rightarrow h + X)={\pi \over 4}
\left( |c_g|^2+ |{\tilde c}_g|^2 \right)
\tau_\ssh \; {d {\cal L}^{gg}
\over d \tau_\ssh} ,
\eeq
where $\tau_\ssh ={m_h^2 / s}$, $\sqrt s$ is the total CM energy
of the proton collider and ${d {\cal L}^{gg} / d \tau_\ssh}$ is the
gluon luminosity ~\ggmn:
\eq
{d {\cal L}^{gg} \over d \tau_\ssh}=\int_{\tau_\ssh}^1
{dx \over x} \; g(x,M^2) \; g(\tau_\ssh/x,M^2) .
\eeq
Here $M$ denotes the factorization scale at which the
gluon structure function, $g(x,M)$, is defined.

Because this process is strongly enhanced by next-to-leading-order
(NLO) QCD corrections ($50-100\%$), these effects must be incorporated
into any realistic calculation. A consistent treatment
of the gluon-gluon parton process
to next order in $\alpha_s$ requires the contributions
of gluon emission from the initial gluon lines and
internal fermion loops, in addition to virtual
gluon exchange between any colour carrying lines.
This must be added to other subprocesses, like
gluon-quark and quark-antiquark collisions, which
can also contribute to scalar production at the same order.

\ref\gsz{D. Graudenz, M. Spira and P.M. Zerwas, \prl{70}{93}{1372}.}

\ref\altpar{G. Altarelli and G. Parisi, \npb{126}{77}{298}.}

Combining all of these contributions, the cross section
at next-to-leading order is then \spira,\spiran,\gsz :
\eq 
\label\ntlosigma
\sigma(pp \rightarrow h+X)=\sigma_1
\tau_\ssh \; {d {\cal L}^{gg} \over d \tau_\ssh}
+ \Delta \sigma_{gg}+\Delta \sigma_{gq}
+ \Delta \sigma_{q {\bar q}}
\eeq
where
\eq
\label\ntlosigmapart
\sigma_1={\pi \over 4} \left(
\left|c_{g}^{(1)} \right|^2 + \left|\tilde{c}_{g}^{(1)}
\right|^2  \right) \left(1+{\alpha_s \over \pi} \; C_{re}
\right) .
\eeq
Here $c^{(1)}_g$ and $\tw{c}^{(1)}_g$ are
defined to include the gluon-loop corrections to the
effective couplings, $c_g$ and $\tw{c}_g$.
The infrared singular part of these virtual-gluon
contributions cancel the infrared singular part of
real gluon emission, which is denoted in the
above by $C_{re}$.
For instance, when integrating out the top
quark in the SM,   ${\tilde c}_g^{(1)}=0$
and
\eq
\label\bfnadimrels
c_g^{(1)} ={1 \over 4 v} \; \left( {\beta(\alpha_s)
\over 1 + \gamma_m(\alpha_s)} \right)
\approx {\alpha_s \over 12 \pi \,v} \;
\left(1 + {11 \over 4}
{\alpha_s \over \pi} \right)
\eeq
where $\beta(\alpha_s)
= {\alpha_s \over 3 \pi} \left( 1 + {19 \alpha_s
\over 4 \pi} + \cdots \right)$ is the heavy quark contribution to the QCD 
beta function and
$\gamma_m(\alpha_s) = {2 \alpha_s \over \pi} + \cdots$
is the anomalous dimension for the quark mass operators.
(Notice that this reproduces eq.~\tqdimfv\ up to second order
in $\alpha_s$.) The SM contribution from real gluon emission is \gsz:
\eq
C_{re}=\pi^2+{(33-2 N_\ssf) \over 6} \; \ln
\left({\mu^2 \over \mh^2} \right)
\eeq
with $\mu$ the renormalization scale.

The remainder of the contributions to
eq.~\ntlosigma\ --- coming from the finite part of
$\hat\sigma_{gg}$, $\hat\sigma_{gq}$
and $\hat\sigma_{q {\bar q}}$) --- are ~\gsz :
\eqa
\Delta\sigma_{gg} &= \int_{\tau_\ssh}^1 d\tau
\; {d {\cal L}^{gg} \over d\tau} \; {\alpha_s
\over \pi} \;  \sigma_0 \left\{-z P_{gg}(z)
\ln\left( {M^2 \over \tau s} \right)
- {11 \over 2} {(1-z)}^3
\right.
\eolnn
&\qquad\qquad \qquad\qquad \left.
+12 \left[{\ln(1-z) \over (1
-z)}-z (2 -z(1-z)) \ln[1-z] \right] \right\} \eol
\Delta\sigma_{gq} &=
\int_{\tau_\ssh}^1 d \tau \;\sum_{q,{\bar q}} {d {\cal L}^{gq} \over d\tau}
\; {\alpha_s \over \pi} \; \sigma_0 \left\{
\left[-{1 \over 2} \ln\left({M^2 \over \tau s}\right)
+\ln(1-z) \right] z P_{gq}(z)-1+2z -{1 \over 3} z^2 \right\}
\eol
\Delta\sigma_{q{\bar q}} &= \int_{\tau_\ssh}^1
d \tau \;\sum_{q} {d {\cal L}^{q {\bar q}} \over d\tau} \;
 {\alpha_s\over \pi} \; \sigma_0 \; {32 \over 27} \,
 {(1-z)}^3 \eeol
 \eeq
where these expressions assume 
that the particles whose loops generate the
effective couplings $c_k$ and $\tw c_k$ are
much heavier than $\mh/2$.
As before, $M$ is the factorization scale of the parton
densities, and $P_{gg}$, $P_{gq}$ are the standard
Altarelli-Parisi splitting functions \altpar.
Finally the remaining collinear singularities
are absorbed into the renormalized
parton densities \gsz.

The beauty of these expressions lie in their generality.
Since the parton expressions use the large-mass limit for
the particle in the loop responsible for the effective
couplings, the decoupling of this particle from lower-energy
partonic QCD is explicit in the appearance of
the new physics contribution only inside the effective
parton-level cross section, $\sigma_0$ and $\sigma_1$, 
which are fixed in terms of $c_g$ and $\tw c_g$ by 
eqs.~\sgma\ and \ntlosigmapart.
As a result, the above expressions hold for any
new physics for which the process $gg \to h$
dominates the $h$ production cross section in hadron
collisions. Different kinds of new heavy particles can
alter their predictions for the gluon-fusion contribution
to the $pp \to hX$ cross section only
through their differing contributions to $\sigma_0$
and $\sigma_1$.

\subsubsection{$W, Z$ Fusion}

\ref\hgw{T. Han, G. Valencia and S. Willenbrock, \prl{69}{92}{3274}.}

To the extent that $W$ or $Z$ fusion processes are
important, the contributions of effective scalar couplings
to these processes may also be incorporated into simulations
using the parton-level cross sections for $WW \to h$ or
$ZZ \to h$.
The differential cross section in the case of W or Z fusion can be
obtained at first approximation directly from eq.(2) of 
ref.~\hgw, by multiplying this equation by a factor 
$\sw^2{|a_\ssw|}^2/e^2 \mw^2$ or $\sw^2 \cw^2
{|a_\ssz|}^2 /e^2 \mz^2$ respectively.

\ref\zepp{D. Rainwater, D. Zeppenfeld, \prd{60}{99}{113004},
 Erratum-ibid.D61:099901,2000;
D. Zeppenfeld, R. Kinnunen, A. Nikitenko, E. Richter-Was,
{\it Phys. Rev.} {\bf D62} (2000) 013009;
 N. Kauer, T. Plehn, 
D. Rainwater, D. Zeppenfeld, {\it Phys. Lett.} {\bf B503} (2001) 113.}

 We do not
explore the detailed implications of
these processes here. For a recent and detailed discussion on how to 
use weak boson fusion to look  for a light Higgs at LHC, by using the 
distinc signal provided by two forward jets, see ~\zepp.

\subsubsection{$Zh$ and $Wh$ Production}
It has been argued \whpreferred, \cmw\ that the parton
processes $\ol{q}q \to Zh$ and $\ol{q} q' \to W h$
may prove to be more important mechanisms for $h$
production at the Tevatron because of the difficulty
in pulling the gluon fusion signal out of the backgrounds.
We record here the cross sections for these parton-level
processes using the effective couplings of \S2.

The cross section for $\ol{q} q \to Zh$ production is 
directly given by eqs.~\dsgmforma\ and 
\dsgmformbc\ (or their integrated versions,
eqs.~\sgmforma\ and \sgmformbc)
provided one uses in them the couplings,
$g_\ssl$, $g_\ssr$, $y_{qq}$ and $z_{qq}$,
appropriate to the quark in question.
The same is true for the contribution to $\ol{q}q' \to
Wh$ of graph
($a$) of Fig.~\prdngraphs\ -- with the scalar
emitted from the $W$ line -- provided one makes
the replacements
\eq
\label\wzreplc
\mz \to \mw , \qquad
\qquad \Gamma_\ssz \to \Gamma_\ssw \qquad
\hbox{and} \qquad { \alpha
\left| a_\ssz \right|^2 (\gl^2 + \gr^2) \over  16
\sw^2 \cw^2 } \to { \alpha
\left| a_\ssw \right|^2 \over 32
\sw^2 }
\eeq
in eq.~\dsgmforma\ or \sgmforma.

The cross section for $\ol{q} q'
\to Wh$ coming from graphs ($b$) and ($c$) of
Fig.~\prdngraphs\ -- with the scalar emitted
from the fermion line -- requires a less trivial
generalization of eqs.~\dsgmformbc\ and \sgmformbc.
The result for the differential cross section is:
\eq
\label\dsgmbcwh
{d\hat\sigma_{bc} \over d\hat{u} d\hat{t}}  =
{ \alpha \over 16 \sw^2 } \;
\left\{  {  |\Scy_q|^2 + |\Scy_{q'}|^2 
\over 2\mw^2 \, \hat s}  +
 \left( {|\Scy_{q'}|^2 \over \hat{t}^2 } +
 {|\Scy_q|^2 \over \hat{u}^2} \right)  \left(
 { \hat{u}\hat{t} - \mh^2
\mw^2 \over \hat{s}^2 } \right) \right\}
\delta(\hat{s} + \hat{t} + \hat{u} - \mh^2 - \mw^2) .
\eeq

\subsubsection{Comparison of production mechanisms}   

\fig\dialeft

\topinsert
\centerline{\epsfxsize=10.5cm\epsfbox{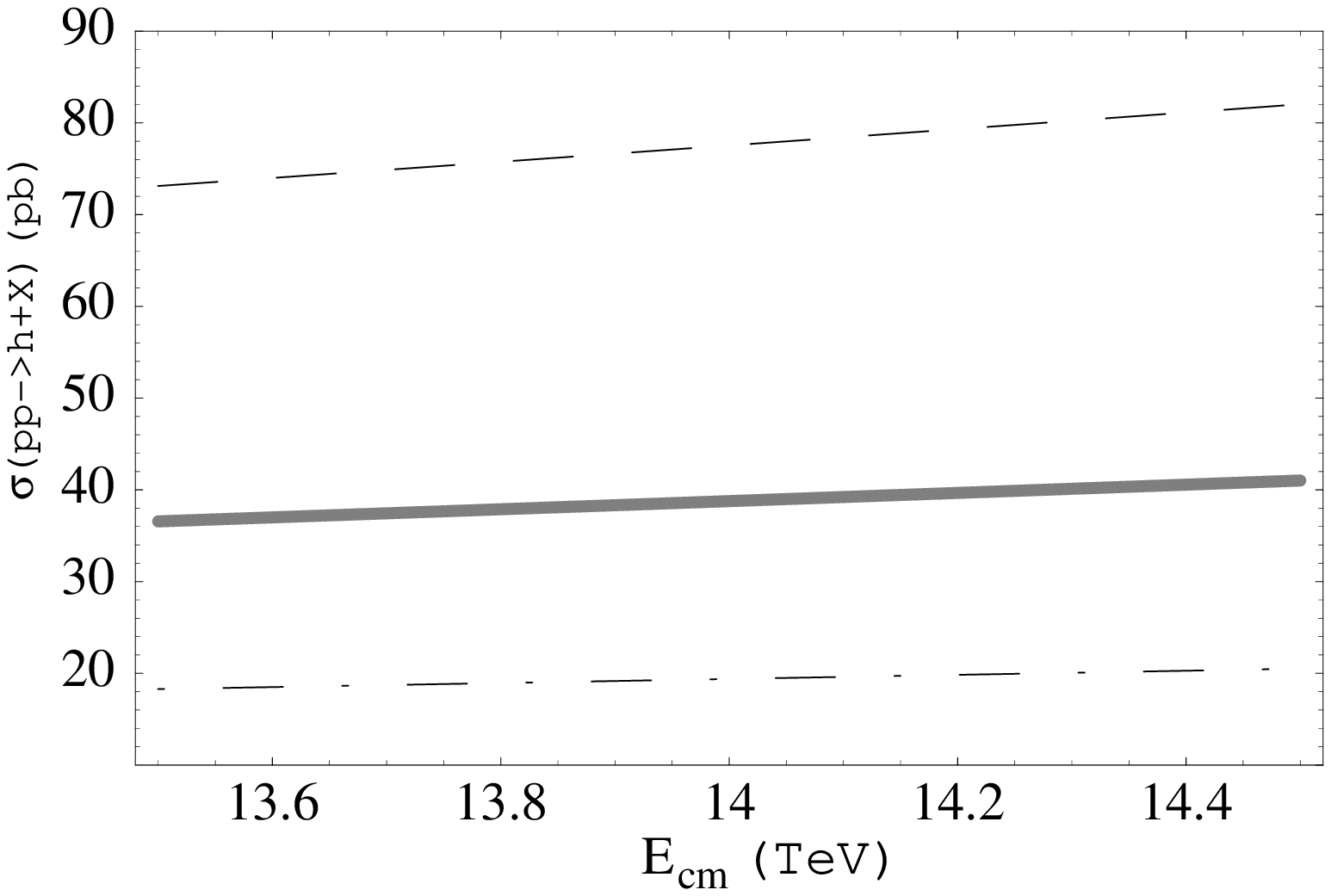}}

\centerline{{\rm Figure \dialeft:}}
\medskip
\caption{
QCD-corrected gluon-fusion contribution to the cross section
$\sigma(pp\rightarrow h + X)$ as a function of the c.m. energy
$E_{\rm cm}=\sqrt{s}$ for a scalar mass of 115 GeV. The thick line
correspond
to $c_g=c_g^{SM}$, $c_{\tilde g}=0$. The dashed line corresponds to the
prediction for a theory
with a 100\% enhancement with respect to the SM, i.e., $|c_g|^2+|c_{\tilde
g}|^2=2 |c_g^{SM}|^2$. And the dashed-dotted line a theory with a
suppression
of 50\% with respect to the SM, i.e., $|c_g|^2+|c_{\tilde
g}|^2=\frac{1}{2} |c_g^{SM}|^2$.\medskip       
}
\endinsert

\fig\diaright

\topinsert
\centerline{\epsfxsize=10.5cm\epsfbox{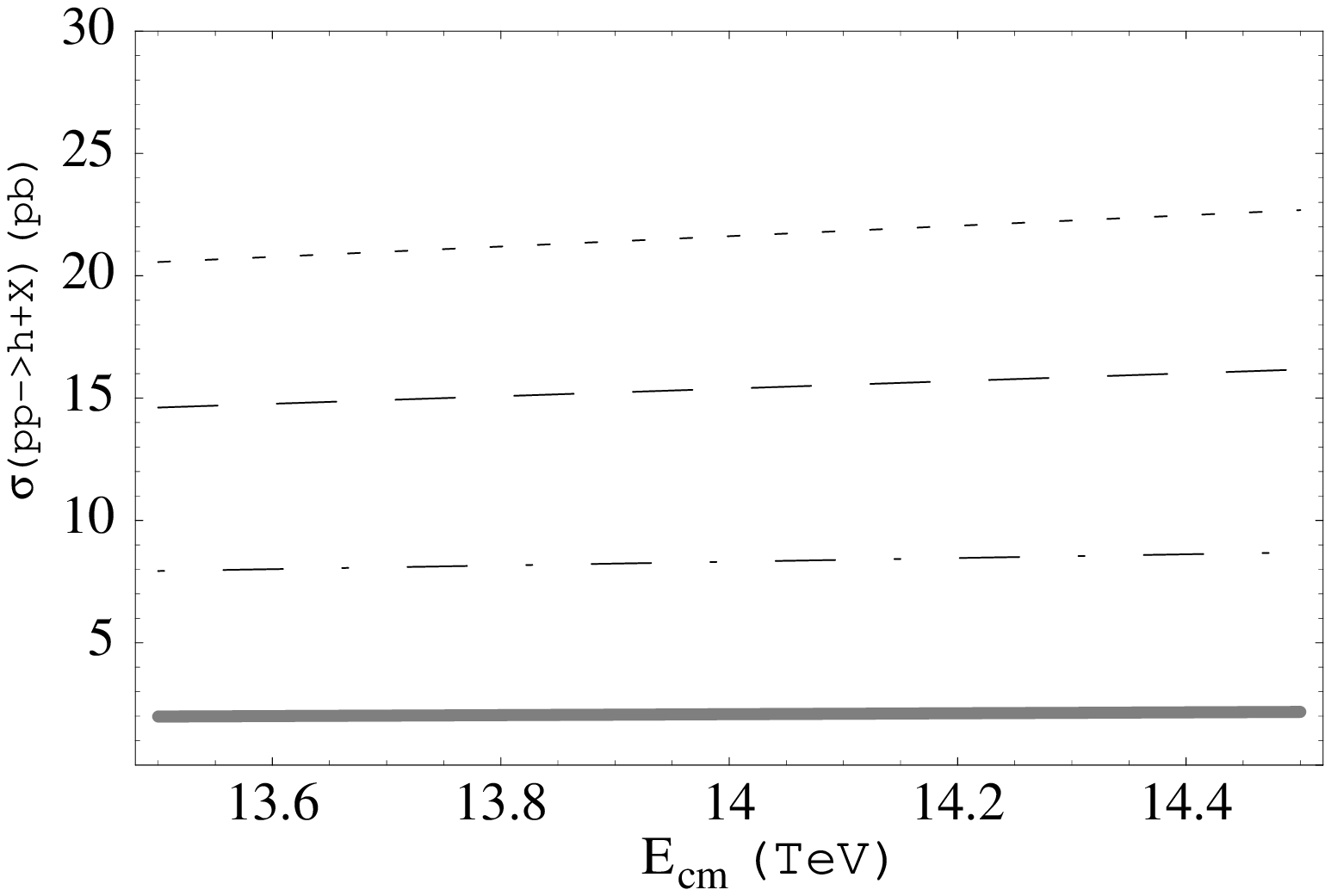}}

\centerline{{\rm Figure \diaright:}}
\medskip
\caption{
 QCD-corrected  parton process $\ol{q} q' \to W
h$ contribution to the cross section
$\sigma(pp\rightarrow h + X)$ as a function of the c.m. energy
$E_{\rm cm}=\sqrt{s}$ for a scalar mass of 115 GeV. The thick line
correspond
to the SM $a_W=a_W^{SM}$ and $|\Scy_q|^2 + |\Scy_{q'}|^2 = 0$. The
dashed-dotted
line
corresponds to the
prediction for a theory
with $a_{W}=2 a_{W}^{\rm SM}$, the dashed line corresponds to a theory
with large yukawas
$|\Scy_q|^2 + |\Scy_{q'}|^2 = 0.01$ and the dotted line to both anomalous
couplings acting together.\bigskip\bigskip                 
}
\endinsert

To close this section we present an explicit comparison between two of 
the prefered 
mechanisms of production of a very light scalar, around 100 GeV,
discussed in the previous subsections.

We have done a parton level calculation using VEGAS of the gluon-gluon
fusion mechanims at NLO~\spiran\  and the $\ol{q} q' \to W h$
~\spiran\ production mechanism
within our effective lagrangian approach. 

In both cases we give
an explicit example of how would affect the presence of an anomalous
scalar coupling to the
prediction for the cross section 
$\sigma(p p \rightarrow h + X)$.
We use the same energy range 
to compare
more easily the two mechanisms.

We present the result of the gluon-gluon fusion case showing
how New
Physics affecting the
coupling between the scalar and the gluons induces a different prediction
for the total cross section $\sigma(p p \rightarrow h + X)$ in two
different situations. First, when  New Physics  is constructive and adds
up to
the SM contribution, in particular, when $|c_g|^2+ |{\tilde c}_g|^2=2
|c_{g}^{\rm SM}|^2$. And second, when New Physics is destructive with
respect to the
SM contribution, for instance, $|c_g|^2+ |{\tilde c}_g|^{2}={1 \over 2}
|c_{g}^{\rm SM}|^{2}$. This is shown in Fig.~\dialeft.

More interestingly,  the higgs production mechanism via the parton process 
$\ol{q} q' \to W h$ allow us to show the effect of an
anomalous 
yukawa coupling between the scalar and fermions.
More precisely, here we are sensitive to two different type of
couplings, the gauge WW-scalar coupling $a_W$ (
graph a) of Fig.~\prdngraphs)  but also to a
possible large
anomalous  yukawa coupling between the scalar and fermions (graph b) and
c) of Fig.~\prdngraphs). We have
computed the effect on the production cross section in three different
cases in Fig.~\diaright. First, if an anomalous and additive large
contribution to the
gauge coupling $a_W$ is present, $a_{W}=2 a_{W}^{\rm SM}$ but no anomalous
yukawa coupling. Second, if we have a very large yukawa coupling 
$|\Scy_q|^2 + |\Scy_{q'}|^2 = 0.01$ between
the discovered scalar and fermions and finally if both situations happens
at the same time. 

It is explicit in Fig.~\diaright\ that the subdominant production
mechanism in the SM $\ol{q} q' \to W h$
can receive an important enhancement as compared to
the
dominant gluon fusion if the scalar couples with
a very large yukawa coupling to fermions or if the gauge coupling gets
enhanced.

\section{Connecting to Observables: The Influence of
Virtual Scalars}

After scalar decays and production, the next most important
class of observables to consider consists of scattering
processes involving only familiar SM fermions as external
states. These have the advantage of often being well
measured, and since they can receive contributions from virtual
scalar exchange, they provide an important source of
constraints on scalar couplings.

Constraints of this type organize themselves into four broad
categories according to whether they involve high- or low-energy
processes (compared to the QCD scale, say), and whether they
do or do not change fermion flavour. Only two of these are
of interest for the present purposes since we have chosen
to restrict our attention to flavour-diagonal processes.
We therefore divide our discussion into two sections, which
respectively describe constraints coming from high- and
low-energy observables.

For the present purposes, it suffices to work at lowest order
in the effective couplings when computing the implications
of the new scalar for high-energy processes. The same need
not be true for the low-energy observables, however.
Since the decoupling of the heavy scalar ensures that its
effects generically become weaker and weaker for
lower energies, only the best measured observables
imply significant bounds on its interactions.
But precisely because these observables are so
well measured, their analysis within the effective theory
proves to be one of those few situations in which it is necessary to
go beyond tree level in the effective interactions.

\subsection{High-Energy Flavour-Diagonal Scattering}

\fig\scalexch

We focus in this section on two-body fermion scattering,
to which virtual scalars may contribute through the Feynman
graphs of Fig.~\scalexch. To these must be added the
usual SM contributions, which at lowest order are also
of the form of Fig.~\scalexch, but involving exchanged
spin-one vector bosons ($W$, $Z$, $\gamma$ and gluons).
Before evaluating these cross sections in detail, we first
draw some general conclusions which follow for all
such processes in the (excellent) approximation in which
external fermion masses are neglected.

\midinsert
\centerline{\epsfxsize=7.5cm\epsfbox{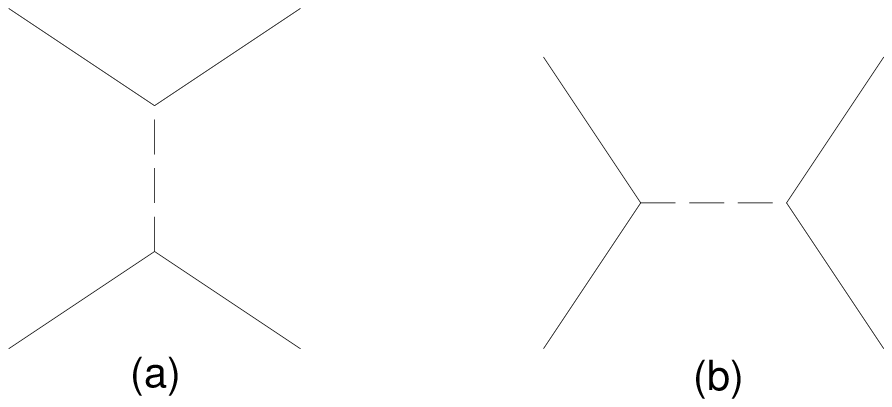}}

\centerline{{\rm Figure \scalexch:}}
\medskip
\caption{The Feynman graphs which
contribute the leading scalar contribution to the reactions
$\ss f \ol{f} \to g \ol{g}$ and $\ss ff \to gg$, for light fermions
$\ss f$ and $\ss g$.\medskip}
\endinsert

\subsubsection{Helicity Considerations}

In the absence of masses for the initial and final
fermions, the scalar-exchange graphs of Fig.~\scalexch\
do not interfere with the vector-exchange graphs
of the SM because of their different helicity
properties. This is because in the absence of fermion masses
the initial and final fermions may be labelled by their helicities,
which are conserved along any fermion line
by vector emission, but which are flipped along
fermion lines by scalar emission.

For example, in $f \ol{f}$
annihilation processes, SM vector-exchange
graphs contribute only to the scattering
of left- (right-) handed fermions with right- (left-) handed
antifermions. The only nonzero scalar-exchange
processes, on the other hand, are to the scattering
of left- (right-) handed fermions with left- (right-) handed
antifermions. Since processes mediated by the exchange
of SM particles and virtual scalars only contribute to
different helicity configurations, it follows that their
contributions to the cross section cannot interfere,
permitting the SM and scalar-exchange cross sections
to be simply summed: \eg\ $\sigma_{\rm tot}
 = \sigma_\SM + \sigma_{h}$.
It therefore suffices to give
explicit expressions here for the cross section due to
scalar exchange only.

We now present these cross sections, doing
so in steps of increasing complexity.
We start by considering the particularly simple case of
scattering of different flavours of fermions, since in this
case only the single Feynman graph $(a)$ or $(b)$ of
Fig.~\scalexch\ contributes to the process. We describe the
$s$-channel process, graph $(a)$ first, followed by the
$t$-channel process, graph $(b)$. We then extend our
conclusions to processes in which {\it both} graphs
$(a)$ and $(b)$ can contribute at once.

\subsubsection{$s$-Channel Scattering: $f \ol{f} \to g \ol{g}$}

\fig\eemumuvs

\topinsert
\centerline{\epsfxsize=6.8cm\epsfbox{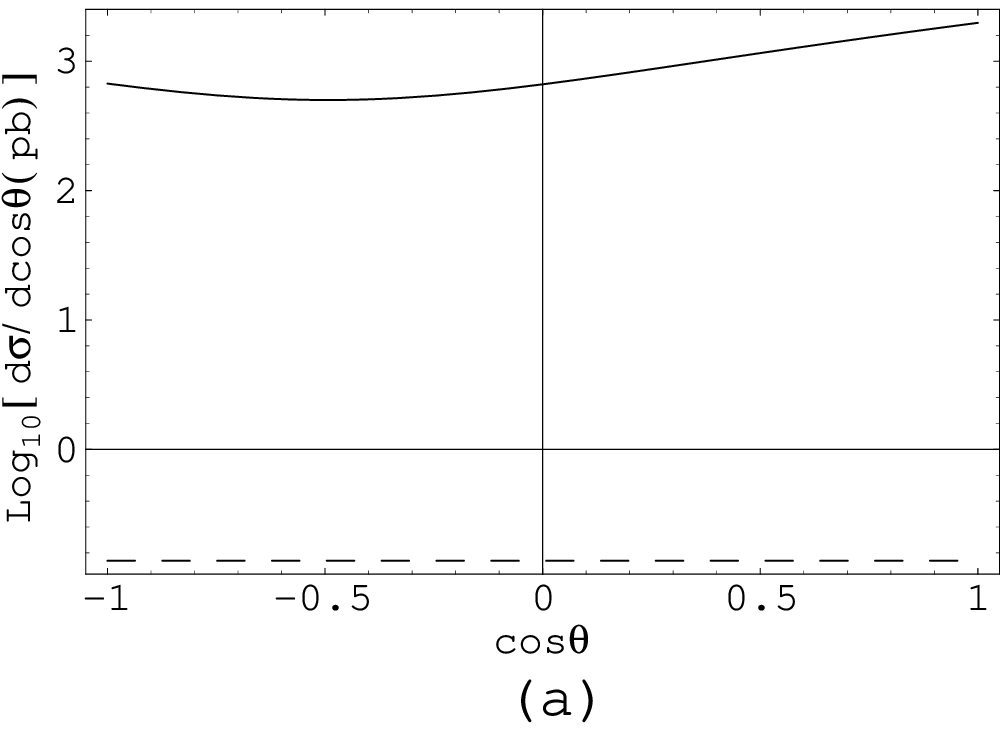}
\epsfxsize=6.8cm\epsfbox{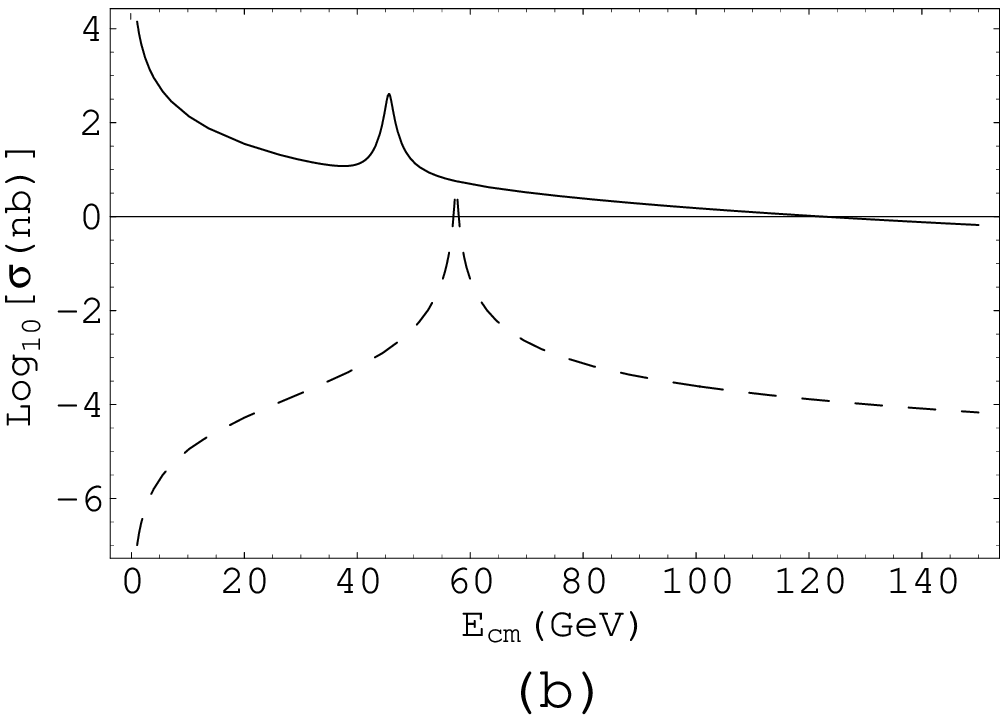}}    

\centerline{{\rm Figure \eemumuvs:}}
\medskip
\caption{A comparison of the tree-level SM (solid line) and
scalar-mediated (dashed line)
contributions to the differential cross section for the
reaction $\ss e^+e^- \to \mu^+\mu^-$, showing in (a)
the dependence on the CM scattering angle, $\ss \theta$.
The figure assumes an electron CM energy
of 100 GeV, a scalar mass $\ss \mh = 115$ GeV,  a width 
$\Gamma_h= 1$ GeV and
the effective couplings $\ss |\Scy_e|^2 = |\Scy_\mu|^2
= 0.01 \; e^2$. And in (b)
the CM energy dependence. 
The cross sections are integrated only over scattering
angles $\ss |\cos\theta| < 0.9$ in order to exclude the
small-angle scattering region, for which SM
radiative corrections are most important.\medskip
}
\endinsert

Consider, then, the reaction $f \ol{f} \to g \ol{g}$ with
$f  \ne g$, and suppose that the flavour-changing Yukawa
couplings, $y_{fg}$ and $z_{fg}$, are too small to
significantly contribute. With these assumptions only
the $s$-channel diagram ($a$) of Fig.~\scalexch\
contributes, giving the result:
\eq
\label\dschan
\left( {d\sigma\over du \, dt} \right)_{s-ch}
= {|\Scy_f|^2 |\Scy_g|^2
\over 16 \pi} \; {1 \over
(s - \mh^2)^2 } \; \delta(s + t + u) ,
\eeq
where $\Scy_f = y_{ff} + i z_{ff}$ and $t = (p_f - p_g)^2$
is the Mandelstam variable defined using the 4-momenta of
the initial and final fermions. Eq.~\dschan\ trivially integrates to
give the total cross section:
\eq
\label\schan
\sigma_{s-ch} = {|\Scy_f|^2 |\Scy_g|^2
 \over 16 \pi} \; {s \over \left|
(s - \mh^2 + i \Gamma_h \mh) \right|^2 } .
\eeq

%
%

Fig.\eemumuvs-a and Fig.\eemumuvs-b compare
these expressions for the reaction $e^+e^- \to \mu^+\mu^-$
with their tree-level SM counterparts.
We restrict the comparison to scattering angles
$|\cos\theta|< 0.9$ --- where $\theta$ is the angle between
the directions of the incoming electron and outgoing
muon --- in order to minimize configurations
for which SM radiative corrections are important.

Since neither cross section grows at high energies, they
must be distinguished in practice by their differing shapes
as functions of both energy and scattering angle. In practice
this requires resolving the potentially narrow scalar
resonance from the energy dependence of the cross
section. Interestingly, since the high-energy
$e^+e^-$ annihiliation cross section is known only for
several energies, it would be easy to miss a narrow
scalar state even if the scalar mass were comparatively
low.

\ref\LEPhbounds{G. Abbiendi et al. (OPAL Collaboration), 
{\it Eur. Phys. J.} {\bf C13} (2000) 553.}

Bounds nevertheless exist, but are comparatively
weak. They are obtained
by examining small-angle scattering processes, for
which radiative corrections can be important because
initial-state radiation can allow scattering at energies
greater than resonance to profit from the large
on-resonance cross section. The best current limits
obtained in this way from the reaction $e^+e^- \to
\mu^+ \mu^-$ \LEPhbounds\ constrain
\eq
\label\scheeffbounds
\eqalign{
\sqrt{|\Scy_e \Scy_\mu|} &\lsim 0.07 \qquad \hbox{if }
\qquad \mh \sim 100 \GeV,\cr
&\lsim 0.02 \qquad \hbox{if }
\qquad \mh \sim 190 \GeV .\cr}
\eeq
These bounds deteriorate sharply with larger $\mh$,
weakening to $\sqrt{|\Scy_e \Scy_\mu|} \lsim 0.3$
when $\mh \sim 300$ GeV.

\subsubsection{$t$-Channel Scattering: $fg \to fg$}

\fig\emuemuvs


\topinsert
\centerline{\epsfxsize=6.8cm\epsfbox{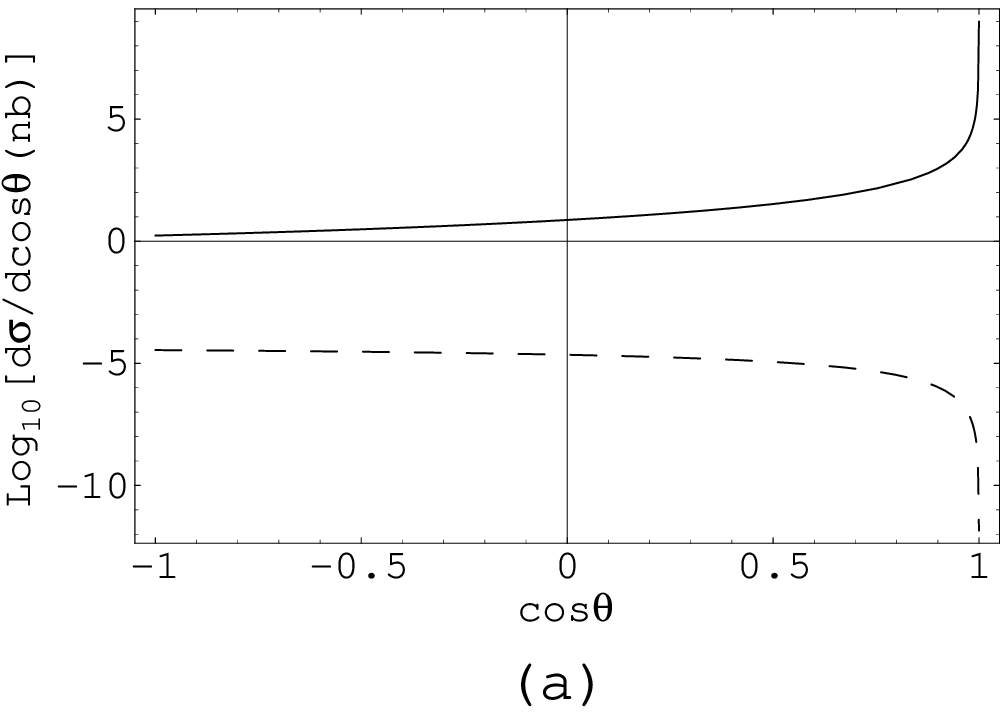}
\epsfxsize=6.8cm\epsfbox{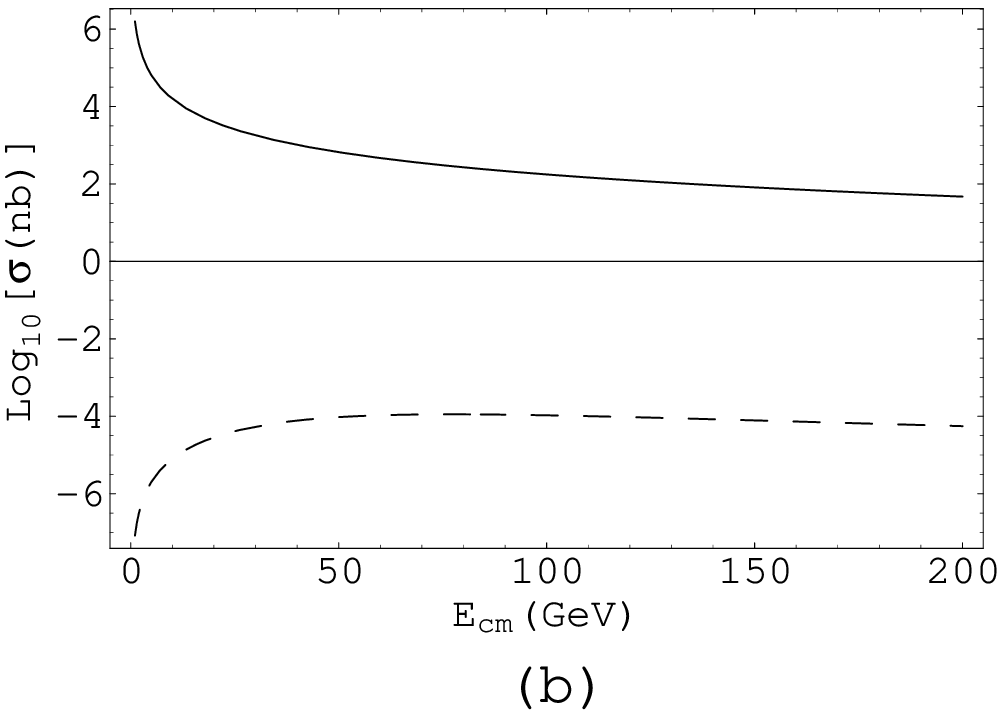}}    

\centerline{{\rm Figure \emuemuvs:}}
\medskip
\caption{Same comparison and input parameters as in Fig.~\eemumuvs\  for
the
reaction $\ss e^-\mu^- \to e^-\mu^-$ (a) CM scattering angle dependence
and (b) CM energy dependence. \medskip}
\endinsert

Consider next the reaction $fg \to fg$, again under the
assumption that the flavour-changing Yukawa couplings,
$y_{fg}$ and $z_{fg}$, are negligibly small. As usual,
neglect of fermion masses permits the SM contribution
to be summed with the cross section due to scalar exchange.
At tree level the scalar-mediated contribution arises
only from the $t$-channel graph, $(b)$, of Fig.~\scalexch.
The result obtained from this graph is easily found
by crossing the external lines of the $s$-channel
result, giving:
\eq
\label\dtchan
\left( {d\sigma\over du \, dt}\right)_{t-ch}
= {|\Scy_f|^2 |\Scy_g|^2 \over 16 \pi \; s^2}
\; {t^2 \over (t - \mh^2)^2 } \; \delta(s + t + u) .
\eeq
The total cross section which follows from this is:
\eq
\label\tchan
\sigma_{t-ch} = {|\Scy_f|^2 |\Scy_g|^2
 \over 16 \pi \; s} \; \left[{s + 2\mh^2  \over s + \mh^2}
-{ 2 \mh^2 \over s} \; \ln \left( {s + \mh^2 \over \mh^2}
\right) \right] .
\eeq


Figure \emuemuvs-a displays the dependence of this cross
section on the CM-frame scattering angle, with the
tree-level SM result plotted for the purposes of
comparison. They differ significantly, with the SM result
showing the characteristically strong forward scattering
due to the exchange of the massless photon.
The energy dependence of the integrated cross section is
similarly displayed in Fig.~\emuemuvs-b.



This kind of scattering would contribute to the interaction
cross section of electron-nucleon collisions, such as
is measured at HERA, by modifying the parton level process
$e q \to e q$ or $e \overline{q} \to e \overline{q}$.
This contribution is in addition to any others, such as
to modifications of the process $\gamma \gamma \to h$
or $g g \to h$ which can be mediated by the effective
couplings $c_g, \tw{c}_g, c_\gamma$ or $\tw{c}_\gamma$.

\ref\heraLQ{C. Adloff \etal\ (H1 Collaboration), 
{\it Eur. Phys. J.} {\bf C11} (1999) 447--471.}

\ref\heraHiggs{B.B.~Levtchenko, preprint SINP-TNP-96-71
(unpublished) ({\tt hep-ph/9608295});
M.Krawczyk, in {\it Future physics at HERA},
the proceedings of the Workshop on Future 
Physics at HERA, Hamburg, pp 244-255,
({\tt hep-ph/9609477}).}

To date, experimental searches for scalars at HERA appear to
have been restricted either to flavour-changing leptoquarks
or to supersymmetry-motivated models with two Higgs doublets
(2HDMs), for which relatively weak
bounds exist \heraLQ, \heraHiggs. Unfortunately,
these analyses are not directly applicable to the general
possible scalar couplings, since they either
presuppose negligible Yukawa couplings to electrons and
light quarks, or assume more dramatic flavour-changing couplings.

Even though their signatures differ, a rough indication
of the strength of the bounds which are
possible for the neutral scalars considered here
may be had by comparing with the results of the
leptoquark searches, for which sensitivity is claimed
to scalars in the mass range of several hundred
GeV if their Yukawa couplings are electromagnetic in
strength: $|\Scy| \sim e$. It would clearly be valuable to have
a more detailed, model-independent study of the effects of
scalar exchange at $ep$ colliders.

\subsubsection{Scattering of Identical Fermions: $f  f \to f  f$}

When the initial and final fermion pairs are all of the same
flavour graph ($b$) of Fig.~\scalexch\ must be summed with
the same graph with the final fermions interchanged.
Besides contributing to electron-electron collisions,
this kind of scattering process also plays a role in
high-energy hadron collisions through parton
process like $q q \to qq$. Unfortunately, as we found for
scalar production at hadron colliders, one cannot simply
adapt existing studies of scalar-mediated scattering
in hadron machines to infer bounds on the quark-scalar
Yukawa couplings, because the light-quark Yukawa couplings
are typically assumed to be negligible in these papers.
We regard a detailed simulation of hadron collisions
to be beyond the scope of this paper, and so do not
pursue these processes more closely in what follows.

\fig\Bhabvs


\topinsert
\centerline{\epsfxsize=6.8cm\epsfbox{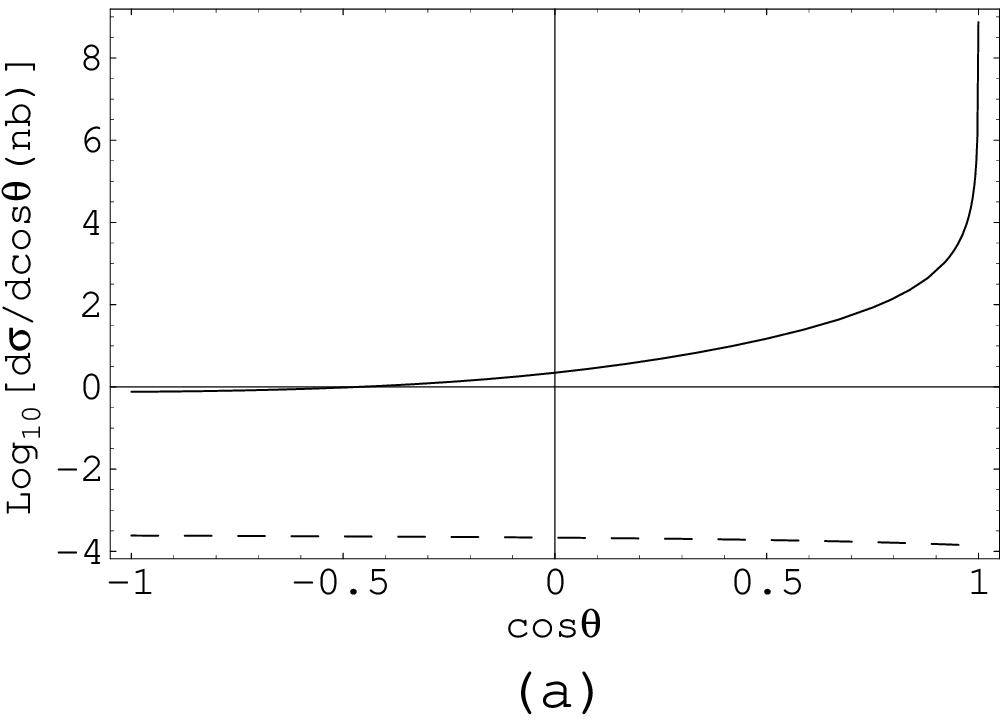}
\epsfxsize=6.8cm\epsfbox{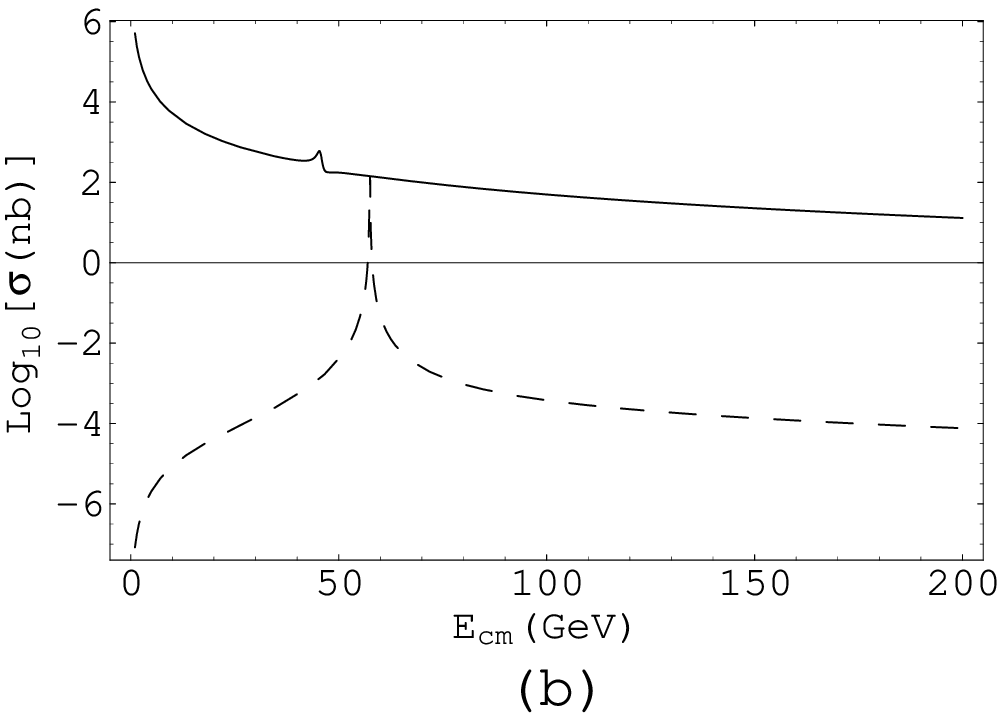}}    

\centerline{{\rm Figure \Bhabvs:}}
\medskip
\caption{Comparison of the tree-level SM (solid line) and
scalar-mediated (dashed-line)
contributions to the differential cross section for
Bhabba scattering: $\ss e^+e^- \to e^+e^-$, versus
(a) the CM scattering angle  $\ss
\theta$ at an electron CM energy of 100 GeV and (b) the CM energy
dependence.
The figure assumes  a scalar mass
$\ss \mh = 115$ GeV and
the effective coupling $\ss |\Scy_e| = 0.1 \; e$.\medskip }
\endinsert

Summing the two relevant graphs gives the contribution of
scalar exchange to elastic $ff$ scattering:
\eq
\label\dsgmff
\eqalign{
{d \sigma_{\rm sc} \over dt \, du}(ff \to ff) &=
{|\Scy_f|^4 \over 16 \pi \; s^2} \left[ {3 \over 4}
\left( {t \over t - \mh^2} + {u \over u - \mh^2} \right)^2
+ {1 \over 4}  \left( {t \over t - \mh^2} -
{u \over u - \mh^2} \right)^2 \right] \delta(s+t+u) \cr
&=
{|\Scy_f|^4 \over 16 \pi \; s^2} \left[
{t^2 \over (t - \mh^2)^2} + {ut \over (u - \mh^2) (t-\mh^2)}
+  {u^2 \over (u - \mh^2)^2} \right] \delta(s+t+u) .}
\eeq
The first expression in eq.~\dsgmff\ has a simple interpretation,
since the interchange $t \leftrightarrow u$ corresponds to exchanging
the momenta of the final-state fermions. Since statistics
dictate that the complete amplitude must be antisymmetric
under the interchange of both momenta and spin, the two
terms in the first of eq.~\dsgmff\ correspond to the scattering
rates with the initial two fermions prepared in an overall
spin-triplet or spin-singlet state.

The total cross section is obtained by integrating eq.~\dsgmff\ over
$t$ and $u$ and dividing by 2 to account for identical fermions
in the final state:
\eq
\label\sgmff
\sigma_{\rm sc}(ff \to ff) = {|\Scy_f|^4 \over 32 \pi \; s} \;
\left[ {3s + 5 \mh^2 \over s+ \mh^2} - \; {2 \mh^2 \over
s} \; \left( {3s +5m_h^2\over s+2 m_h^2} \right)\;\ln \left( {s + \mh^2
\over \mh^2} \right) \right] .
\eeq

\subsubsection{Scattering of Identical Fermions: $f \ol{f} \to f \ol{f}$}

The final high-energy process we consider is the reaction
$f\ol{f} \to f \ol{f}$, whose cross section may be obtained
from the result of the previous section by crossing symmetry.
One finds in this way  $\sigma(f \ol{f} \to f \ol{f})
= \sigma_\SM + \sigma_{\rm sc}$, with:
\eq
\label\dsgmffb
{d \sigma_{\rm sc} \over dt \, du}(f \ol{f} \to f \ol{f}) =
{|\Scy_f|^4 \over 16 \pi \; s^2} \left[
{t^2 \over (t - \mh^2)^2} + {st \over (s - \mh^2) (t-\mh^2)}
+  {s^2 \over (s - \mh^2)^2} \right] \delta(s+t+u) .
\eeq

The integrated cross section then becomes:
\eq
\label\sgmffb
\eqalign{
\sigma_{\rm sc}(f\ol{f} \to f\ol{f}) &= {|\Scy_f|^4 \over 16 \pi \; s} \;
\left[ {s + 2 \mh^2 \over s+ \mh^2} +
{s^2 \over (s - \mh^2)^2} + {s \over s - \mh^2} \right. \cr
&\qquad\qquad \left. - \; {\mh^2 \over s} \;
\left({3 s - 2\mh^2 \over s - \mh^2}
\right)  \ln \left( {s + \mh^2 \over \mh^2} \right) \right] .
\cr}
\eeq




The dependence of these results on CM scattering
angle and energy are summarized in Figs.~\Bhabvs-a
and \Bhabvs-b. Searches for scalar contributions to the process
$e^+e^- \to e^+e^-$ at LEP furnish
the best existing bounds on the strength of the electron
Yukawa couplings, which depend in detail on the scalar mass
\LEPhbounds. The constraints inferred for the size of potential
electron Yukawa couplings improve as $\mh$ increases from
100 to 190 GeV, and then decrease with still higher
scalar masses. The 95\% CL exclusion limits at a few
representative energies are
\eq
\label\scheebounds
\eqalign{
|\Scy_e| &\lsim 0.12 \qquad \hbox{if }
\qquad \mh \sim 100 \GeV,\cr
&\lsim 0.01 \qquad \hbox{if }
\qquad \mh \sim 190 \GeV , \cr
&\lsim 0.16 \qquad \hbox{if} \qquad
\mh \sim 300 \GeV.\cr}
\eeq

\subsection{Expressions for Low-Energy Observables}

Besides contributing directly to high-energy scattering,
scalar particle exchange can induce various observable
effects at energies, $E$, much smaller than $\mh$.
Even though the size of these effects are typically
suppressed by small powers like $E/\mh$, the high
precision of low-energy measurements can sometimes
lead to strong constraints on combinations of couplings.

Precisely because the applications envisaged are to energies much
smaller than $\mh$, it is again fruitful to employ an
effective-lagrangian approach. To do so we imagine successively
integrating out all particles, starting with the new scalar and
proceeding down to very low energies. Of the many effective
interactions amongst the ordinary SM particles which are generated
by integrating out the scalar, those of lowest dimension are
most important, being least suppressed by powers of $1/\mh$.
A complete list of these is easily made, and here we follow the
notation of ref.~\bigfit\ (which also gives expressions for how
these effective couplings contribute to observables, as well
as numerical bounds which experiments imply for their size).

The following three classes of operators emerge as being of
most interest (and are examined in more detail in the remainder
of this section):
\item{1.}
{\it Dimensions 2, 3 and 4:} The lowest-dimension effective
couplings which are well measured describe $W$ and $Z$
propagation (most notably through the parameter $\rho = \mw^2/
\mz^2 \cw^2$)
as well as $W$ and $Z$ couplings to fermions, and so are constrained
by precision electroweak data. Because these receive contributions
at loop level in the scalar effective theory, the bounds they imply
for the scalar effective couplings are comparatively weak. Other
operators like fermion masses and trilinear and quartic gauge-boson
self-couplings also arise at this dimension, but do not provide
practical constraints on scalar couplings
because they are much less
accurately measured.
(For example, to see possible contributions of $h$ loops
to $m_e$ would require independent precise measurements of the
electron mass at low and high energies, and the latter 
of these is practically impossible.)
\item{2.}
{\it Dimensions 5:}
Of the larger number of dimension-5 effective interactions,
anomalous photon and gluon couplings to fermions are the best
measured, and so furnish the best bounds on the scalar couplings.
Of these the CP-violating electric- and chromoelectric-dipole operators
provide the best limits, because they cannot be confused with
SM CP violation.
\item{3.}
{\it Dimensions 6:}
Some four-fermi interactions are the best measured of the many
dimension-six operators which are possible. This is particularly
true for those operators which break exact or approximate SM
conservation laws, like flavour conservation or CP symmetry.

Before turning to the more detailed discussion of these
categories, a general remark is required concerning the nature
of the bounds which can be obtained in this way.
Since the generic Lagrangian presented in \S2\ is
effective, it is imagined to have been obtained after
integrating out other heavier degrees of freedom having
mass $\Lambda$, say. In general this integration generates
many effective couplings which do not involve the scalar
$h$ at all, in addition to the $h$-dependent ones which are
our focus here. The main point to be made is that
we disregard any possible conspiracy
between these two classes of contributions to low-energy
observables, and treat the experimental limits as
applying solely to the $h$-mediated contributions.
This is reasonable so long as the bound obtained is
quite strong, since then the cancellation which would
be required between the scalar-mediated and direct
contributions in order to evade this bound would be unnatural.
It is a less believable assumption if the bounds are weak,
since then partial cancellations between these contributions
to observables could be plausible.

\subsubsection{Integrating Out the Scalar (Tree Level)}

We now integrate out the scalar at tree level.
There are two types of new interactions
which are induced by scalar particle exchange: effective quartic gauge boson
self-couplings and effective four-fermion interactions. We find
the following quartic boson couplings:
\eq
\label\Boseint
\Scl^{\rm (tree)}_b  = {a_\ssw^2 \over 2 \, \mh^2}
\; (W^*_\mu \, W^\mu)^2 + {a_\ssz^2 \over 8 \, \mh^2}
\; (Z_\mu \, Z^\mu)^2 .
\eeq
Unfortunately deviations of these couplings from their SM
values are not strongly constrained experimentally, and so the
limits derived from them are very weak.

The four-fermion interactions arising at tree level are
\eq
\label\Fermiint
\eqalign{
\Scl^{\rm (tree)}_f &= - \; {1 \over \mh^2} \sum_{ff'gg'}
\bar f(y_{ff'}+ i\gamma_5 z_{ff'}) f' \;
\bar g(y_{gg'}+ i\gamma_5 z_{gg'}) g', \cr
&= - \; {1 \over \mh^2} \sum_{fg}
\bar f(y_{f}+ i\gamma_5 z_{f})
f \; \bar g(y_{g}+ i\gamma_5 z_{g}) g +
\hbox{(flavour-changing
terms)} , \cr}
\eeq
where $y_f = y_{ff}, z_f = z_{ff}$ and
the sum runs over light fermions having masses
$m_f \ll \mh$.  The terms not written explicitly in the
second equality involve the flavour-changing couplings, $y_{fg}$
and $z_{fg}$, which we assume to be negligibly small.

Low-energy loop corrections to this Lagrangian can be important,
largely due to the strength of the strong interactions at low energies.
Thus, the interactions of eqs.~\Boseint\ and \Fermiint\ should be
evolved using the renormalization group (RG) down to the energy
scales at which a specific low-energy process occurs (e.g. $m_B$, $m_D$,
$m_K$, $m_n$, etc.). This can be done perturbatively unless the
low-energy scale lies below the QCD scale.
In what follows we incorporate the resulting leading-order
perturbative anomalous dimensions into our expressions when
computing implications for observables.

\fig\vacpolfigs

\midinsert
\vskip 5mm
\centerline{\epsfxsize=7.5cm\epsfbox{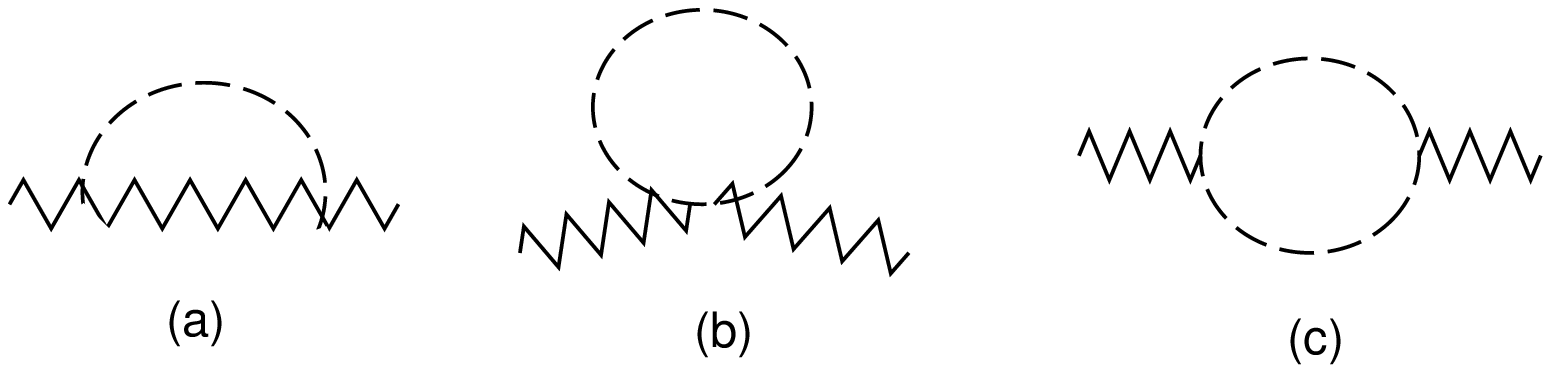}}

\centerline{{\rm Figure \vacpolfigs: }}
\medskip
\caption{The Feynman graphs which
contribute at one loop to gauge boson vacuum polarizations.\medskip}
\endinsert

\subsubsection{Integrating Out the Scalar (One Loop)}

At one-loop level a much wider variety of effective operators
is possible, although with effective couplings which
necessarily involve small loop-suppression factors.
These can nevertheless be important if the
reactions they mediate can be cleanly predicted in the
SM and are very well measured. With a few
exceptions -- such as precision tests of QED -- this
requirement focuses our attention onto low-dimension
effective operators which violate exact or
approximate SM selection rules or conservation laws.

Since we do not consider flavour-changing scalar
Yukawa couplings, we become restricted to violations of
other approximate conservation laws (like mass suppressions
due to approximate helicity conservation, or
CP violation), and to precision measurements
like anomalous magnetic moments.
We are led in this way to consider loop-generated contributions to
gauge-boson vacuum polarizations as well as to
effective fermion-photon and
fermion-gluon effective interactions.

\topic{Oblique Corrections}

The graphs of fig.~\vacpolfigs\ give the one-loop scalar-mediated
contributions to the electroweak vacuum polarizations. All
three graphs are ultraviolet divergent, implying that the scalar/gauge-boson
couplings mix under renormalization with effective spin-one
mass and kinetic operators. This mixing introduces a logarithmic
dependence on the renormalization point, $\mu$, of these effective
couplings, which are the effective theory's way of keeping track of
potentially large logarithms of mass ratios in any underlying theory.

For instance, integrating out a single scalar field, $h$, gives
the following contributions to the $W$ and $Z$ effective mass terms,
renormalized at $\mu = \mh$:
\eq
\label\deltavmass
\delta m_\ssv^2  =  {1 \over 16 \pi^2}
\; \left[ - \; a_\ssv^2 + b_\ssv \; \mh^2 +
{g_\ssv^2 \over 4}  \; \left(\mh^2 + M_\ssv^2 \right) \right] \;
\ln \left({\Lambda^2 \over \mh^2} \right) ,
\eeq
where we neglect all masses in comparison with $\mh$,
$V = W,Z$ and $\Lambda$ is the scale at which the
original scalar effective lagrangian of \S2\ is defined. (In practice
$\Lambda$ can be taken as the mass of the lightest particle
which generated $\Scl_\eff$ once it was integrated out.)
The quantity $g_\ssv$ requires some explanation:
$g_\ssz$ denotes the trilinear coupling of eq.~\dimfouropsv\
between the new scalar, $h$, and the unphysical scalar, $z$, which
would have been absorbed into the longitudinal $Z$ polarization in
unitary gauge, but which arises in the renormalizable gauges we
use to evaluate figs.~\vacpolfigs. $g_\ssw$ denotes the
analogous coupling between $h$, $W$ and the electrically charged
unphysical scalar, $w$. The SM contribution is obtained from
\deltavmass\ by substituting the couplings of eqs.~\SMdthree\
and \SMdfourb, and using $g_\ssz = e/( \sw\cw)$ and
$g_\ssw = e/\sw$.

\ref\obliquea{M.E. Peskin and T. Takeuchi, \prl{65}{90}{964};
\prd{46}{92}{381};
W.J. Marciano and J.L. Rosner, \prl{65}{90}{2963};
D.C. Kennedy and P. Langacker, \prl{65}{90}{2967};
G. Altarelli and R. Barbieri, \plb{253}{91}{161}.}

\ref\obliqueb{M. Golden and L. Randall, \npb{361}{91}{3};
B.Holdom and J. Terning, \plb{259}{91}{329}.}

\ref\stuvwx{I. Maksymyk, C.P. Burgess and D. London,
\prd{50}{94}{529}.}

Although these mass shifts are not themselves detectable, they
contribute to measurable `oblique' corrections to $W$ and $Z$
vacuum polarizations \obliquea, \obliqueb.\foot\alphafoot{If powers
of $\ss \mz^2/\mh^2$ cannot be neglected then a generalization of the usual
parameterization of these oblique corrections is required \stuvwx.}
For example, they contribute to the parameter 
$\rho = \mw^2/(\mz^2\cw^2)$ as follows:
\eq
\label\deltarholoop
\eqalign{
\Delta \rho = \rho -1  &= - \; {1 \over 16 \pi^2}
\; \left\{  \left[ \left( {a_\ssw \over \mw} \right)^2 -
 \left( {a_\ssz \over \mz} \right)^2 \right] 
- \mh^2 \left[ \left( {b_\ssw \over \mw^2} \right) -
 \left( {b_\ssz \over \mz^2} \right) \right]   \right. \cr
&\left.
- {1 \over 4} \; \left[ \left(\mh^2 + \mw^2 \right)
\left( {g_\ssw \over \mw} \right)^2 -
\left(\mh^2 + \mz^2 \right)
 \left( {g_\ssz \over \mz} \right)^2 \right] \right\}
\; \ln \left(
{\Lambda^2 \over \mh^2} \right) .\cr}
\eeq
Although all terms proportional to $\mh^2$ in this equation 
vanish when specialized
to the couplings of the SM Higgs, the others do not. The
resulting nonzero value precisely cancels the divergent
($\ln\Lambda^2$) contributions to $\Delta \rho$ which arise from
other SM graphs, not involving the physical Higgs, $h$. Once the
$\Lambda$ dependence of these other graphs has cancelled in this way,
what remains is the well-known SM logarithmic $\mh$ dependence
of $\Delta \rho$ \obliquea:
\eq
\label\lnmtinrho
\Delta \rho = - \;  {3 \alpha \over 16 \pi \cw^2} \; \ln \left(
{\mh^2 \over \mz^2} \right) + (\hbox{non logarithmic terms}).
\eeq
Contributions to the other oblique parameters may be computed
along similar lines, although we do not do so here because the bounds
which follow on the effective scalar couplings are not particularly strong.

\fig\dipoles

\midinsert
\vskip 5mm
\centerline{\epsfxsize=7.5cm\epsfbox{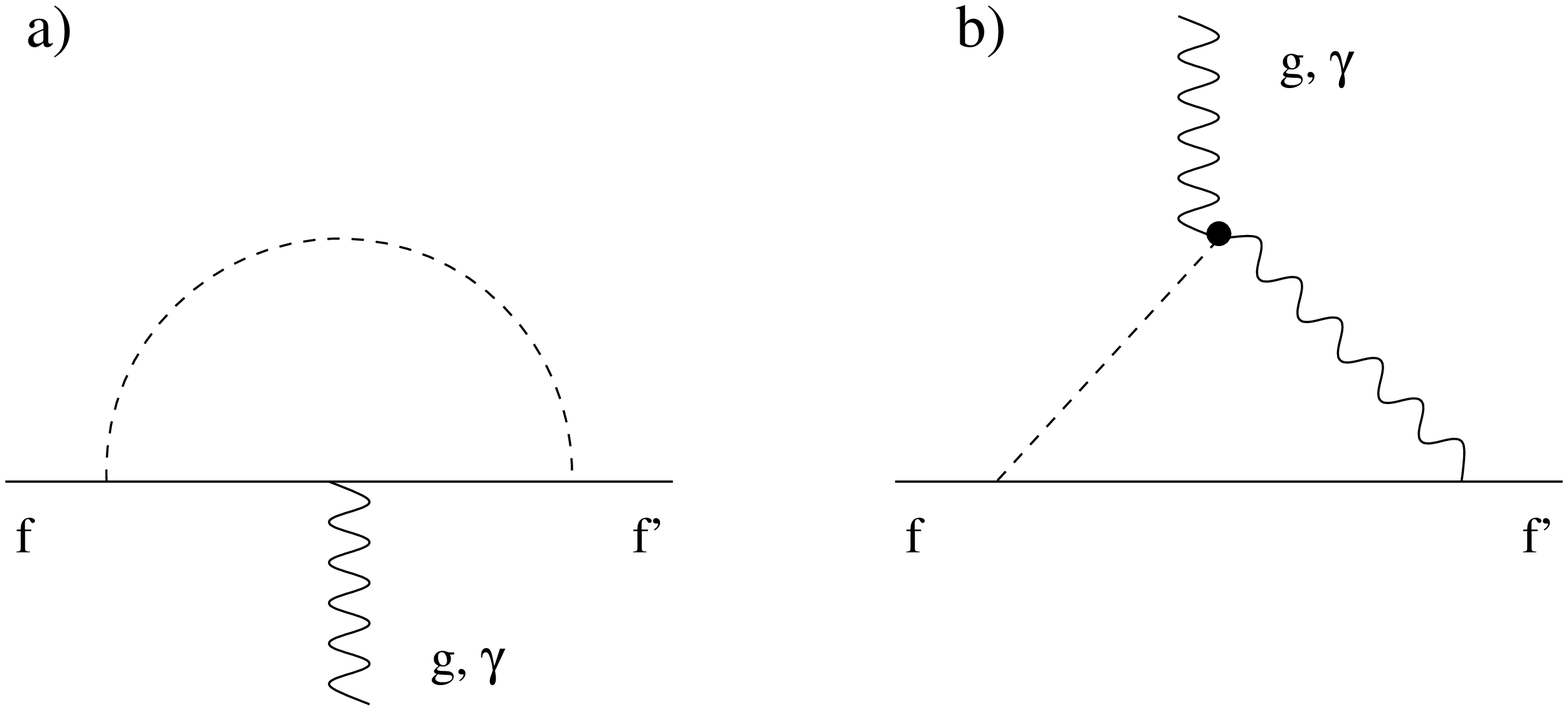}}

\centerline{{\rm Figure \dipoles: }}
\medskip
\caption{The Feynman graphs which
contribute at one loop to low energy effective
interactions between light fermions and
massless gauge bosons.\medskip}
\endinsert

\topic{Vertex Corrections}

Another class of well-measured effective operators are the
dimension-five modifications to the fermion-photon and
fermion-gluon interactions, which we consider in detail next.

Evaluating the scalar-fermion and scalar-gauge boson
exchange diagrams of Fig.~\dipoles, we
consider separately the contribution of graphs ($a$)
and ($b$).
It is useful to group the results of evaluating graph ($a$)
according to whether they are proportional to the internal
fermion mass, the external fermion mass, or neither.
That is: $\Scl^a = \Scl^a_{\rm int} + \Scl^a_{\rm ext} +
\Scl^a_{\rm neith}$. Neglecting all external momenta
and masses in comparison with $\mh$ we find,
for flavour-changing Yukawa couplings:
\eq
\label\lowenergy
\eqalign{
\Scl^{a}_{\rm int} &= \frac{eF_{\mu\nu} }
{32\pi^2\mh^2} \sum_{fkf'} Q_k
 m_k G_1 \left( {m_k^2 \over \mh^2} \right)
 \bar f \sigma^{\mu\nu} [y_{fk} y_{kf'} -
 z_{fk} z_{kf'} + i\gamma_5
(y_{fk} z_{kf'} + z_{fk} y_{kf'})] f' ,\cr
\Scl^a_{\rm ext} &= \frac{eF_{\mu\nu}}
{32\pi^2\mh^2}  \sum_{fkf'} Q_k
G_2 \left( {m_k^2 \over \mh^2} \right)
 \bar f \sigma^{\mu\nu} [(m_f+m_{f'})
 (y_{fk} y_{kf'}+ z_{fk} z_{kf'}) \cr
 &\qquad\qquad \qquad\qquad \qquad\qquad +
 i\gamma_5 (m_f-m_{f'}) (y_{fk} z_{kf'}
- z_{fk} y_{kf'})] f' \cr
\Scl^a_{\rm neith} &= \frac{ e\partial^\mu F_{\mu\nu}}
{32\pi^2\mh^2}
\sum_{fkf'} Q_k G_3 \left({m_k^2 \over \mh^2}\right)
\bar f\gamma^\nu
[y_{fk} y_{kf'}+ z_{fk} z_{kf'}+
i\gamma_5 (y_{fk} z_{kf'}
-z_{fk} y_{kf'})] f' ,\cr}
\eeq
The result for the case of external gluons is obtained
from this by a simple substitution:
$e Q_f F_{\mu\nu}\rightarrow g_s
G^a_{\mu\nu}t^a$. Furthermore, since one is free
to simplify an effective lagrangian using the
lowest-order equations of motion, the derivative $\partial^\mu
F_{\mu\nu}$ can be eliminated in favour
of the electromagnetic (or colour) current by
using the equations of motions for
the electromagnetic (gluon) field. Expressions \lowenergy\ follow
for light external fermions if only terms linear in $m_f$ and
$m_{f'}$ are retained. The (possibly
large) mass, $m_k$, of the intermediate fermion is not
neglected, and appears through the invariant
functions $G_k(a)$, given by
\eq
\label\integrals
\eqalign{
G_1(a)&={1\over 2(1-a)^2}( - 3+4a-a^2-2\ln a ) \cr
G_2(a)&={1\over 12(1-a)^4}(20-39a+24a^2-5a^3+12\ln a - 6a \ln a)\cr
G_3(a)&=-{1\over 18(1-a)^4}(38-81a+54a^2-11a^3+24\ln a - 18a \ln a)
. \cr}
\eeq
For flavour-diagonal couplings, $f=f'=k$
we have significant simplifications
and the results are given by
\eq
\label\lowenergyfda
\Scl^{a} = e \sum_{f} \left[
 -{1\over 2}\; F_{\mu\nu} \; \bar f \sigma^{\mu\nu}
\Bigl(d^a_f +i\gamma_5 \tw{d}^a_f \Bigr) f
+  A^a_f \; \partial^\mu F_{\mu\nu} \bar f
\gamma^\nu  f \right],
\eeq
with the leading contributions to these
effective couplings being given (for large $\mh^2/m_f^2$) by:
\eq
\label\coefdefs
\eqalign{
d^a_f &= \frac{Q_f m_f}{16\pi^2\mh^2} \;
\ln\left({m_h^2 \over m_f^2}\right) \;
\left( y_f^2 + 3z_f^2 \right),\cr
\tw{d}^a_f &= -\frac{Q_f m_f  }{8\pi^2\mh^2} \;
\ln\left({m_h^2 \over m_f^2}\right) \;
 y_f z_f\,\cr
A^a_f &= \frac{ Q_f m_f }{24\pi^2\mh^2} \;
\ln\left({m_h^2 \over m_f^2}\right) \;
\left(y_f^2 + z_f^2 \right) .\cr}
\eeq

Unlike graph ($a$), graph ($b$) of Fig.~\dipoles\ diverges
(logarithmically) in the ultraviolet implying a mixing between the operators
of eq.~\dimfiveops\ and the (chromo-) magnetic- and electric-
dipole moment operators we are considering here. As was the
case for the vacuum polarization graphs considered earlier, this mixing
tracks potentially large logarithms, through the logarithmic
dependence on the renormalization point which it implies
for these effective couplings. For example,
if in the SM we regard $c_g$ and $c_\gamma$ to be generated
by integrating out the $t$ quark, then the effective couplings
are defined by matching conditions at scale $\mu = m_t$.
Renormalization of the effective couplings down to $\mu = \mh$,
where the scalar particle is integrated out, generates the
logarithmic factor $\ln(m_t^2/\mh^2)$, which directly appears
in a full SM calculation.

To the extent that such logarithms are large, it makes sense to keep
track of only the divergent term, and this is most easily done
simply by cutting off the integral at the scale of new physics 
responsible for the generation of the $c_i$. Although this misses
terms which are subdominant to the logarithm, which can be dangerous
if the mass ratio is not enormous (as $m_t/\mh$ is not for a
light SM Higgs), it suffices for our present purposes wherein we
look only for order-of-magnitude limits. For the same reasons,
it suffices for numerical purposes to evaluate the large logarithm
itself as being order unity, which we do in what follows.
Thus we have
\eq
\label\lowenergyb
\eqalign{
\Scl^{b} &= \frac{eF_{\mu\nu}}{16\pi^2}\;
\ln\left( {\Lambda^2 \over \mh^2} \right)
\sum_{ff'} Q_f \bar f \sigma^{\mu\nu}
\left(y_{ff'}c_\gamma +  z_{ff'}\tilde c_\gamma +
i\gamma_5(y_{ff'}\tilde c_\gamma +
z_{ff'}c_\gamma) \right )f' \cr
&= -{e\over 2}\; F_{\mu\nu}
\sum_{f} \bar f \sigma^{\mu\nu}
\left(d^b_f + i\gamma_5 \tw{d}^b_f \right )f  +
\hbox{(flavour-changing terms)} ,\cr}
\eeq
with
\eq
\label\dfrunning
d_f^b = - \; \frac{ Q_f}{16 \pi^2}
\; \left( y_{f} c_\gamma +
z_{f}\tilde c_\gamma \right )
\; \ln\left( {\Lambda^2 \over \mh^2} \right)
,\qquad
\tw{d}_f^b = - \;\frac{ Q_f}{16 \pi^2}
\;
\left(y_{f}\tilde c_\gamma +
z_{f}c_\gamma \right )
\; \ln\left( {\Lambda^2 \over \mh^2} \right) .
\eeq
These expressions have the noteworthy feature
of not being suppressed by inverse powers of $\mh$,
making them more sensitive in the limit of
large scalar masses.

\topic{Four-Fermion Interactions}

Effective four-fermion interactions receive non-trivial contributions
at one loop level which may contain more information about various
scalar-fermion couplings. In particular, the box diagrams, involving
either $hh$ or $hW$ exchange, may induce some structures not
already contained in the tree-level terms. Since we neglect
flavour-changing couplings for $h$, it turns out that only the
$hW$-exchange diagram produces significant bounds:

\fig\boxes

\midinsert
\vskip 1cm
\centerline{\epsfxsize=3.5cm\epsfbox{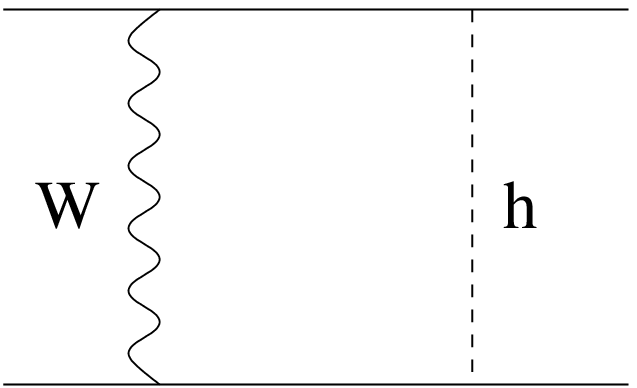}}
\centerline{{\rm Figure \boxes: }}
\medskip
\caption{The Feynman graph (plus those obtained through
various permutations of the bosonic lines) which
contribute into a low energy four-fermion interaction.\medskip}
\endinsert
\eq
\label\Fermbox
\eqalign{
\Scl_{box} &=
-{\GF \over\sqrt{2}}\;{ V_{ff'}V^*_{gg'}
\over 16\pi^2}\;{\mw^2 \over\mh^2- \mw^2}\;
\ln\left({m_h^2 \over \mw^2}\right) \cr
&\qquad\qquad
\times \sum \Bigl[ (\bar f\Gamma_{ff'}f') \; ( g\Gamma_{gg'}g' )
+ (\bar f\sigma_{\mu\nu}\Gamma_{ff'}f')\; ( g\sigma^{\mu\nu}
\Gamma_{gg'}g' ) \Bigr] + h.c. .\cr}
\eeq
In this expression $ V_{ff'}$ are the elements of the
Kobayashi-Maskawa mixing matrix and $\Gamma_{ff'}$
denotes the following combination of couplings:
\eq
\Gamma_{ff'}=y_{f}-y_{f'}-iz_{f}-iz_{f'}-
\gamma_5(y_{f}+y_{f'} -iz_{f}-iz_{f'})
\eeq
As in the previous cases, the box graphs are computed in the limit
of small external masses and momenta. We also use an explicitly
renormalizable gauge for the $W$-propagator.

\subsubsection{Comparison with Precision Measurements}

We now turn to constraints on the couplings
of the effective low-energy Lagrangians we
have just constructed.

\topic{Anomalous Magnetic Moments}

\ref\pdg{Particle Data Group: {\it Review of Particle
Properties}, \ejp{C3}{98}{280}.}

The quantities $d_e = d^a_e + d^b_e$ and
$d_\mu = d^a_\mu + d^b_\mu$ contribute
directly to the anomalous magnetic moment of
the electron and muon, with
\eq
\delta a_i = 2 m_i  \; d_i \eeq
where the anomalous moment is defined as
$a_i = (g_i -2)/2 = (\mu_i/\mu_{\ssb_i}) - 1$, 
where $\mu_{\ssb_i} \equiv e_i/2 m_i$ is
the Bohr magneton (in units for which $\hbar=c=1$)
and $i = e, \mu$.

A bound is obtained by requiring $\delta a_\mu$ to
be smaller than the 1.6$\sigma$
experimental error of the present experimental value
\pdg, $a_\mu^{\rm exp} = 1~165~923~0(84) \times 10^{-10}$.
The strong $(m_\mu/\mh)^2$ dependence of $m_\mu \,
d^a_\mu$ implies that only a weak bound on the scalar Yukawa
couplings is obtained from this constraint:
\eq
\label\amubounda
\left( {100 \GeV \over \mh} \right)^2
\Bigl(y_\mu^2 + 3 z_\mu^2 \Bigr)
< 0.07.
\eeq
For this limit we have used $\mh =
100$ GeV to evaluate the logarithm of eq.~\coefdefs.
The same bound as applied to $d^b_\mu$ in eq.~\dfrunning\
(evaluated at $\mu = m_\mu$) gives a stronger limit,
due to its weaker dependence on $\mh$:
\eq
\label\amuboundb
\Bigl(y_\mu \, c_\gamma + z_\mu \,
\tw{c}_\gamma \Bigr)
< 5 \pwr{-4} .
\eeq
Although eq.~\amuboundb\ looks quite restrictive,
in order to limit $y_\mu$ and $z_\mu$
some information about the effective couplings
$c_\gamma$ and $\tw{c}_\gamma$ is necessary. To get
a rough estimate we use the SM $t$-quark value of
eq.~\tqdimfv: $\tw{c}_\gamma^\SM = 0$ and $c_\gamma^\SM(t)
= 2 \alpha/(9 \pi v) = 2.25 \pwr{-6} \GeV^{-1}$,
to learn that
\eq
\label\bestamubd
y_\mu \; \left( {c_\gamma \over
c^\SM_\gamma(t)} \right) \lsim 2.
\eeq
Clearly only weak limits are obtained from these observables.

\topic{Electric Dipole Moments}

\ref\book{  I.B. Khriplovich and S.K. Lamoreaux, {\it ''CP
Violation Without Strangeness''}, Springer, 1997.}

The presence of both scalar and pseudoscalar couplings of
$h$ to fermions breaks CP invariance. In the
flavour-conserving channel this should  lead to the
generation of electric dipole moments (EDMs) for
elementary particles and atoms. This can be used to
usefully constrain the effective couplings because
SM CP-violation due to the Kobayashi-Maskawa matrix
is known to produce EDMs many orders smaller
than are the current experimental limits
(see, e.g. ref.~\book).

\ref\nEDM{ K.F. Smith {\it et al.}, \plb{234}{90}{191};
I.S. Altarev {\it et al.}, \plb{276}{92}{242}.}

\ref\eEDM{ E.D. Commins {\it et al.},
\pra{50}{94}{2960}.}

\ref\mEDM {J.P. Jacobs
{\it et al.}, \prl{71}{93}{3782}.}

The best experimental constraints come from the searches
for the EDMs of the neutron ~\nEDM, paramagnetic \eEDM\
and diamagnetic atoms \mEDM. Although the results of the
experiments with paramagnetic atoms are often interpreted
as limits on the electron EDM, there are often
many other possible contributions to the EDM of
paramagnetic atoms from other effective low-energy
couplings, such as from CP-odd semi-leptonic
four-fermion operators like $\bar e \gamma_5 e \;
\bar qq$ and the like. Because of the potentially
large number of effective operators in eqs.~\Fermiint\
and \lowenergy\ which can contribute to the EDMs of
the neutron and diamagnetic atoms, closed form expressions
for these moments are problematic, if not impossible, to
obtain.

At the same time, there could be direct
contributions to EDMs generated at the scale $\Lambda$,
without direct participation of the scalar $h$. These cannot
be taken into account in our approach unless the underlying
theory is more fully specified. In fact, for some
models these direct contributions can be
more important than $h$-mediated terms. In view of this,
we follow the strategy outlined earlier of treating
every operator as giving an independent contribution
to EDM, which is separately constrained from experiment,
assuming no cancellations with other terms.
We must keep in mind when so doing not to trust
weak bounds that arise in this way.

\ref\fourfermi{V.M. Khatsimovsky, I.B. Khriplovich and A.S.
Yelkhovsky, \anp{186}{88}{1};
V.M. Khatsimovsky and
I.B. Khriplovich,  \plb{296}{94}{219};
C. Hamzaoui, M.  Pospelov, \prd{60}{99}{036003}.}

\ref\Edmhg{T. Falk, K. Olive, M. Pospelov
and R. Roiban, \npb{560}{99}{3}.}

To obtain constraints we combine the theoretical
analysis performed in refs.~\fourfermi\ and \Edmhg\ with
the experimental results of \nEDM, \eEDM\ and \mEDM.
The limits found in this way are presented in Table 1.
The rows of this table are labelled by flavour-diagonal parity
odd (pseudoscalar) couplings, while its columns are labelled by
parity even (scalar) ones. All CP-violating
effective couplings involve products of one of each of these.
For example, the quantity $\tw{d}_e$ represents
a contribution to the electron EDM, which is
strongly constrained by atomic EDM measurements.
Eq.~\coefdefs~ gives this quantity in terms
of the product $z_{ee} \, y_{ee}$, and the bound
which follows on this product may be read from
the $z_{ee} - y_{ee}$ element of the table.

\ref\GGG{S. Weinberg, \prl{63}{89}{2333}.}

Electron-quark cross terms in the table
represent limits imposed by atomic EDMs on the operators of the
form $\bar e \gamma_5 e \; \bar q q$ and $\bar e e\;\bar q\gamma_5 q$,
while those involving cross terms of the form $z_f c_\gamma$
and $y_f \tw{c}_\gamma$ come from bounds on
effective couplings like those of eq.~\dfrunning.
Finally those involving products of the form $c_g \tw c_g$,
$c_\gamma\tilde c_\gamma$, etc.,
are obtained by estimating their contribution either to the CP-odd three gluon
operator \GGG\ or to quark dipole moments, induced
by these combinations at two loops.

All limits are given for $\mh=100$ GeV and should be
quadratically rescaled for different values of  $\mh$
except for the limits in two last rows and columns which
are sensitive to $\mh$ only logarithmically. The effective
coupling constants $c_{\gamma(g)}$ and $\tilde
c_{\gamma(g)}$ are taken in units of $(100 \GeV)^{-1}$, and
one-loop QCD RG evolution coefficients have been taken
into account where necessary, with all couplings taken
at the renormalization point of 100 GeV.

\midinsert
\vbox{\offinterlineskip
\def\tablerule{\noalign{\hrule}}
\halign{&\vrule#&
  \strut\quad\hfil#\quad\cr
\tablerule
height3pt&\omit&&\omit&&\omit&&
\omit&&\omit&&\omit&&\omit&\cr
&-&&$y_{ee}$&&$y_{uu}$&&
$y_{dd}$&&$y_{ss}$&&$c_\gamma$ &&$c_g$&\cr
height3pt&\omit&&\omit&&\omit&&
\omit&&\omit&&\omit&&\omit&\cr
\tablerule
height3pt&\omit&&\omit
&&\omit&&\omit&&\omit&&\omit
&&\omit&\cr
&$z_{ee}$&&$
1\cdot 10^{-5}$&&$3\cdot10^{-9}$&&
$3\cdot10^{-9}$ &&$6\cdot10^{-9}$
&&$1.5\cdot10^{-9}$&&$1.5\cdot10^{-6} $& \cr
height3pt&\omit&&\omit&&\omit&&
\omit&&\omit&&\omit&&\omit&\cr
\tablerule
height3pt&\omit&&\omit&&\omit&&
\omit&&\omit&&\omit&&\omit&\cr
&$z_{uu}$&&$5\cdot10^{-9}$&&
$1.5\cdot10^{-7}$&&$1.5\cdot10^ {-7}$
&&$3\cdot10^{-7}$ &&$2\cdot 10^{-7}$
&&$1\cdot 10^{-8}$& \cr
height3pt&\omit&&\omit&&\omit
&&\omit&&\omit&&\omit&&\omit&\cr
\tablerule
height3pt&\omit&&\omit&&\omit&&
\omit&&\omit&&\omit&&\omit&\cr
&$z_{dd}$&&$5\cdot10^{-9}$&&
$1.5\cdot10^{-7}$&&$1.5\cdot10^ {-7}$
&&$3\cdot10^{-7}$&&$1\cdot10^{-7}$
&&$1\cdot10^{-8}$&
\cr
height3pt&\omit&&\omit&&\omit&&
\omit&&\omit&&\omit&&\omit&\cr
\tablerule
height3pt&\omit&&\omit&&\omit&&
\omit&&\omit&&\omit&&\omit&\cr
&$z_{ss}$&&$1\cdot 10^{-6}$&&
$7\cdot10^{-7}$&&$7\cdot10^{-7}$
&&$\sim10^{-6}$&&$\sim 10^{-5}$
&&$\sim10^{-6}$& \cr height3pt&\omit
&&\omit&&\omit&&\omit&&\omit&&
\omit&&\omit&\cr
\tablerule
height3pt&\omit&&\omit&&\omit&&
\omit&&\omit&&\omit&&\omit&\cr
&$\tilde c_{\gamma}$&&$1.5\cdot10^{-9}$
&&$2\cdot 10^{-7}$&&$1\cdot 10^{-7}$
&&$\sim10^{-5}$&&$\sim1$&&$\sim10^{-1}$& \cr
height3pt&\omit&&\omit&&\omit&&\omit&&
\omit&&\omit&&\omit&\cr\tablerule
height3pt&\omit&&\omit&&\omit&&\omit&&
\omit&&\omit&&\omit&\cr &$\tilde c_{g}$
&&$1\cdot 10^{-6}$&&$1\cdot 10^{-8}$&&
$1\cdot 10^{-8}$ &&$\sim 10^{-6}$&&
$\sim10^{-1}$&&$\sim10^{-4}$& \cr
height2pt&\omit&&\omit&&\omit&&
\omit&&\omit&&\omit&&\omit&\cr\tablerule}}

\eject
\centerline{{\rm Table 1: }}
\smallskip
\caption{The limits on the couplings of
scalar $\ss h$ with  light fermions, photons and gluons imposed
by various EDM experiments.}
\endinsert

A few comments about these limits are in order. First, we implicitly
assume the absence of dangerous CP-violating operators like
$G_{\mu\nu} \twi G^{\mu\nu}$, such as might be assured by
existence of a PQ symmetry, for example.  Second, several
entries marked with the symbol `$\sim$' represent at best an
order-of-magnitude estimate, since their extraction from the observables
involves hadronic matrix elements which cannot be reliably 
estimated. The same symbol marks the entries which 
give rise to EDMs at two-loop level, which we estimate here 
simply on dimensional grounds, although more refined
calculations are feasible. This is the case for the
$c_\gamma\tilde c_\gamma$ entry, for example, which generates
a quark EDM at two loops. The combination $c_g\tilde c_g$, 
which we bound by estimating the three-gluon CP-odd operator, 
$\Tr\, \Bigl( GG\twi G\Bigr)$ \GGG, which it generates, has in addition
a large matrix-element uncertainty which further
complicates the estimate
for the size it implies for nucleon and atomic EDMs.

These constraints can be further generalized to give bounds on
the $yz$-combinations of heavy-quark/scalar couplings.
Thus, $y_{bb}z_{bb}$ contributes at two-loop order to the
operator, $\Tr\, \Bigl(GG\twi G\Bigr)$, leading
to the approximate bound, $y_{bb}z_{bb} <
10^{-2}-10^{-3}$.

What might be the implications of these limits for collider physics?
Let us imagine a situation in which the scalar-electron coupling
$y_{ee}$ is taken at the edge of the limit coming from
Bhabha scattering, eq.~\scheebounds, and so which conceivably
might be seen in the decay of the new scalar. Then the limit
$y_{ee}z_{ee} \lsim 10^{-5}$ coming from the electron EDM data
is so strong that it requires very high statistics to detect any
spin correlations in the decay $h\rightarrow e^+e^-$.
The EDM constraints cannot similarly exclude measurable
spin correlations in $h$ decays into heavier fermions, however.
For example, even the bound $y_{bb}z_{bb} \lsim 10^{-2}$
quoted earlier is not sufficient to rule out spin correlations in
$h\rightarrow \bar bb$, because it permits $y_{bb}\sim z_{bb}$
to be as large as $|y_{bb}|\sim 0.1$.

\topic{Pion Decays}.

\ref\gstudy{ V. Barger, K. Cheung, K. Hagiwara and  D.
Zeppenfeld, \prd{57}{98}{391}. }

In this subsection we consider the decays of the neutral and
charged pions induced by scalar exchange at tree or
one-loop level. Leptonic pion decay is sensitive to scalar couplings because
of the well-known suppression its SM contribution receives from powers
of the final-state lepton masses. Here we study what may be learned from
these decays about the couplings of  $h$-particles to light fermions.

We start with the decay $\pi^0\rightarrow e^+e^-$, whose amplitude  is
tremendously suppressed in the SM by electromagnetic  loop
factors and the electron mass. It has been measured
only recently to have branching ratio $B = (7.5 \pm 2.0)
\pwr{-8}$. The effective coupling of eq.~\Fermiint, induced by
scalar exchange, contributes to this decay with the rate
\eq
\Gamma_{\pi_0\rightarrow ee} = \eta^2\; {m_\pi\over 8\pi}
\left({m_{\pi}\over \mh}\right )^4
\left({F_\pi\over m_u+m_d}\right )^2
(z_{uu}-z_{dd})^2(z_{ee}^2+y_{ee}^2),
\label\piee
\eeq
where $\eta \simeq 5$ is a QCD RG coefficient. Here we
evaluate  $\bra{\hbox{vac}}
\left( z_{uu} \, \bar u \gamma_5 u 
+ z_{dd} \, \bar d \gamma_5 d \right)
\ket{\pi^0}$, which is the relevant strong matrix element, using chiral
perturbation theory, with $F_\pi = 92$ MeV
and current quark masses defined at the renormalization point
$\mu = 1$ GeV. 

Although the observed branching ratio agrees in order of
magnitude with SM predictions, a detailed comparison
is complicated by theoretical uncertainties in the SM amplitude.
We therefore obtain our bound by requiring
eq.~\piee\ to be of order the SM process, which we take
to be  $\lsim 10^{-7}$ of the total width. This 
produces the following limit:
\eq 
\label\pibound
\left({100 {\rm GeV}\over \mh}\right)^4
(z_{uu}-z_{dd})^2(z_{ee}^2+y_{ee}^2) \lsim 10^{-4}  .
\eeq

The decays of charged pions can be even more restrictive.
In this case the hadronic uncertainties can be cancelled
by considering a ratio of two branching ratios, such as
$R = \Gamma_{\pi^+ \rightarrow \nu_e e^+}/
\Gamma_{\pi^+\rightarrow \nu_\mu \mu^+}$. Although this process
proceeds at tree level if charged scalars exist, it cannot do so given
only neutral scalars, such as is the present focus. Instead it
is mediated by the effective interaction, eq.~\Fermbox, induced
by the box graph involving $hW$ exchange. This effective Lagrangian
generates a decay amplitude which interferes with the SM
result, giving the following correction to the $\nu_e e$ partial width:
\eq
\delta\Gamma_{\pi^+\rightarrow \nu_e e^+} = \eta\;
{m_\pi^3 F_\pi^2  \GF^2 |V_{ud}|^2 \over 16\pi^3}
\left({\mw^2 \over\mh^2- \mw^2}\right)
\ln\left({m_h^2 \over \mw^2}\right) {m_e\over m_u+m_d}
\left(y_e(z_{d}-z_{u})+z_e(y_{d}+y_{u})\right)
\label\piplusee
\eeq
This correction can be quite significant since the SM decay width is
suppressed by the small electron mass.

Measurements of the ratio $R$
constrain the Yukawa coupling combination in
eq.~\piplusee\ provided that these couplings break lepton
universality (as is often true in models having large
Yukawa couplings). If so, then the experimental bound
$\delta R \approx \delta \Gamma_{\pi^+\rightarrow \nu e^+}/
\Gamma_{\pi_+\rightarrow \nu_\mu \mu^+} < 8\cdot 10^{-7}$,
implies an important limit:
\eq
y_e(z_{d}-z_{u})+z_e(y_{d}+y_{u})
<4\cdot 10^{-5}.
\label\pipluslim
\eeq

In summary, we see that while there are important limits on
scalar/fermion and scalar/gauge-boson couplings coming from
low- and high-energy processes, they are relatively weak in that
they do not cover much of parameter space. In particular, they are
insufficient as yet to significantly constrain the best-motivated
models of new scalar physics, and they are very far from
being sensitive to the Higgs-fermion couplings predicted by the SM.
This is true even for electron couplings, about which the strongest
limits are known.

\section{Predictions for Effective Couplings in Specific
Models}

Whereas the previous sections are dedicated to identifying
the possible low-energy scalar interactions and their
implications for experiment in as model-independent a
way as possible, in this section we change gears and
explore what form the effective couplings take
in a number of representative models.
Because it is more model-specific, this section more closely follows
results that appear in many places throughout the literature.

Our main motivation for making contact with models is to
profit from the interplay between the generality of the
effective lagrangian approach, and the predictiveness
of the underlying models. By comparing how different
classes of models contribute to the effective couplings,
and combining this with the previous sections' expressions
for their implications for observables, we intend to
learn which observables best discriminate between
the various models on the market. Indeed, the
detailed discussion of this discrimination is the
climax of this paper, and is the topic of the next
section.

Partially motivated by the next section's comparison,
the models are discussed in this section in the order
of increasing similarity with the Standard Model.
We start with a very brief discussion of a particular
kind of composite scalar which does not carry
a linear realization of the electroweak gauge
group, but then devote most of our attention to
models involving scalars in various electroweak
representations.

\subsection{Strong Electroweak Symmetry Breaking}

\ref\tcreviews{E. Farhi, L. Susskind, \prep{74}{81}{277--321};\bk
T. Appelquist, in {\it Beijing 1993, Particle physics at the Fermi scale},
proceedings of the CCAST Symposium on Particle Physics at the Fermi Scale, 
Beijing, China, 27 May - 4 Jun 1993.}

Models for new physics broadly divide into two classes,
according to whether or not the sector which breaks
the electroweak gauge symmetries is strongly or weakly
coupled. Unfortunately, the predictions of models in which
this sector is strongly coupled \tcreviews\ are difficult to
accurately obtain, precisely because of these strong
couplings. Partly because of this there is no canonical
reference model which  involves well-understood
dynamics and which is known to satisfy all experimental
constraints. This makes a general survey of their
properties more difficult to perform, since these
properties are less definitively understood.

\ref\tcpseudos{
S.W. Weinberg, \prl{29}{72}{1698};
L. Susskind, \prd{20}{79}{2619}
}

For the present purposes we content ourselves with
focussing on some of the properties of the lightest
scalar states, which typically arise as
pseudo-Goldstone bosons for spontaneously broken
approximate flavour symmetries \tcpseudos. Since the
low-energy couplings of such particles are more
constrained by the symmetry-breaking pattern itself,
their properties may be predicted more robustly.

\ref\GBreview{For a recent review see, for example,
C.~Burgess, {\it Phys. Rept.} {\bf 330} (2000) 193.}

In particular, since the Goldstone boson for any
exact symmetry must decouple at low energies, variables
can always be chosen to ensure that these bosons
are derivatively coupled in the low-energy action
\GBreview. Of course, strictly massless bosons are
not of realistic interest in the present context, so the relevant spontaneously
broken symmetries are usually taken to be only approximate.
Nonzero masses and nonderivative couplings are then permitted,
but they are suppressed by the small symmetry-breaking
parameters which must be present if the symmetry limit is to
be a good approximation. Unfortunately, to the extent that
the low-energy scalar couplings do depend on these parameters
they also become more model dependent.

\ref\technireview{
R. Casalbuoni, A. Deandrea, S. de Curtis, D. Dominici, F. Feruglio,
R. Gatto, M. Grazzini, \plb{349}{95}{533};
R. Casalbuoni, A. Deandrea, S. de Curtis, D. Dominici,
R. Gatto, M. Grazzini, \prd{53}{96}{5201};
R. Casalbuoni, S. de Curtis, D. Dominici, M. Grazzini
 \prd{56}{97}{2812};
  R. Casalbuoni, A. Deandrea, S. de Curtis, D. Dominici,
R. Gatto, J.F. Gunion, {\it JHEP} {\bf 9908} (1999) 011.}

\ref\barger{V. Barger, W.Y. Keung, \plb{185}{87}{431};
V. Barger, R. Phillips, Collider Physics, ed. by Addison-Wesley 1987.
}

One feature which often distinguishes such pseudo-Goldstone
bosons (pGBs) \technireview\ is their absence of direct
tree-level couplings to electroweak gauge bosons.
This is because the electroweak gauge group is
typically a subgroup of the larger flavour group
whose spontaneous breaking produces the pGBs in
the first place. But because pGBs quite generally
transform in the adjoint representation of this
group, they typically must transform as either
singlets or doublets of $SU_\ssl(2)$. Since electroweak
interactions typically raise the masses of the doublet
states, one often finds the lowest-energy neutral
scalar to be an electroweak singlet, which does not
couple to $W$'s and $Z$'s at tree level \barger.

$W$ and $Z$ couplings can be generated once
loops are taken into account, however. For instance,
an analogy with QCD suggests one class of effective
couplings which can be phenomenologically significant,
and yet also is reasonably model independent.
This class consists of the parity-violating
dimension-five interactions of eqs.~\dimfiveops,
since these are often determined purely by the
anomalies of the global symmetries of interest.
For instance, in QCD pions are
pGBs for approximate chiral $SU_\ssl(2)
\times SU_\ssr(2)$ transformations, and the $U(1)$
subgroup to which the $\pi^0$ corresponds has an
electromagnetic anomaly. In the effective
theory of pions obtained at energies below the
chiral-symmetry breaking scale, $\Lambda_\chi \sim 4 \pi \,
F_\pi \sim 1$ GeV, this anomaly is represented
by an effective interaction of the form
\eq
\label\qcdcvalues
\Scl_{\rm eff} = - \, \tw{c}_{\gamma\pi} \;
\pi^0 \; F_{\mu\nu} \, \twi{F}^{\mu\nu}
\qquad \hbox{with}\qquad
\tw{c}_{\gamma\pi} = {\alpha \, N_c \over 12 \pi \;
F_\pi} = {\alpha  \over 4 \pi \;
F_\pi},
\eeq
which is well known to correctly describe neutral
pion decay once inserted into eq.~\twophdecay.

For models with strongly-coupled electroweak-breaking
sectors, electromagnetic (or QCD) anomalies amongst the
approximate global symmetries generically arise, so
interactions of the above form coupling light scalar
states to photons (or gluons and electroweak bosons)
are expected to arise,
but with
\eq
\label\tccvalues
\tw{c}_\gamma \sim O\left( {\alpha \over 4 \pi \;
v} \right)
\qquad \hbox{and/or} \qquad
\tw{c}_g \sim O\left({\alpha_s  \over 4 \pi \;
v} \right) .
\eeq
$c_\gamma$ and $c_g$ might also be expected to be of the same
order of magnitude.

\subsection{Linearly Realized Electroweak Multiplets:
Generalities}

The rest of the models we consider contain scalars which
acquire their coupling to gauge bosons because they arise
as explicit members of (linearly-realized) representations
of the electroweak gauge group. We will call scalars
of this type {\sl elementary} scalars. Although scalars
are elementary (in this sense) in models
where the electroweak symmetry breaking
is due to perturbative physics, they need not be restricted
to this case. For instance electroweak symmetry
breaking might arise dynamically within a
strongly-coupled supersymmetric model
which also contains other `elementary' scalars.

It proves to be useful divide models with elementary
scalars into two subcategories when discussing how
experiments can differentiate them from one another.
We therefore distinguish between: ($i$) light scalars
which arise as part of a multiplet whose nonzero
{\it v.e.v.} breaks the electroweak symmetries, and so
contributes to the mass to ordinary SM particles; and
($ii$) light scalars arising within multiplets which do not
acquire {\it v.e.v.}s, and so whose couplings are not
significantly connected to particle masses. In this paper
we reserve the name {\it Higgs} particle only for scalars
in class $(i)$.

Before quoting expressions for effective couplings in
some specific models, we first concentrate on what features
are generic to the tree-level couplings in
all models where the light scalar is elementary. We do
so for the scalar/gauge-boson and scalar/fermion interactions.

Consider, then, a collection ($k=1,\cdots,n$)
spinless particles transforming under various
representations of the $SU_\ssl(2) \times U_\ssy(1)$
electroweak gauge group. We may without loss of
generality choose to represent these particles using
real scalar fields, $\phi_k$, in which case the
gauge generators represented on the scalars,
$T_a, Y$, are imaginary and antisymmetric $n \times n$
matrices. With these choices the tree-level scalar-gauge boson
couplings come from the scalar kinetic terms
\eq
\label\exthiggs
{\cal L}_{\rm kin} = - \; \hf \sum_{k=1}^n {\cal D}_\mu
\phi_k  {\cal D}^\mu \phi_k
\eeq
where the covariant derivative is defined in the
usual way:
\eq
{\cal D}_\mu=\partial_\mu-i g W_\mu^a
T^a - {i \over 2} \; g^\prime \; Y  \, B_\mu .
\eeq
Since tree-level scalar-gauge boson couplings
arise purely from the scalar kinetic term in
the lagrangian, their form may be relatively
cleanly specified in terms of the electroweak
representation content of the scalar sector.

\subsubsection{Gauge Boson Masses}

The electroweak gauge bosons acquire their masses
once some of the scalars aquire nonzero {\it v.e.v.}s,
$v_k$. If we assume negligible mixing between the
$SU_\ssl(2)\times U_\ssy(1)$ gauge bosons and
other spin-one particles, then the
$W^a_\mu$ and $B_\mu$ are related to the physical
gauge bosons through the usual rotation
$A_\mu = W_\mu^3 \sw + B_\mu \cw$,
$Z_\mu = W_\mu^3 \cw - B_\mu \sw$ and $W_\mu^{\pm}
=(W_\mu^1\mp i W_\mu^2)/\sqrt{2}$, where (as
before) $\cw$ and $\sw$ are the cosine and sine of
the weak mixing angle, $\theta_\ssw$, which relates
the couplings $g$ and $g'$ to the electric
charge $e$ via the relations $g = e/\sw$ and
$g^2 + (g')^2 = e^2/(\sw^2 \cw^2)$. In terms of
these quantities (and, as stated earlier,
neglecting any mixing with
a $Z'$ or $W'$) the tree-level gauge boson
masses become:
\eq
\label\gbmasses
M_\ssw^2 = {e^2 \over 2 \sw^2} \;  v^\sst \left(
\vec{T}^2 - T_3^2 \right) v
\qquad \hbox{and} \qquad
M_\ssz^2 = {e^2 \over \sw^2 \cw^2} \;  v^\sst
T_3^2  v.
\eeq
In these expressions the superscript `$T$' denotes the
transpose of the $n$-component column vector of
scalar fields, and the charge neutrality of the
vacuum, $Q \; v = 0$, has been used, where in
our conventions the electric charge is given
by $Q = T_3 + {Y \over 2}$. The vector
sum $\vec{T}^2$ denotes the quadratic Casimir
operator for $SU_\ssl(2)$, with eigenvalues
$t(t+1)$ for non-negative integer or half-integer
$t$. A second observable combination of these quantities
is given by the tree-level expression for Fermi's constant:
\eq
\label\gfgendef
{1 \over \sqrt2  \GF} = {4 \sw^2 \mw^2 \over e^2}
= 2 \;  v^\sst \left(\vec{T}^2 - T_3^2 \right) v .
\eeq

A basic constraint which must be imposed on any
realistic model is the success of the SM tree-level
prediction, $\rho = \mw^2 / (\mz^2 \, \cw^2) = 1$,
since deviations from this prediction are currently
limited by experiment to be $\lsim O(10^{-3})$ \pdg.
Inspection of eq.~\gbmasses\ shows that
this constrains the scalar expectation values,
through their connection to $\rho$:
\eq
\label\rhoconstr
\rho = 1 + \Delta \rho = {v^\sst \Bigl( \vec{T}^2 - T_3^2
\Bigr) v \over 2 \, v^\sst \, T_3^2 \, v}.
\eeq

We may read off the contribution of each multiplet
to this expression: $\Delta \rho = \sum_k (\Delta \rho)_k$
with
\eq
\label\rhobymult
\Delta_k \rho = {\left( t_k \, (t_k +1) - 3 t_{3k}^2 \right) v^2_k
\over 2 \Scv} \; ,
\eeq
where $t_k$ and $t_{3k}$ are the weak isospin and
the eigenvalue of $T_3$ on scalar field $\phi_k$,
and $\Scv = \sum_k t_{3k}^2 \; v^2_k$, with
all multiplets included in the sum.

At tree level the experimentally `safe'
representations are therefore those for
which the neutral scalars satisfy
$t_{3k}^2 = \nth3 \; t_k \, (t_k+1)$ \outb.
Clearly this only has solutions with integer
or half-integer $t_{3k}$ for particular values
for $t$. Besides the standard simple solutions --  $t = 0$ or
$t = \hf$ with $t_3 = \pm \hf$ -- the next smallest solutions
are $t = 3, y = 4$, or $t = {25\over 2}, y= 15$, \etc. with $y$ the 
hypercharge ($y/2=-t_3$ for a neutral scalar).
Barring unnatural cancellations, for any other representations,
$\Delta \rho \ne 0$ at tree level, and so
agreement with experiment implies an upper
bound on each scalar's {\it v.e.v.}.

\subsubsection{Scalar-Gauge Boson Vertices}

With these preliminaries in hand, the tree-level
scalar-gauge couplings involving the electrically
neutral scalars may be written in terms of
the mass eigenfields quite generally, as follows.

The physical scalar states are obtained by
first shifting all fields by their {\it v.e.v.}s,
and then rotating the result to diagonalize the
scalar mass matrix:
\eq
\label\scfredef
\phi_k \to v_k + A_{kl} \; h_l ,
\eeq
where $A$ is an $n \times n$ orthogonal matrix.
For electrically neutral scalars, and in the
absence of $W$ and $Z$ boson mixing, the dimension-three
scalar-vector couplings of eq.~\dimtwoops\ then
become
\eq
\label\dimthreeform
\eqalign{
a_\ssw^i &= {e^2 \over \sw^2} \; \Bigl[ A^\sst \; \left(
\vec{T}^2 - T_3^2 \right) \; v \Bigr]^i, \cr
a_\ssz^i &= {2 \, e^2 \over \sw^2 \cw^2} \;
\Bigl[ A^\sst \; T_3^2  \; v \Bigr]^i . \cr}
\eeq

The dimension-four scalar-vector couplings of eq.~\dimfouropsv\
are similarly given by:
\eq
\label\dimfourforms
\eqalign{
b_\ssw^{ij} &= { e^2 \over \sw^2} \; \Bigl[
A^\sst \left( \vec{T}^2 - T_3^2 \right) A \Bigr]^{ij}\cr
b_\ssz^{ij} &= {2 e^2 \over \sw^2\cw^2} \; \Bigl[
A^\sst T_3^2  A \Bigr]^{ij} \cr
g_\ssz^{ij} &= {i e \over \sw \cw} \; \Bigl[
A^\sst T_3 A \Bigr]^{ij} .\cr}
\eeq

These expressions clearly reduce to the usual SM formulae,
eqs.~\SMdthree\ and \SMdfourb, in the special case of
a lone doublet scalar field, for which $n=1$, $A=1$,
$t=\hf$ and $t_3 = - \, \hf$.
They also lend themselves to the derivation of
general relations amongst the coupling constants, which
can be used to discriminate amongst the various
kinds of models.

For instance, suppose $k=1$ represents
the lightest scalar state, which we assume has been observed.
Comparison of eqs.~\gbmasses\ and \dimthreeform\ show that
the $WWh$ trilinear couplings for this
scalar are related to the $W$ mass by:
\eq
\label\svsumrule
{a_\ssw^1  \over e \, M_\ssw/\sw}
= { \sqrt2 \sum_k A_{k 1} \lambda_k
v_k \over \sqrt{\sum_l \lambda_l v^2_l }},
\eeq
where the sum is over all neutral scalars in the
electroweak basis, and $\lambda_k$ denotes the eigenvalue,
$\lambda_k = t_k \, \left(t_k +1 \right) - t_{3k}^2$,
for each multiplet.

The ratio of coupling to mass of eq.~\svsumrule\ is
unity at tree level in the Standard Model:
$\left( a_\ssw \sw/e \mw\right)_\SM=1$.
By contrast, in a model containing only doublets
and singlets we have $\lambda_k = 0$ or $\hf$ for {\it all}
electrically neutral scalars, and so
eq.~\svsumrule\ reduces to
\eq
\label\dbsumrule
{a_\ssw^1  \over e \, M_\ssw/\sw}
= { \sum_k A_{k1} v_k \over \sqrt{\sum_l v^2_l}} \le 1 ,
\eeq
where the final inequality follows because eq.~\dbsumrule\
may be interpreted as
the inner product of two unit vectors, $\vec{e} \cdot
\vec{f}$, with components $e_k = A_{k1}$ (which has
unit length because $A$ is
orthogonal) and $f_k = v_k/\sqrt{\sum_k v_k^2}$. As a
result the ratio $a_\ssw^1\sw /(e  \mw)$ is
always smaller than one for any model involving
only doublets and singlets.

The same need {\it not} be true once other representations
are included, as we make clear in detail with examples
once we examine individual models in
subsequent sections. In particular, the lower
bound $t(t+1) - t_3^2 \ge t$ implies that $\lambda_k >
\hf$ for all representations other than singlets or
doublets. The general statement is this:
the ratio $a_\ssw^1 /(e  M_\ssw/\sw)$ {\it must} be
smaller than one for a model containing only doublets
and singlets, but can
be either larger or smaller than unity when other
multiplets also appear, depending on the details
of the scalar {\it v.e.v.}s, and their mixing matrix, $A_{ij}$.
Although no definite conclusions
are possible if this ratio should be experimentally
found to be smaller than one, singlet/doublet models
may be ruled out if it should be greater than unity.

Comparing the trilinear $ZZh$ vertex with the $Z$ mass
only permits somewhat weaker conclusion. The argument
is identical to the one presented above
\eq
\label\szsumrule
{a_\ssz^1  \over e \, M_\ssz/\sw\cw}
= {2 \sum_k A_{k1} \, t^2_{3k}
v_k \over \sqrt{\sum_l t^2_{3l} v^2_l }},
\eeq
We see that the conclusion $a_\ssz^1 \sw\cw/e \mz \le 1$
follows inevitably for any model having neglible $W$ and $Z$ mixing,
and for which the only scalars with nonzero {\it v.e.v.}s have
$t_3 = \pm \hf$ or $t_3 = 0$ (regardless of the total isospin, $t$,
carried by each multiplet).

\subsubsection{Fermion-Scalar Vertices}

Since the overwhelming majority of experiments are
performed using light fermions as initial or final
particles, the fermion Yukawa couplings are perhaps
the most crucial interactions to measure once a scalar
particle is discovered. In this section we collect
general tree-level results concerning these couplings.

Imagine coupling an arbitrary collection of
scalar fields, $\phi_i$, to a general set of
fermion fields, $\chi^r$. We are free to choose,
without loss of generality, all scalar fields to be
real and all spinor fields to be majorana.
(With this choice electrically charged particles
like electrons are represented by two spinor
fields -- \eg: one for $e^-_\ssl$ and one for
$e^-_\ssr$ -- just as electrically charged
spinless particles are represented by one complex, and so
two real, scalar fields.) If the particles are
organized into linear representations of the
electroweak gauge group, then we must require
their interactions to be invariant under the
variation $\delta \phi = i \omega^a \, T_a\,
\phi$ and $\delta \chi = i \omega^a \, \Sct_a
\, \gamma_\ssl \chi - i \omega^a \,
\Sct_a^* \gamma_\ssr \chi$, where $\gamma_\ssl$
and $\gamma_\ssr$ denote the projection matrix
onto left- and right-handed spinors.

The most general fermion mass terms and
renormalizable scalar/fermion interactions
which are possible are given by:
\eq
\label\yukgenform
- \Scl_{\rm yuk} = {m_{rs} \over 2} \; \ol\chi^r
\gamma_\ssl \chi^s +{\Gamma^i_{rs} \over
2} \; \ol\chi^r  \gamma_\ssl \chi^s \;
\phi_i + \hc,
\eeq
where the coefficient matrices, $\Gamma^i_{ab}$
are required by electroweak gauge invariance to
satisfy $\Gamma^j_{rs} {(T_a)_j}^i +
\Gamma^i_{rt} {(\Sct_a)^t}_s +
{(\Sct_a)^t}_r \Gamma^i_{ts} = 0$.
Once the scalars acquire nonzero {\it v.e.v.}s, $v_i$,
the left-handed fermion mass matrix therefore becomes
$M = m + \Gamma^i \; v_i$.

The rotation from an interaction basis to a mass
basis is accomplished by performing an orthogonal
rotation, $A_{ij}$, amongst the scalars, as well
as a unitary rotation, ${U^r}_s$, amongst the
left-handed fermion fields: $\chi^r = {U^r}_s
\gamma_\ssl \eta^s + {U^{r*}}_s \gamma_\ssr
\eta^s$. The matrix $U$ is chosen to ensure
that the fermion masses are diagonal and
nonnegative: $U^\sst M \, U = \hbox{diag}(m_1,
m_2,m_3,\dots)$.

The Yukawa couplings amongst the physical
propagation eigenstates therefore become:
\eq
\label\yukpreig
y^i_{rs} = \Bigl(U^\sst \Gamma^j \, U \Bigr)_{rs}
{A_j}^i.
\eeq
The main observation concerning these couplings,
which drives much of the model building, is
that they need not be diagonal in their fermionic
indices ($r,s$) when expressed in terms of mass
eigenstates. If not, then scalar emission changes
fermion flavour, a prospect which in many circumstances
is very strongly precluded by experiment.

Of course, the SM provides an important example where such
flavour changes do not arise. They do not because
the model has only one Higgs particle and its {\it
v.e.v.} is the only source of mass in the entire
model. As a result we have $M = \Gamma \; v$ and
so the same rotation which diagonalizes the fermion
masses automatically also diagonalizes $y =
U^\sst \Gamma \, U = \hbox{diag}(y_1,y_2,\dots)$.

Electroweak gauge invariance has further implications
for such Yukawa couplings. Since all of the SM fermions
transform as doublets or singlets under $SU_\ssl(2)$,
they can form Yukawa couplings (without also requiring
new exotic fermions) only with a scalar multiplet which
transforms as a triplet, doublet or singlet. These
representations therefore play a special role in
the models which follow.
$U_\ssy(1)$ hypercharge invariance further restricts
which fermions can couple to which scalars, as we
explore in more detail for various models.

\subsubsection{Contributions to $c_k$ and $\tw c_k$}

The effective couplings, $c_k$ and $\tilde c_k$ vanish
at tree level in all renormalizable models of elementary
scalars. Because these couplings can nevertheless contribute
significantly to observables, we record here general expressions
for their one-loop contributions, which we will use for
particular models in subsequent sections.

The general result for the CP-invariant photon and
gluon effective couplings due to a loop of spinless,
spin-half or spin-one particles may be written:
\eq
\label\cgsumform
c_k = {\alpha_k \over 6 \pi}
\sum_{s=0,\hf,1} (-1)^{2s+1} \; \sum_{R_s}
{\Scc_k(R_s) y(R_s) \over m(R_s)} \; I_s\left[
{m^2(R_s) \over \mh^2} \right] ,
\eeq
where $R_s$ denotes the various representations
on spin-$s$ fields of the colour or electromagnetic
gauge group which is carried by the particle content of the
model. $m(R_s)$ represents the mass of the particle
in the loop, and $\Scc_k(R_s)$ is the quadratic Casimir for
this particle's representation, defined by $\Tr (T_a T_b) = \Scc(R_s)
\; \delta_{ab}$. The spin-dependent functions, $I_s(r)$, are given
explicitly by expressions \gcfnresults\ of section 2.
Specializing this result to the SM particle content and
couplings reproduces eq.~\SMcforms.

In eq.~\cgsumform\ $y(R_s)$ is the relevant trilinear coupling
to scalars enjoyed by the particle circulating in the loop.
These couplings are normalized so that $y$ is
the Yukawa coupling, $y_f$, for spin-half particles;
$y_\ssw = a_\ssw/(2\mw)$ for spin-one particles,
and $y_\ssS = \nu_{{\sss SS}h}/m_\ssS$ for spin-zero
particles (here denoted by $S$).

The result for the CP-odd effective couplings,
$\tw{c}_k$, is obtained by omitting
the bosons ($s=0,1$) and simply replacing the axial coupling,
$z(R_s)$ for $y(R_s)$ in the fermion contribution to
eq.~\cgsumform.

We now turn to the examination of some particular
models in more detail. The models we present are
chosen either because they are theoretically
well motivated (and hence popular) or because
they illustrate particular points we wish to
emphasize.

\subsection{Models With Higher Scalar Multiplets}

The main feature which distinguishes the various
models of elementary scalars is the representation
content which is chosen for the scalar fields. Although
the majority of the best-motivated models involve
only doublets, our goal here is to assess the extent
to which the field content can be inferred from experiments.

As just discussed, scalars which are not electroweak
triplets, doublets or singlets cannot form Yukawa
couplings involving just the known SM fermions.
All scalars of this type can therefore be distinguished
in principle by their prediction of vanishing
tree-level fermion-scalar couplings, but nonvanishing
scalar-gauge boson interactions \outb.

\subsubsection{The $3-4$ Model}

We pause here to examine one such model
in slightly more detail. We do so --- even though
the model we consider is not particularly attractive --- for two
reasons. First, it illustrates within a simple context most
of the issues which arise later when we discuss better-motivated
models. Second, it furnishes the existence proof for models
having $a_\ssw \sw/e \mw$ potentially significantly
larger than unity.

In this model we supplement the usual SM particle
content, including its doublet Higgs, $\phi$, with a single complex
scalar multiplet, $S$, transforming in a $T=3$ multiplet
of $SU_\ssl(2)$ and carrying hypercharge $y = 4$.
These quantum numbers are chosen to ensure the
vanishing of the new scalar's tree-level
contribution to $\Delta \rho$,
as given by eq.~\rhobymult.

There are two new neutral scalar particles
in this model, and for the present purposes
it is sufficient to assume that the scalar
potential does not break CP spontaneously
(renormalizability and gauge invariance
in this model together preclude explicit CP
violation in the scalar potential). In this
case the CP-odd state neither mixes nor acquires
a nonzero {\it v.e.v.}, while both of these are possible
the two CP-even states. Denoting the
mixing angle by $\alpha$, we have
\eq
\label\Afposp
\pmatrix{A_{1h} \cr A_{2h} \cr} =
\pmatrix{\cos\alpha \cr
-\sin\alpha\cr} ,
\eeq
where 1 denotes the doublet scalar, 2 labels the CP-even
neutral scalar from the $S$ multiplet
and $h$ is the mass eigenstate which is assumed to have been
observed.

\topic{Yukawa Couplings}

Because of its unorthodox charge assignments, the
two neutral scalars in the new multiplet cannot
couple renormalizably with SM fermions at all.
So if the light Higgs is the CP-odd state it does not
couple to SM fermions, and if it is one
of the CP-even ones its coupling is as given
by the usual SM expressions up to an overall mixing-angle
factor:
\eq
\label\ykmposp
y_{fg} = \delta_{fg} \;
{m_f \over v_\phi} \; \cos\alpha ,
\qquad
z_{fg} = 0 .
\eeq
\vfill
\eject

\topic{Gauge Boson Couplings}

Of more direct interest here is the tree-level
prediction: $\Delta \rho = 0$, as well as the
prediction for $\GF$ and the trilinear $WWh$ and $ZZh$
couplings. These vanish for the CP-odd scalar
but take the following nonzero values for the
CP-even case:
\eq
\label\awzposp
\eqalign{
{1\over \sqrt2 \, G_\ssf} &=
v^2_\phi \left[ 1 + 16 r^2 \right] ,\cr
{a_\ssw^h \over e \mw/\sw} &= {a_\ssz^h \over e\mz/\sw\cw} =
{ \cos\alpha - 16 r \sin\alpha \over
\sqrt{ 1 + 16 r^2 }}, \cr}
\eeq
where $r = v_\ssS/v_\phi$. Clearly, because the condition
$\Delta \rho \ll 1$ does not require $r$ to be small (unlike
what we shall find in some of the other models we consider), in
this model both of the ratios $\left|{a_\ssw^h \over
e \mw/\sw}\right|$ and $\left|{a_\ssz^h \over e
\mw/\sw}\right|$ can be much larger than one.

\topic{Loop Effects}

It is important to ask whether predictions for small effective couplings
are not merely artifacts of the tree approximation. For this theory
tree-level Yukawa couplings are very small because they are suppressed
by small fermion masses. It is easy to see that this remains
true once radiative corrections
are included, for the same reasons as were given in \S2\
for the SM.

As noted earlier, if the observed light scalar should be
CP-odd, then it cannot mix with the SM doublet, and so
its Yukawa couplings vanish. The absence of Yukawa
couplings then rules out nonzero values for $\tw c_g$
and $\tw c_\gamma$.  CP itself also directly forbids
a nonzero $c_g$ and $c_\gamma$, so all four of these 
loop couplings are zero.

On the other hand, should the observed scalar be one of the CP-even
states, then the predicted $c_k$'s differ from the SM results for two
reasons. First, the scalar/fermion and scalar/$W$ couplings are
modified by mixing in the scalar sector. Next, $c_\gamma$
acquires a new contribution
due to the
existence of new charged scalars. In this case we find $\tw c_g
= \tw c_\gamma = 0$ and
\eq
\label\pospceven
\eqalign{
c_g &= c_g(\hbox{SM}) \; \cos\alpha, \cr
c_\gamma &= c_g\left (\hbox{SM-spin $\hf$}\right)
\; \cos\alpha \cr
&\qquad + c_g(\hbox{SM-spin 1}) \; { \cos\alpha - 16 r \sin\alpha \over
\sqrt{ 1 + 16 r^2 }} + {\alpha \over 24 \pi}
\sum_{S} Q_\ssS^2 {\nu_{{\sss SS}h} \over m_\ssS^2}
.\cr} \hbox{if CP($h$) = +1}
\eeq

\subsection{Models with Extra Scalar Singlets}

The simplest extensions of the SM in the scalar
sector would add electroweak singlet scalars, $S$, to
the usual SM field content. Our assumption of
electric neutrality for these scalars then further
implies that their hypercharge is zero, so they
have no tree-level couplings at all to the $W$,
$Z$ or photon.

\ref\singletmodels{G.B. Gelmini and M. Roncadelli,
\plb{99}{81}{411}.}

\topic{Tree Level Couplings}

Electroweak gauge invariance also precludes forming
tree-level Yukawa couplings involving these scalars
and only ordinary SM fermions, making them not
particularly interesting for experiments unless
more light fermions beyond those of the SM are
also included. In fact, there is an important class of
such fermions which arise in theoretically well-motivated
models, and could plausibly have avoided detection to date:
right-handed neutrinos (\ie\ electroweak
singlet fermions, $N$) \singletmodels. Such singlet
fermions can couple
to scalar singlets, and transfer these couplings
to SM fermions through mixing with left-handed
neutrinos in the neutrino mass matrix.

In this scenario only the new singlet fermions acquire tree-level
Yukawa couplings to the new scalars, and so ordinary fermions
only learn about these through neutrino mixing with these new
fermions, which arise due to Yukawa couplings with the 
SM-type Higgs. As a result the scalar phenomenology is intimately
tied up with neutrino properties, and depends crucially on
whether or not any of the lepton numbers, $L_i$, remain
conserved.

If the interactions conserve all lepton numbers, then the singlet scalar
must carry $L_i = 2$ (if it couples at all to generation `$i$'). If
$S$ does not acquire an expectation value, then neutrinos cannot
oscillate and there are no observable new-scalar effects unless the
new scalar's mass is small enough (less than roughly the QCD scale)
to contribute to big-bang nucleosynthesis.

More interesting possibilities arise if some or all of the lepton
numbers are broken, since then both the neutrinos and the
scalars can mix in more interesting ways. If the symmetries
are broken spontaneously, strong
bounds arise on the resulting Goldstone boson, and so we
disregard this possibility here. If they are instead explicitly
broken and all scalar masses are reasonably heavy (larger
than $\Lambda_{\sss QCD}$, say), then the main constraints
come from experiments on the light neutrino states, such as
searches for oscillations or double-$\beta$ decay.

In these types of models, the expected properties of an observed
scalar depend on how much overlap this state has with the
SM Higgs and with the new singlet states. If there is only
a single, real new singlet then there are two neutral scalar
states, and these can mix according to eq.~\Afposp. If, on
the other hand, the new scalar is complex, there are 
three neutral scalar states. If CP is conserved
then the pseudoscalar doesn't mix with the two scalars, and
so the mixing proceeds as for the $3-4$ model, with
the CP even states again mixing according to eq.~\Afposp. If
CP is broken, then all three neutral scalars can mix,
generalizing eq.~\Afposp\ to
\eq
\label\Aftrip
\pmatrix{A_{1h} \cr A_{2h} \cr A_{3h}\cr} =
\pmatrix{\cos\alpha\cos\beta \cr
-\sin\alpha\cos\beta \cr
-\sin\beta \cr} ,
\eeq
where $h$ denotes the observed light scalar state,
and we take `1' to label the doublet scalar,
`2' to be what would have been the CP-even scalar
in the absence of CP violation.  `3'
denotes what becomes the CP-odd state in this limit.

If $\sin\alpha$ and $\sin\beta$ are both small, then
the observed scalar shares the usual properties of the SM
Higgs. On the other hand, if the observed scalar has much
singlet overlap, then its Yukawa couplings to all electrically-charged
fermions remain proportional to mass, although with
mixing-dependent strength. Its couplings to neutrinos,
however, need not be small, and so this kind of $h$
could be relatively long-lived, and dominantly decay through
invisible neutrino modes.  Of course for the same reasons
it would be also difficult to detect within accelerator
experiments.

\topic{Loop Effects}

The size of $c_k$ in these models depends dramatically
on whether the observed scalar is dominantly SM Higgs
or singlet. If dominantly a Higgs, the $c_k$ differs from
the SM result purely through the effects of scalar
mixing on the coupling strengths to charged fermions and
the $W$. Notice there is no scalar contribution
to the $c_k$'s because there are no new electrically-charged
scalars in these models.
If the new scalar is dominantly a singlet, then all of the
$c_k$'s vanish at one loop, regardless of whether it is
CP odd or even or if CP is broken.
\bigskip

\subsection{Triplet Models}
\ref\tripletrefs{For a recent review of triplet models,
with references, see: H. E. Logan, Ph.D. thesis, UC Santa Cruz
(hep-ph/9906332).}
Consider next a model containing one
standard isodoublet, $\phi$, (with hypercharge
$y_2 = 1$) and one scalar isotriplet, $\psi$ \tripletrefs.
In order that the triplet contain electrically neutral
components, its hypercharge eigenvalue must be
$y_3 =0$ or $y_3 = 2$ (with the neutral scalar
then having $t_3 = 0$ or $-1$ respectively).

Since the perturbative spectrum and couplings of the
model depend in detail on whether $y_3 = 0$ or
$y_3 = 2$, we consider each case separately.

\subsubsection{The Case $y_3 = 0$}

If $y_3 = 0$ then the triplet may be chosen
to be real. In this case there are only two neutral
spinless particles, and one scalar with charge
$q  = 1$. 

\topic{Tree-level Couplings}

Since hypercharge conservation precludes coupling
the triplet to fermions, only the doublet has
Yukawa couplings, which therefore take the same
form as in the Standard Model. Both neutral
scalar mass eigenstates can participate in these
interactions inasmuch as the triplet mixes with the doublet.
Since the scalar mixing matrix is a two-by-two rotation,
it is given by eq.~\Afposp, implying the following
relations between the Yukawa couplings, fermion masses
and scalar mixing angle:
\eq
\label\yukmastrip
y_{fg} = \delta_{fg} \;
{m_f \over v_\phi} \; \cos\alpha ,
\qquad
z_{fg} = 0 .
\eeq

Triplets contribute to $\Delta \rho$ at tree
level, so we must ensure the ratio of triplet to
doublet {\it v.e.v.}s, $r = v_\psi/v_\phi$,
is sufficiently small. From eq.~\rhobymult:
\eq
\label\drhotrip
\Delta \rho = 4 r^2  \lsim O(10^{-3}),
\eeq
so $v_\phi$ is to good approximation related to Fermi's constant
in the same way as in the SM:
\eq
\label\GFtrip
{1\over \sqrt2 \, G_\ssf} = {4 \, \sw^2 \, \mw^2 \over e^2}
= v^2_\phi \left[ 1 + 4 r^2 \right] \approx v_\phi^2 .
\eeq

The trilinear $WWh$ and $ZZh$ couplings become:
\eq
\label\awztrip
\eqalign{
{a_\ssw^h \over e \mw/\sw} &= { \cos\alpha - 4 r
\, \sin\alpha \over \sqrt{ 1 + 4 r^2}}, \cr
&= {\cos\alpha - 2 \sin\alpha \sqrt{\Delta \rho} \over
\sqrt{1 + \Delta \rho} } ,\cr
&\approx \cos\alpha  - 2 \sin\alpha \,
\sqrt{\Delta \rho} + O(\Delta\rho) \cr
\hbox{and} \qquad
{a_\ssz^h \over e\mz/\sw\cw} &=  \cos\alpha . \cr}
\eeq
Clearly the SM expression, $a_\ssw \sw/e \mw =
a_\ssz \sw\cw/e \mz$ holds here, to within an accuracy
of a few percent, even though both of these quantities
can differ appreciably from the corresponding SM predictions.

Although $a_\ssz^h \sw\cw \le e \mz$
quite generally in this model, $a_\ssw^h \sw$ can be
larger or smaller than $e \mw$ depending on the
value of $\alpha$. For example, it is larger if
$\alpha$ lies in the range
$|\sin\alpha| \lsim \sqrt{3 \Delta\rho}$.
Even though the ratio $a_\ssw^h \sw/e \mw$ can be
bigger than 1 in this way, the constraints on
$\Delta \rho$ do not permit it be much bigger.
In this model it is never larger than
$\sqrt{(1 + 4 \Delta \rho)/(1 + \Delta \rho)}
\approx 1 + {3 \over 2} \, \Delta \rho$. This
is not likely to be measurable.
Of course this conclusion is not generally true for
all models, as the above discussion of the $3-4$
model explicitly illustrates.

\topic{Loop Effects}

The same arguments as were given for the SM in \S2\ imply
that loops do not ruin the tree-level prediction that Yukawa
couplings are suppressed by small fermion masses.

In this model the contributions to the effective couplings
$c_k$ and $\tw c_k$ are similar to what was found earlier for
the 3 -- 4 model. Fermion and gauge boson contributions are
identical to those of the SM, weighted by the corresponding
effective couplings --- $y_f/y_f(\hbox{SM})$ and
$a_\ssw/a_\ssw(\hbox{SM})$ --- as required by scalar mixing.
To these must be added the contribution of the heavy $Q=1$
scalar, in the case of $c_\gamma$.

\subsubsection{The Case $y_3 = 2$}

If $y_3 = 2$ then the triplet must be complex,
implying a total of three neutral scalars, plus one
charged and one doubly-charged state. In
general all three neutral scalars can get
{\it v.e.v.}s --- which we denote $v_\phi,
v_\ssr, v_\ssi$ --- and mix with one another, but if the
scalar interactions are CP-preserving then the
three neutral states break up into two CP-even ones which
do not mix with the third, which is CP-odd.
Mixing in the general case is therefore given by
eq.~\Aftrip, with the CP-conserving limit corresponding
to either $\sin\beta = 0$ or $\cos\beta = 0$.

\ref\tripletneut{Y. Chikashige, R.N. Mohapatra and R.D. Peccei,
\prl{45}{80}{1926}; \plb{98}{81}{265}.}

In this case hypercharge conservation only
permits couplings of the neutral component of the triplet
to the left-handed neutrinos \tripletneut, and these
preserve overall lepton number
provided $L(\psi) = -2$. In the presence of
such couplings there are three possibilities:
($i$) $L$ is broken by the scalar potential
$V$  (\eg\ through a term
of the form $\phi^\sst \sigma_2
\vec{\sigma}\phi \cdot \vec{\psi}^*$);
($ii$) $L$ is not broken by $V$ and $\psi$
gets a {\it v.e.v.}, so one of the neutral scalars is a massless
Goldstone mode; or, ($iii$) $L$ is neither broken by
$V$ nor spontaneously. Each of these options presents
its own potential phenomenological difficulties (although
arguably, none are fatal), whose
pursual goes beyond the scope of the present paper.

\topic{Tree-level Couplings}

All fermions apart from neutrinos have only
Yukawa couplings to the doublet scalar, making
their description similar to the $y_3 = 0$ case.
In terms of the fermion masses and scalar mixing
angles we have (for all fermions except neutrinos):
\eq
\label\ykmstrip
y_{fg} = \delta_{fg} \;
{m_f \over v_\phi} \; \cos\alpha \cos\beta,
\qquad
z_{fg} = 0 .
\eeq

In this model $\Delta \rho$ can depend on the {\it v.e.v.}s
of both of the (real) neutral scalar fields of the
triplet. Defining $r_\ssr = v_\ssr/v_\phi$ and $r_\ssi =
v_\ssi/v_\phi$ we have:
\eq
\label\drhotrip
\Delta \rho = {-2 (r_\ssr^2 + r_\ssi^2)  \over
1 + 4 (r_\ssr^2 + r_\ssi^2)}
 \lsim O(10^{-3}) ,
\eeq
Fermi's constant is:
\eq
\label\GFtrip
{1\over \sqrt2 \, G_\ssf} =
v^2_\phi \left[ 1 + 2 (r_\ssr^2 + r_\ssi^2) \right]
\approx v_\phi^2 ,
\eeq
and the trilinear couplings with gauge bosons are:
\eq
\label\awztripp
\eqalign{
{a_\ssw^h \over e \mw/\sw} &=
{ \cos\alpha \cos\beta- 2 (r_\ssr \sin\alpha\cos\beta
+ r_\ssi \sin\beta) \over \sqrt{ 1 + 2
(r_\ssr^2 + r_\ssi^2)}}, \cr
{a_\ssz^h \over e\mz/\sw\cw} &=  {\cos\alpha \cos\beta
- 4 (r_\ssr \sin\alpha\cos\beta
+ r_\ssi \sin\beta) \over 1 + 4 (r_\ssr^2 + r_\ssi^2)} .\cr}
\eeq

The limit of conserved lepton number corresponds to the limit
$r_\ssr = r_\ssi = 0$ and absolutely no mixing amongst
the three scalars --- and so $(\alpha,\beta) = (0,0),
\left({3\pi \over 2},0 \right)$ or $\left(\alpha,
{3\pi \over 2} \right)$ depending on which scalar is
the one observed. Similarly, if CP
is not broken explicitly or spontaneously
then $r_\ssi = 0$. If it happens that the observed
scalar is one of the model's CP-even states in this
limit, then we may take $\beta = 0$ in the above, in which case
our results reduce to two-by-two mixing, as they must.
Alternatively, the case where the observed light
state is CP-odd corresponds to the limit $\beta =
{3\pi \over 2}$, in which case all of the above
couplings vanish (as they again must).

In the excellent approximation that we neglect
quantities of order $O(\sqrt{|\Delta \rho}|)$,
the model predicts relations among the
fermion and gauge-boson couplings of the
light scalar: $y_f^2 = \sqrt2 \GF \, m_f^2
(a_\ssz^h)^2$. The same prediction also holds
to good approximation with $a_\ssz^h$ replaced
by $a_\ssw^h$. This kind of prediction also holds
for the $y_3=0$ model, and is
reasonably robust, relying only on the absence
of direct Yukawa couplings to the new scalar
multiplet in the model of interest.

\topic{Loop Effects}

As for the previously-studied models, 
the suppression of tree-level Yukawa
couplings by small fermion masses survives loop corrections.

The contributions to $c_k$ and $\tw c_k$ in the
$y_3 = 2$ triplet models
resemble those in the singlet models discussed above, both because
of the potential for three-scalar mixing, and because of
the triplet having tree-level Yukawa couplings
exclusively to neutrinos. Fermion and gauge boson
contributions get rescaled compared to the SM
by the appropriate effective couplings, and both the
$Q=1$ and $Q=2$ scalars contribute to $c_\gamma$.

\subsection{Two-Higgs-Doublet Models (2HDM)}

The scalar sector of a great many of the theoretically
most plausible models, including prominently the
Minimal Supersymmetric Standard Model (MSSM),
fall into the general class of multi-doublet theories,
consisting of several copies of the basic SM
$Y=1$ doublet.
Besides being very well motivated, these models enjoy
many attractive properties, such as having naturally
vanishing tree level contributions to $\Delta \rho$.

\ref\grant{A.K. Grant, \prd{51}{95}{207}.}

In this section we consider the basic Two-Higgs Doublet Model
(2HDM) ~\outb,\grant, which in many ways is the minimal archetype for
the rest of the multi-doublet models. It consists of
the usual SM particle content, supplemented only by
a second Higgs doublet.

The model predicts one $Q=1$ electrically-charged
spinless particle state in addition to a total of three electrically-neutral
mass eigenstates. As usual, two of these -- called
$h$ and $H$ -- are CP-even and one -- called $A$ -- is
CP-odd, if CP is not broken. (Conventionally $h$ denotes
the lighter of the two CP-even states.) Finally, if we
write the {\it v.e.v.}s of
the two would-be CP-even states by $v_1$ and $v_2 \cos\xi$,
and denote the third {\it v.e.v.} by $v_2 \sin\xi$. These
must satisfy the tree-level prediction
for Fermi's constant, $1/(\sqrt2 \, G_\ssf)
= v_1^2 + v_2^2 $, but neither the ratio $v_2/v_1$
nor $\xi$ are constrained by $\Delta\rho$. $\xi = 0$
if the scalar sector conserves CP.

In general all three of the neutral eigenstates mix
among themselves (and with the would-be Goldstone mode
which is eaten by the $Z$ boson, in a non-unitary gauge),
and so the interactions of the lightest scalar state are
describable in terms of two mixing angles, as in eq.~\Aftrip.
In this section we do not adopt this earlier notation for
the mixing angles, since the angles $\alpha$ and $\beta$
are conventionally chosen differently when describing
these models. Since it is not our purpose here to
exhaustively explore the parameter space of the model,
in the interests of brevity we quote expressions here
for the effective couplings in the
CP-conserving limit, where $\xi = 0$ and
the CP-odd state $A$ does not mix with $h$ and $H$.
We emphasize that we do {\it not} use this
assumption in our later discussion of the properties
of the low-energy scalar in these models.

Under the assumption of CP invariance, only two angles
turn out to be required to diagonalize all of the scalar masses
in a general gauge. This is because the CP-odd states ($A$
and the would-be Goldstone boson, $z$) do not mix
with the CP-even states ($h$ and $H$) in the electrically neutral
sector, and because the same rotation angle, $\tan\beta = v_2/v_1$,
required to diagonalize the $A-z$ mass matrix also diagonalizes
the mixing between the electrically-charged state and the
charged Goldstone mode, $w$. The rotation required to diagonalize
the two-by-two CP-even mass matrix defines the second mixing
angle, $\alpha$.

\topic{Gauge Boson Couplings}

With these conventions the CP-odd state, $A$,
does not participate in trilinear interactions with two gauge
bosons, and the trilinear gauge couplings
to the light CP-even state, $h$, can be written:
\eq
\label\thdmzzh
\eqalign{
a_\ssw^h &= {e^2 \over 2\sw^2} \;
( v_2 \cos\alpha - v_1 \sin\alpha) =
{e \mw \over \sw} \; \sin(\beta - \alpha) ,\cr
a_\ssz^h &= {e^2 \over 2\sw^2\cw^2} \;
(v_2 \cos\alpha - v_1 \sin\alpha) =
{e \mz \over \sw\cw} \;\sin(\beta - \alpha) ,\cr}
\eeq
which are clearly both bounded above by their SM counterparts,
in accordance with the general results of the previous
sections.

We record here also the tree-level trilinear $Z$-scalar-scalar
coupling, the $ZAh$ vertex, which is given by:
\eq
\label\gzthdm
g_\ssz^{Ah} = {e \over 2 \sw\cw} \; \cos(\beta - \alpha) ,
\eeq
and the tree-level quartic scalar/gauge-boson interactions, which are
completely unchanged from the SM:
\eq
\label\bsthdm
b_\ssw^{ij} = {e^2 \over \sw^2} \; \delta^{ij}, \qquad
b_\ssz^{ij} = {e^2 \over 2 \sw^2 \cw^2} \; \delta^{ij} .
\eeq

\topic{Yukawa Couplings}

We next turn to the neutral scalar Yukawa couplings.
Because gauge invariance
cannot rule out the couplings of either of these doublets
to SM fermions, in the generic case we expect to find
tree-level couplings between neutral scalars and
fermions which change fermion flavour.

The desire to avoid these kinds of dangerous
flavour-changing couplings motivates the definition of two
particular subclasses of models for which these
couplings are naturally forbidden by a discrete symmetry.
That is, adopting the nomenclature of ref.~\outb:

\item{1.} {\it Type I}  models
are defined in such as way as to arrange for one of the
two doublets to not couple at all to fermions. This can be
done, for instance, by imposing the discrete symmetry
$(\phi_1,\phi_2) \to (-\phi_1,\phi_2)$, with none of the
fermions transforming.

With this choice it is only the doublet $\phi_2$ which
generates all fermion masses, and so the resulting
mass-eigenstate Yukawa couplings are diagonal and independent
of $\gamma_5$
for the same reason they are in the SM. The Yukawa couplings
to the light CP-even state $h$ are then related to fermion
masses by the tree-level formulae: $z^h_{fg} =  0$ and
\eq
\label\toneyuk
y^h_{fg} =  \delta_{fg} \; {m_f \over v_2} \; \cos\alpha
=  \delta_{fg} \; \Bigl(\sqrt2 \GF m_f^2 \bigr)^\hf
\; {\cos\alpha \over \sin \beta} .
\eeq
This implies the well-known tree-level prediction
that all {\it ratios} of Yukawa couplings are the same as
they would be in the SM: $y_f/y_g = m_f/m_g$.

\item{2.} {\it Type II}  models
are defined in such a way as to ensure that only one
doublet couples to `up-type' $\left( t_3 = +\hf \right)$ fermions,
with the other doublet coupling only to `down-type'
$\left( t_3 = - \hf \right)$ fermions. This can also be ensured
using a discrete symmetry of the form
$(\phi_1,\phi_2 ) \to (\phi_1,- \phi_2)$, although
this time with all right-handed up-type fields
also changing sign under the symmetry.

Since the fermions in any one charge sector couple only
to one kind of scalar, all tree-level Yukawa couplings
are again flavour diagonal, with $z^h_{fg} = 0$ and
\eq
\label\ttwoyuk
\eqalign{
y^h_{uu'} &=  \delta_{uu'} \; {m_u \over v_2} \; \cos\alpha
=  \delta_{uu'} \; \Bigl(\sqrt2 \GF m_u^2 \bigr)^\hf
\; {\cos\alpha \over \sin \beta} ,\cr
y^h_{dd'} &= -\delta_{dd'} \; {m_d \over v_1} \; \sin\alpha
= -\delta_{dd'} \; \Bigl(\sqrt2 \GF m_d^2 \bigr)^\hf
\; {\sin\alpha \over \cos \beta} .\cr}
\eeq

Although ratios involving only up-type (or only
down-type) Yukawa couplings agree with the
SM predictions -- $y_u/y_{u'} = m_u / m_{u'}$ --
mixed ratios differ from their SM predictions
by a fixed factor: $y_u/y_d = (m_u/m_d)
(\tan\alpha/\tan\beta)$.

\topic{Loop Effects}

For the same reasons as for the other models examined to
this point, the suppression of tree-level Yukawa couplings
by small fermion masses also survives loop corrections in
this model.

The contribution to $c_\gamma$ in a  2HDM is obtained from
eq.~\cgsumform\ by adding to the $W$ and fermion contributions
(using the $a^h_\ssw$ and $y^h_{fg}$ of the 2HDM)
the contribution of the new charged Higgs \outb,\spira. That is,
we have for Type I models:
\eq
\label\typeIck
\eqalign{
c_g^{\sss 2HDM}(\hbox{Type I}) &= {\cos\alpha \over \sin\beta}
\; c_g^{\sss SM} , \cr
c_\gamma^{\sss 2HDM}(\hbox{Type I})  &=
{\cos\alpha \over \sin\beta} \; c_\gamma^{\sss SM}(
\hbox{fermions}) - \sin(\beta - \alpha) \;
c_\gamma^{\sss SM}(W) + {\nu_{cch} \over 4 m_c^2}, \cr}
\eeq
where the last term is due to the charged scalar, labelled `$c$',
for which the large-mass limit has been used.
For Type II models, these quantities are instead:
\eq
\label\typeIck
\eqalign{
c_g^h(\hbox{Type II}) &= {\cos\alpha \over \sin\beta}
\; c_g^{\sss SM}(\hbox{up-type fermions}) -
{\sin\alpha \over \cos\beta} \;
c_g^{\sss SM}(\hbox{down-type fermions}) , \cr
c_\gamma^h(\hbox{Type II})  &=
{\cos\alpha \over \sin\beta}
\; c_\gamma^{\sss SM}(\hbox{up-type fermions}) -
{\sin\alpha \over \cos\beta} \;
c_\gamma^{\sss SM}(\hbox{down-type fermions}) , \cr
&\qquad\qquad - \sin(\beta - \alpha) \;
c_\gamma^{\sss SM}(W) + {\nu_{cch} \over 4 m_c^2}, \cr}
\eeq
For a comparatively light charged Higgs (around 200 GeV) we get a
negative contribution of the order of $0.02 \nu_{cch}/\mh^2$.

The CP-violating couplings, $\tw c_\gamma^h$ and $\tw c_g^h$,
are obtained by everywhere substituting $z^h_{fg}$ for $y^h_{fg}$, and
so vanish for Type I and II models, but need not do so for a generic
2HDM.

\subsection{Left-Right Symmetric (LR) Models}

\ref\LRmodels{J.C.Pati and A. Salam, \prd{10}{74}{275};
R.N. Mohapatra and J.C. Pati, \prd{11}{75}{566};
G. Senjanovic and R.N. Mohapatra, \prd{12}{75}{1502};
F. Feruglio, L. Maiani and A. Masiero, \plb{233}{89}{512};
A. Deandrea, F. Feruglio and G.L Fogli, \npb{402}{93}{3};
A. Pilaftsis, \prd{52}{95}{459} and references therein.}

\ref\dfmz{A. Donini, F. Feruglio, J. Matias and F. Zwirner,
\npb{507}{97}{51}.}

Left-right symmetric models have also been extensively discussed
in the literature \LRmodels, \dfmz. Their gauge group is
$SU(2)_L\times SU(2)_\ssr \times U(1)$, and so their
treatment requires a generalization of
our previous discussion because of the
possibility which arises here of mixing amongst the LH and RH
electroweak gauge bosons. The usual scalar
sector of a left-right symmetric model  (LRSM) contains a
bidoublet and two triplets:
\eq
\label\newway
\phi = \pmatrix{ \phi_1^0 & \phi_1^+ \cr
\phi_2^- &\phi_2^0  \cr},  \quad
\Delta_\ssl = \pmatrix{
  \frac{1}{\sqrt{2}} \delta_\ssl^+
& \delta_\ssl^{++} \cr   \delta_\ssl^0 &
- \frac{1}{\sqrt{2}} \delta_\ssl^+
\cr}, \quad \Delta_\ssr = \pmatrix{
\frac{1}{\sqrt{2}} \delta_\ssr^+
& \delta_\ssr^{++} \cr   \delta_\ssr^0 & -
\frac{1}{\sqrt{2}} \delta_\ssr^+
 \cr}.
\eeq
\ref\egre{G. Beall, M. Bander and A. Soni, \prl{48}{82}{848};
G. Ecker and W. Grimus, \npb{258}{85}{328};
G. Barenboim, J. Bernabeu and M. Raidal, \npb{478}{96}{527};
G. Barenboim, J. Bernabeu, J.Matias and M. Raidal,
\prd{60}{99}{016003};
P. Ball, J.M. Frere and J. Matias, {\it Nucl.Phys.} {\bf
B572} (2000) 3, (hep-ph/9910211).}
whose neutral components acquire a non-zero vacuum
expectation value,
$\langle \phi_1^0\rangle=k_1/\sqrt{2}$, $\langle
\phi_2^0\rangle=k_2/\sqrt{2}$ and $\langle \delta_{\sss L,R}^0
\rangle=v_{\sss L,R}/\sqrt{2}$. A large mass for the `other'
charged gauge boson, $W'$, is required
on phenomenological grounds \egre, and this requires a
large value for $v_\ssr$. The preference is for vanishing
$v_\ssl$, on the other hand, to avoid possible generation
of large Majorana masses for left-handed neutrinos.

\topic{Gauge Boson Couplings}

The kinetic part of the lagrangian from which our scalar/gauge-boson
vertices arise is
\eq
\label\exthiggs
-{\cal L} = {\rm Tr}{({\cal D}_\mu
\phi)}^\dagger({\cal D}^\mu \phi)+
{\rm Tr}{({\cal D}_\mu
\Delta_\ssl)}^\dagger({\cal D}^\mu \Delta_\ssl)+
{\rm Tr}{({\cal D}_\mu
\Delta_\ssr)}^\dagger({\cal D}^\mu \Delta_\ssr)
\eeq
where the covariant derivatives are defined by
\eq
\eqalign{
&{\cal D}_\mu \phi = \partial_\mu \phi+
{i g_\ssl \over 2} \, {\vec W}_{\ssl \mu}
\cdot {\vec  \sigma}\, \phi
 -{i g_\ssr \over 2} \, \phi \, {\vec W}_{\ssr \mu}
\cdot {\vec \sigma}  \cr
&{\cal D}_\mu \Delta_{\sss L,R}=\partial_\mu
\Delta_{\sss L,R}+{i g_{\sss L,R} \over 2} \;
[{\vec W}_{\ssl,\ssr \mu} \cdot {\vec \sigma},
\Delta_{\sss L,R}]  + i g^{\prime} \,
B_{\mu} \, \Delta_{\sss L,R}} .
\eeq
Here $\sigma^i $ are the Pauli matrices, while $g_\ssl$, $g_\ssr$
and $g'$ are the coupling constants for the factors of the
gauge group $SU_\ssl \times SU_\ssr \times U(1)$.

The most important difference from our previous discussions
arises when we rotate the gauge fields to diagonalize their mass
matrix. There are three mixing angles which are generated
in this way. In order to recover the SM relation
$e = g_\ssl \sin\theta_\ssw$, the weak mixing angle is
conventionally defined as $\sin\theta_\ssw =
xy/\sqrt{x^2+z^2+x^2z^2}$, where $x=g_\ssr/g_\ssl$
and $z=g^\prime/g_\ssl$ denote ratios of
the gauge couplings. The other two mixing angles
are $\alpha_c$, used to diagonalize the charged
gauge-boson sector, and $\alpha_0$ which describes
the rotation amongst the two massive neutral gauge bosons,
$Z$ and $Z^\prime$ (see, for instance, \dfmz).
In what follows we liberally use the small-mixing-angle
approximation, $\alpha_c, \alpha_0 \ll 1$, since typically
$\alpha_c \sim \mw^2 / M_{\ssw^\prime}^2$
when $\mw \ll M_{\ssw^\prime}$, and similarly for $\alpha_0$.

The $\rho$ parameter in this model gets two contributions at leading order
in the mixing angles: $\rho=1+\Delta \rho_\ssw+\Delta \rho_\ssz$,
where
\eq
\Delta \rho_\ssw = - \alpha_{c}^2 \; {M_{\ssw^{\prime}}^2-
M_{\ssw}^2 \over M_\ssw^2} \quad \quad
\Delta \rho_\ssz= \alpha_{0}^2 \; { M_{\ssz^{\prime}}^2-
M_{\ssz}^2 \over M_\ssz^2} .
\eeq
The expression for the Fermi coupling constant, defined at tree level
from muon decay, also differs in these models from our earlier
treatment, because it receives contribution from both charged
gauge bosons
\eq
\label\gfexplr
{\GF \over \sqrt2} = {g_\ssl^2 \over 8} \left( {\cos^2 \alpha_c
\over \mw^2 } + {\sin^2 \alpha_c \over M_{\ssw^\prime}^2 }\right)
\approx {g_\ssl^2 \over 8\mw^2} \left( 1+ \alpha_c^2 \;  {\mw^2
\over M_{\ssw^\prime}^2 }\right) .
\eeq

\ref\gunionlr{J.F. Gunion, J. Grifols, A. Mendez, B. Kayser and F. Olness,
\prd{44}{89}{1546}.}
\ref\fcnclr{F.G. Gilman and M.H. Reno, \plb{127}{83}{426};
M. Pospelov, \prd{56}{97}{259}.}

The composition of the lightest
Higgs in terms of initial neutral fields $\phi^0_1$, $\phi^0_2$,
$\delta^0_\ssl$ and $\delta^0_\ssr$ can be absolutely
arbitrary, since the scalar potential may have very many different
terms with unspecified coefficients \gunionlr.
It should be noted at this point that an unattractive
feature of these types of left-right models is the tree-level
FCNC --- mediated by a linear combination of the $\phi^0_1$ and
$\phi^0_2$ fields --- which they generically have.
As a result, limits on the masses of scalars having a strong overlap with
these FCNC states are generally stronger than are the limits
for $M_{\ssw^\prime}$ \fcnclr. Unfortunately there is no natural way
of giving these scalars such a large mass, apart from simply fine tuning
the appropriate scalar couplings \gunionlr.
We assume here that this problem
is solved in some way (a possible solution was
proposed in \dfmz\ by considering 
a fermiophobic model), so that all FCNC scalars are very heavy,
with the lightest scalar having flavour-conserving couplings.

With these assumptions and notational choices,
the $WWh$ and $ZZh$ vertices, in the limit of
very small mixing angle \dfmz, are
\eq
\label\hwwlr
\eqalign{
a_\ssw^h \approx & \; {g_\ssl} \left( \mw - \alpha_{c}^2 \;
{3 M_{\ssw^\prime}^2 \over 2 \mw} \right) \cr
a_\ssz^h \approx & \; {g_\ssl \over c_w } \left(
\mz - \alpha_0^2 \; {3 M_{\ssz^\prime}^2 \over 2 \mz }\right) .
}
\eeq
We see that left-right symmetric models always reduce $a^h_\ssw$
and $a^h_\ssz$ relative to the SM result, as was the case for extra doublets.
Notice also that the second term in each of these expressions
is directly related to the corresponding contributions $\Delta \rho_\ssw$ and
$\Delta \rho_\ssz$, so (assuming no fine-tuned cancellations in $\rho$)
they can be at most of order $\Delta \rho \sim 10^{-3}$.

\topic{Yukawa Couplings}

The Yukawa couplings of the model are similar to
the triplet model discussed above. The neutral components
of the bidoublet can have Yukawa couplings to all
fermions except neutrinos, and so their
{\it v.e.v.}s are responsible for the masses
of all of these particles. As mentioned earlier, we suppose these
couplings to be flavour diagonal for the lightest scalar,
assumed to be the one observed.
The triplet has ordinary weak hypercharge $y=2$
and so can only have tree-level
Yukawa couplings with the left-handed lepton doublet,
implying a coupling of its neutral component only
to neutrinos.

\subsubsection{Loop Effects}

The suppression of small Yukawa couplings persists into
the loop expansion for the model considered here, but for
reasons which differ from those of the models previously
considered. In those models each fermion acquires a new
chiral symmetry in the limit of vanishing mass, and it is
this symmetry which forces corrections to mass-suppressed Yukawa
couplings to remain proportional to the same small masses.

It turns out that the argument differs for the LR models because of
the presence of both LH and RH fermion-$W$ couplings in the model.
(The connection between these couplings and the chiral symmetry
argument is explained in detail in \S6, below.) In this case $W$
loop corrections to fermion self-energies and fermion-scalar vertices
can give corrections to down-type fermions which are proportional
to up-type fermion masses. Since they are also proportional to
the RH $W$-fermion coupling, which is of order $\alpha_c \ll 1$
in these models, these mass corrections are nonetheless negligible in practice.

Other corrections which take advantage of flavour-changing scalar
interactions are also possible, but can be ignored here because we
assume the absence of flavour-changing neutral-scalar couplings,
and because the flavour-changing Yukawa couplings to charged
scalars are themselves small.

\ref\mpt{R. Martinez, M.A. Perez and J.J. Toscano \prd{40}{89}{1722}}

In the absence of flavour-changing Yukawa couplings for the
light scalar, $h$, the one-loop contributions to $c_g^h$
are proportional to the SM results, rescaled
by the mixing angle between $h$ and the bidoublet components,
$\phi_1^0$ and $\phi^0_2$. The contributions to
$\tw c_g^h$ and $\tw c^h_\gamma$ are given by the same
expressions with $z^h_{ff}$ replacing $y^h_{ff}$.

The contributions to $c_\gamma^h$ are more complicated
since this coupling receives contributions from boson loops.
In addition to the usual SM terms, with fermion and $W$
contributions rescaled by $y^h_{ff}/y_f^\SM$ and
$a^h_\ssw / a^\SM_\ssw$, we have contributions from the
electrically charged scalars and from the heavy gauge
boson, $W^\prime$. Using the large-mass limit for both
the charged scalars and $W'$ we find:
\eq
\label\LRcgam
\eqalign{
c_{\gamma}^{\sss LR} &= {\alpha \over 6 \pi \, v}\; \left[
\sum_{q} 3 Q_q^2 \; \eta_q  \;
I_\hf\left({m^2_q \over \mh^2} \right)
+ \sum_{\ell} Q_\ell^2  \; \eta_\ell  \;
I_\hf\left({m^2_\ell \over \mh^2} \right) \right. \cr
& \qquad\qquad \left.
- \left( 1 -\alpha_c^2 {3 M^2_{\ssw^\prime} \over 2\mw^2}
\right)  \; I_1 \left( {\mw^2 \over \mh^2} \right)
 - \; { a_{\ssw^\prime}^h 
I_1\left( {M_{\ssw^\prime}^2 \over \mh^2} \right)}
 + \nth4 \; \sum_c {Q_c^2 \nu_{cch}
\over m^2_c} \right] .\cr
}
\eeq
Here $\eta_f = y_{ff}^h/\left(\sqrt 2 \GF m_f^2 \right)^\hf$
denotes the relevant fermion Yukawa coupling, normalized
by the SM result, and the final sum over $c$
runs over all of the heavy charged scalar states.

Naively, since $I_1(r)$ is a slowly varying function of $r$
one might expect a large contribution to $c_\gamma$
from $W'$, however this is would only be true if its coupling
to $h$, $a^h_{\ssw^{\prime}}$, were proportional to the gauge boson
mass, $M_{\ssw^\prime}$, as is the case for $W$ in the SM.
In left-right models, however, $a^h_{\ssw^{\prime}}$
is instead proportional to $\mw$ \mpt, implying a
$W'$ contribution approximately proportional to
\eq
a_{\ssw^\prime}^h 
I_1\left( {M_{\ssw^\prime}^2 \over \mh^2} \right) 
\sim
{\left(M_\ssw \over M_{\ssw^\prime}\right)}^2
I_1 \left( {\mw^2 \over \mh^2} \right)
\approx {21\over 4}\left({M_\ssw \over M_{\ssw^\prime}}\right)^2 .
\eeq
This suppression of the $W'$ contribution, whose mass is
constrained to be heavier than $1.6$ TeV \egre\
in standard left-right models, makes it numerically insignificant.

\subsection{The Minimal Supersymmetric Standard Model}

Supersymmetric extensions of the SM are probably the best motivated
of the theoretical alternatives we discuss, because of the well-known
merits of supersymmetry for protecting the heirarchy between the
electroweak and higher scales. The minimal such variant of the
SM --- the so-called Minimal Supersymmetric Standard Model (MSSM) ---
has accordingly been the subject of considerable theoretical and experimental
efforts \outb, \pdg. Here we wish only to emphasize several features
of the model, relating to the properties of its neutral scalars.

In fact, a variety of models are all called `the' MSSM, and although
all such models agree on the supersymmetric generalization of the
SM particle content, they can differ on the precise way in which
supersymmetry is spontaneously broken. Luckily enough, these
differences do not have much impact on the interactions
of the most interest here, such as Yukawa and
trilinear gauge-boson/scalar couplings. The same cannot be said
for other couplings, such as trilinear and quadratic interactions
in the scalar potential, and so these (in general) imply some
model dependence for the scalar mixing matrices.

\ref\rparity{S. Dimopoulos and L. Hall, \plb{207}{88}{210};
             V. Barger, G.F. Giudice and T. Han, \prd{40}{89}{2987}.}

The model requires two doublet chiral Higgs supermultiplets,
$H_1$ and $H_2$, with opposite hypercharges. The three neutral
scalar states which emerge from these multiplets after the breaking
of the electroweak gauge group are denoted $h$, $H$ and $A$,
as in the 2HDM. There are, in addition, scalar $y=-1$ doublets
arising as superpartners to the lepton doublets, the neutral elements
of which contain three more (complex) sneutrinos, denoted
$\tw\nu_k$, for $k=1,2,3$. The sneutrinos
share the lepton number of their spin-half superpartners, and so they
cannot mix with the Higgs sector unless lepton number is broken
(as is the case in some $R$-parity violating models, for instance
\rparity\ \foot\rpfoot{The formula $\ss R = (-)^{3B+L+F}$
relates $\ss R$ parity  to baryon
number, $\ss B$, lepton number, $\ss L$ and fermion number, $\ss F$,
and so its conservation is automatic if baryon
and lepton number are both conserved.}).
All of these scalars can be produced at colliders, in principle.

\ref\habhempf{H.E. Haber and R. Hempfling, \prl{66}{91}{1815};
Y.Okada, M. Yamaguchi, T. Yanagida, {\it Prog. 
Theor. Phys. }{\bf 85} (1991){1}; 
J.R. Espinosa and M. Quiros, \plb{302}{93}{51-58}, (hep-ph/9212305);
J.A. Casas, J.R. Espinosa, M. Quiros and A. Riotto, \npb{436}{95}{3-29}, 
erratum \npb{439}{95}{466-468}, (hep-ph/9407389);
H.E.Haber, R. Hempfling, A.H.Hoang, {\it Z. Phys.} {\bf C75} (1997) 539;
S. Heinemeyer, W. Hollik, G. Weiglein, \prd{58}{98}{091701} 
\plb{455}{99}{179};{\it Eur.Phys.J.} {\bf C9} (1999) 343;  
R-J. Zhang, \plb{447}{99}{89}; 
M. Carena, H.E. Haber, S. Heinemeyer, W.Hollik, C.E.M. Wagner, G. Weiglein,
{\it Nucl. Phys.}{\bf B580}(2000) 29;
J.R.Espinosa, R-J.Zhang, {\it Nucl.Phys.}{\bf B586} (2000)3; JHEP 
0003:026,2000;
S. Abdullin et al. hep-ph/0005142;
S. Heinemeyer, G. Weiglein, {\it Nucl.Phys.Proc.Supp.} {\bf 89} (2000) 216;
J. R. Espinosa and I. Navarro, {\it Nucl. Phys.}{\bf B615} (2001) 82;
G. Degrassi, P. Slavich, F. Zwirner, {\it Nucl. Phys.}{\bf B611} 
(2001) 403; 
A. Brignole, G. Degrassi, P. Slavich and F. Zwirner,  hep-ph/0112177 
}

The validity of our effective Lagrangian
approach presupposes that the masses of the
at-present-undiscovered superpartners are sufficiently large.
We might assume here, for instance, that squarks, sleptons and
gauginos are all as heavy as several hundred GeV.
To the extent that sneutrino masses must be as large as are
those of their charged slepton partners, they would also be
too heavy to include within the effective Lagrangian as well.
However, since scalar masses are among the parameters
which depend in detail on supersymmetry breaking, we
entertain the possibility that the hypothetically-observed new
particle could be a sneutrino.

On the other hand, since the
Higgs scalars must acquire nonzero expectation values, their
masses are connected to quartic scalar interactions, and so are
more robustly constrained. In particular, an important
MSSM prediction is that the mass of the lightest Higgs boson
cannot be made too large. At the tree level it must be smaller than
$\mz$, with this upper bound rising to 130 GeV once radiative
corrections to $m_h$ are taken into account \habhempf.
Given this upper limit, a betting man's prejudice would 
be to expect a single
light scalar to be a Higgs, within the MSSM framework.

\topic{Scalar/Gauge-boson Couplings}

The general expression for the trilinear and quartic scalar/gauge-boson
couplings as functions of the neutral-scalar mixing matrix, $A_{ij}$,
are found by specializing eqs.~\dimthreeform\ and \dimfourforms\ to
the pure-doublet case:
\eq
\label\thdmabk
a_\ssw^i = a_\ssz^i \cw^2 = {e^2 \over 2 \sw^2} \; A_{ji} v_j ,
\qquad
b_\ssw^{ij} = b_\ssz^{ij} \cw^2 = {e^2 \over 2 \sw^2} \; \delta_{ij} .
\eeq
It follows that the quartic couplings are unchanged from
the SM, regardless of how the observed scalar mixes with
Higgses and sneutrinos. The trilinear couplings depend on
this mixing in more detail, and this is the only place where
supersymmetry-breaking details enter.

An important special case, where statements can be made
independent of the details of supersymmetry breaking, is
when lepton number is conserved, since then
none of the sneutrinos can acquire a lepton-number breaking
{\it v.e.v.}. In this case all of the tree-level trilinear sneutrino/gauge-boson
couplings vanish, and those of the remaining three neutral
Higgses,  $h$, $H$, $A$, can be expressed in terms of the
scalar mixing angles of the 2HDM. These are particularly
simple if CP is also unbroken, since they then reduce to the dependence
on two angles, $\alpha$ and $\beta$. Decomposing the neutral
components of the doublets, $H_1$ and $H_2$, as in the previous
section
\eq
\eqalign{
H_{2}^{0} &=v_{2}+H \sin \alpha +h \cos \alpha +iA \cos \beta
+i z \,\sin \beta \cr
H_{1}^{0} &=v_{1}+H \cos \alpha -h \sin \alpha +iA \sin \beta
-iz\, \cos \beta ,}
\eeq
leads to the prediction, eqs.~\thdmzzh\ and \gzthdm, for these trilinear
couplings.
\vfill
\eject

\topic{Tree-Level Yukawa Couplings}

Because Yukawa couplings are dimension-four interactions they
get no direct contribution from soft supersymmetry-breaking
interactions, making them sensitive to the details of supersymmetry
breaking only to the extent that they depend on the mixing amongst
scalars (or amongst neutral fermions). In particular, the tree-level
Yukawa couplings of the Higgs states are determined by the model's
superpotential, which is required to be a holomorphic function
of the (complex) superfields. For instance, the terms relevant to
Yukawa couplings are
\eq
\label\superpot
W = \mu \; H_1 \, H_2 + \Scy^d_{ij} \; Q_i^\sst \, D_j \; H_1
+ \Scy^u_{ij} \; Q_i^\sst \, U_j \; H_2,
\eeq
where $i = 1,2,3$ is a generation label, $Q_i = {u_i \choose d_i}$
represents the left-handed quark supermultiplets, while $U_j$
and $D_j$ are left-handed antiquark supermultiplets.
The important point about eq.~\superpot\ is that 
each Higgs doublet  couples to either up-type or down-type 
fermions,
but not to both.  The tree-level Yukawa couplings of the scalar doublets
which emerge from this therefore take the form of a Type II 2HDM,
eq.~\ttwoyuk.

Sneutrino Yukawa interactions with SM fermions are more
model dependent than are those of the Higgs states, even though
they are also determined at tree level by the holomorphic
superpotential. (Unlike the SM, the symmetry and field content of
the MSSM admit renormalizable lepton- and baryon-number
violating interactions.) The model dependence arises because the sneutrino
couplings necessarily break lepton number, and so the corresponding
couplings are only bounded by experiment to be consistent with zero,
and are not related to other known parameters, like SM fermion
masses.  The experimental constraints on the size of these couplings
originate from fermion-fermion scattering and from low energy
precision measurements \rparity. If parameters are chosen to ensure
the proton is stable, then the limits obtained from these experiments
are numerically similar to the bounds of eq.~\scheeffbounds\ and
eq.~\scheebounds.

\topic{Loop Effects (Yukawa Couplings)}

\ref\largtan{R. Hempfling, \prd{49}{94}{6168}}
\ref\bak{K.S. Babu and C. Kolda, \plb{451}{99}{77}.}
\ref\nonhol{F. Borzumati, G.R. Farrar, N. Polonsky and
S. Thomas, \npb{555}{99}{53--115}.}

Supersymmetric models represent a case where loop corrections
to Yukawa couplings can be important \mref{\largtan}{\nonhol}.
This is because the
superpartner particle content permits the existence of strong
interactions which couple light flavours to heavy flavours, and
so can ruin the chiral symmetry protecting Yukawa couplings
to light fermions. Indeed, these radiative corrections
can be dangerous, since if squark masses
have non-trivial flavour dependence, then Higgs exchange
generates effective FCNC interactions. (In this paper we assume
no such FCNC contributions to be present.)

In the most likely scenario, the lightest scalar in the MSSM
is a Higgs particle. At tree level it was supersymmetry itself
which forced the Yukawa couplings of this scalar
to be of Type II, but since supersymmetry is broken,
the loop-corrected effective Yukawa couplings need not
respect this Type II form. The size of these corrections as well as their
flavour structure strongly depends on the assumptions that are made about
how supersymmetry breaks. In the absence of FCNCs, we may
write the one-loop-corrected effective Yukawa couplings to
neutral scalars as
\eq
\eqalign{
y^h_{uu'}&= \delta_{uu'} \Bigl[ \cos\alpha
\left( y^{(0)}_u +  y^{(1)}_u + \cdots \right)
- \sin\alpha \left( \tilde y^{(1)}_d + \cdots \right) \Bigr] \cr
y^h_{dd'}&=\delta_{dd'}\Bigl[ - \sin\alpha \left(
y^{(0)}_d + y^{(1)}_d + \cdots \right)
+\cos\alpha \left( \tilde y^{(1)}_u + \cdots \right) \Bigr] , \cr}
\label\yukdiag
\eeq
where $y_{u,d}^{(0)}$ denote the tree level Yukawa couplings
(which are found by diagonalizing the couplings $\Scy^{(u,d)}$ of the
superpotential, eq.~\superpot)
and $y_{u,d}^{(1)}$ represent their one-loop corrections.
$\tilde y_{u,d}^{(1)}$ denotes a new type of radiatively
generated correction which couples up-type 
quarks to the $H_1^*$ field and down-type quarks
to $H_2^*$. These couplings do not exist in the
initial formulation of the theory because they break supersymmetry.
However, they do arise below the scale of superpartners, as the
supersymmetry is effectively broken there.

Now comes the main point. The quark Yukawa couplings
can receive QCD-strength radiative corrections, due to graphs
in which the quark splits into a squark-gluino pair. The order of magnitude
for down-type quarks is given by
\eq
y_d^{(1)} \simeq {\alpha_s\over 3\pi}~~{m_\lambda A_d
\over m_{squark}^2} ,
\label\susycorr
\eeq
where $A_d$ is a trilinear scalar coupling appearing
amongst the soft supersymmetry-breaking interactions.
Equation \susycorr\ has two noteworthy features. First,
$y_{d}^{(1)}$ remains finite even in the limit of
vanishing $y_d^{(0)}$, because the trilinear breaking parameter,
$A_d$, breaks protective chiral symmetry. Second, the corrections
do not `decouple' in the sense that they do not go to zero in the
limit of heavy superpartners, provided that  the ratios
$A_d/m_{squark}$ and $m_\lambda/m_{squark}$ are kept fixed.
This property is natural as the Yukawa coupling corresponds to an
operator of dimension four which is protected by supersymmetry,
and so does not have to be suppressed as
the splitting in a supermultiplet gets large.
In many of the versions of the supersymmetry-breaking terms of
the MSSM this kind of chiral-symmetry breaking is suppressed by hand,
by setting $A_i\equiv A y_i^{(0)}$ when writing
the soft supersymmetry-breaking terms.

A different type of coupling,
$\tilde y_{u,d}^{(1)}$, can be generated due to the same diagram,
when the $\mu$-dependent
interaction of the Higgs field with squarks is taken into account:
\eq
\tilde y_d^{(1)} \simeq y_d^{(0)} ~~{\alpha_s\over 3\pi}~~{\mu \, m_\lambda
\over m_{squark}^2} ,.
\label\susycorrb
\eeq
Even though this result is proportional to the
``initial'' tree-level couplings $\simeq y_d^{(0)}$,
it is still important as it gives a new type of interaction,
which does not exist in the type II 2HDM.
The loop corrections to
fermion masses in the down-quark sector become numerically
significant in the large-$\tan\beta$ regime \largtan, when
the loop suppression of
$\tilde y^{(1)}_d$ is compensated by large ratio $v_2/v_1$, so that
$\tilde y^{(1)}_d \; v_2 $ and $y^{(0)}_d \;  v_1$ are comparable.
In this regime the interaction of the lightest Higgs $h$ to
the down-type quarks is also significantly modified with respect
to the usual type II 2HDM  expectations of
$y_d/y_{d'} = m_d / m_{d'}$ \bak.
This picture requires the other neutral scalars, $A$ and $H$,
to be in an intermediate range, as the relation
of masses and couplings becomes indistinguishable 
from the SM ones in the limit $m_\ssa, m_\ssh \to \infty$.
Large corrections to masses
and Higgs-fermion couplings may also arise at $\tan\beta\sim 1$ if
the trilinear soft-breaking terms include new non-holomorphic
structures \nonhol.


Thus a large value of  $\tan\beta$ and/or the presence of
unorthodox trilinear soft-breaking terms
(of unusual form or unusual size) may be detected through the deviation
from the $y_d/y_{d'} = m_d / m_{d'}$ relation.
This is one way to distiguinsh
SUSY models from other models having a similar low-energy particle
content and scalar sector, like the Type II 2HDM.
It is clear, however, that the MSSM cannot be
distiguished from a completely generic 2HDM,
simply based on measurements of Higgs-fermion couplings,
simply because the Yukawa couplings of the low-energy scalar
are essentially arbitrary in a general 2HDM.

\subsubsection{Loop Effects (Contributions to $c_k$)}

If the light scalar is one of the neutral Higgs states, then
these effective couplings receive loop contributions which are
very much like those of the 2HDM. There are some differences,
though, due to the contributions of the various superpartners
which might be hoped to distinguish the MSSM from
the 2HDM. We focus here on contributions which are beyond
those already in the 2HDM.

Only the coloured superpartners -- \ie\ squarks and gluinos -- can
contribute to $c_g$ and $\tw c_g$, but since light scalars don't
couple at tree level to gluinos only the squarks actually contribute,
by an amount
\eq
\label\squarkcg
\Delta c_g^h(\hbox{squarks}) \approx {\alpha_s \over 48 \pi}
\sum_{\tw q} {\nu_{\tw q \tw q h} \over m_{\tw q}^2},
\eeq
in the large-squark-mass limit. The trilinear coupling in this
expression gets contributions both from the supersymmetric
part of the lagrangian, $\nu_{\tw q \tw q h} \sim y_q^h \; \mu$,
and from the soft supersymmetry-breaking terms, $\nu_{\tw q
\tw q h} \sim A_q$.

Similarly, the electromagnetic vertex receives contributions from
both charginos, higgsinos and charged squarks and sleptons. For
instance the contributions of heavy squarks are given
by \outb, \spira:
\eq
\label\squarkcgam
\Delta c^h_{\gamma}(\hbox{squarks}) \approx
{\alpha \over 8 \pi \, v}\; \sum_{\tilde q}  {Q_{\tw q}^2 \;
\nu_{\tw q \tw q h} \over m_{\tw q}^2} .
\eeq
Heavy charginos
similarly give
\eq
\label\chginocgam
\Delta c_\gamma(\hbox{charginos}) \approx
{\alpha \over 6 \pi} \sum_\chi \eta^h_{\chi} ,
\eeq
where $\eta_\chi = {2 \mz} \; \left( S_{ii} \cos \alpha - Q_{ii}
\sin \alpha \right)/m_\chi$, with $S_{ii}$ and $Q_{ii}$
defined as in ~\spira.

\ref\djouadi{A. Djouadi, V. Driesen, W. Hollik and J.I. Illana,
Eur. Phys. J. {\bf C1} (1998) 149--162.}

One could hope that  $c_\gamma$ and $c_g$ are indeed
the most important couplings of the lightest scalar, $h$,
since then a measurement of its size would be very sensitive
to the existence of new charged particles.
More detailed anlysis shows, however, that
even a mild decoupling of superpartners ($m_{chargino} \sim 250$ GeV)
leads to SUSY corrections which are less than 10\% of the SM
values for $c_\gamma$ and $c_g$ \djouadi.

\section{Summary: The Decision Tree}

We now come to the nub of this paper, which is a discussion of what
experiments involving a newly discovered scalar can ultimately
tell us about its underlying nature. Provided the framework we have
assumed in this paper indeed holds -- {\it viz} all other
hitherto undiscovered particles are sufficiently massive to justify
their neglect in a low-energy effective-lagrangian description -- then
we have seen that there is nothing more we can hope to learn about
the properties of the new scalar than what are its effective couplings.

Clearly, then, any model-independent discussion of the properties of
a new scalar must therefore come in two parts. First, we must
turn to experiments to learn the values of the
effective couplings; and second, we can ask what
these inferred values tell us about the nature of new physics which
has thrown this new scalar up for our inspection. We
discuss these two issues separately in the following sections.

\subsection{Experimentally Determining the Effective Couplings}

Immediately after the discovery of a new scalar, only limited
information can be expected to be available. The first information
will undoubtedly be the kind of reaction in which the scalar
was produced ({\it e.g.:} Is it produced in $e^+ e^-$ or at hadron
collisions? At what CM energy? What other particles are produced
in the production and decay reactions?), and into what kinds of
particles it decays. Only later can we be expected to have more
detailed information ({\it e.g.:} What are the detailed angular distributions
of the production process? What are the precise branching ratios
into each daughter particle?), from better measurements of the various
effective coupling constants. Because most of the interesting
conclusions are relatively obvious, but impossible to draw until
the data arrives, we confine ourselves to comparatively few
remarks in this subsection.

In principle, those effective couplings which contribute to the
production process may be disentangled from those appearing in
the decay by examining the total reaction rate as a function of
the independent variables, such as CM energy, scattering angle,
\etc. Some qualitative conclusions may be drawn simply by considering
the kinds of reactions which are involved. For example, if the
 new scalar is only
pair-produced, then perhaps it carries an approximately conserved
quantum number. If it is singly produced in $e^+ e^-$
collisions, then couplings to electrons, photons and/or
$Z$ bosons are indicated. Which of these is actually present, or
dominates, depends on which other kinds of particles are produced
in association with the new scalar. Comparison with the cross sections
of Section 3 permits the inference of which effective couplings
are involved from the measurement of the dependence on CM energy
or the angular distribution of the produced scalar in the CM frame.

Observation of the decay products gives immediate information
about the relative strength of the scalar couplings to its potential
daughter particles. For example, decay into $W$ pairs likely indicates
that the coupling $a_\ssw$ is more important than are the Yukawa
couplings, an hypothesis which might be tested by looking for $W-h$
production in hadron colliders.

\subsection{Using Effective Couplings to Discriminate Amongst Models}

Once the existence of a new scalar is established, the
most interesting interpretational issue is to discover what
it is trying to tell us. That is, given the values for the
effective couplings which are inferred from its production and
decay (as well as from the absence of  other  lower-energy processes),
what can be said to distinguish the relative
likelihood that it is a harbinger of supersymmetry, the
first of many strongly-coupled resonant bound states,
or an indication of something else?

Organizing the discussion using the effective lagrangian
permits a systematic approach to the problem of distinguishing
models. Since the only possible information which may be used
to distinguish them is the size of the effective couplings
they predict, we may use the relative size of couplings to
broadly characterize models. Sixteen categories immediately
suggest themselves, depending on the answers a model gives
to the following four yes/no questions:

\item{1.} Can the effective couplings to the $W$
and $Z$ be at least $O(e)$ in size?

\item{2.} Are there any fermions for which the effective
Yukawa couplings can be $O\left(e\right)$ in size?

\item{3.} Can the effective couplings to two photons
be at least $O\left( \alpha/2 \pi v\right)$ in size?

\item{4.} Can the effective couplings to two gluons
be at least $O\left( \alpha_s/2 \pi v\right)$ in size?

\noindent  In items 3 and 4, $v$ denotes
the quantity $(\sqrt2 \GF)^{-1/2} = 246$ GeV.

\midinsert
$$\vbox{\tabskip=0pt \offinterlineskip
\halign to \hsize{
\strut#& #\tabskip 1em plus 2em minus .5em&
#\hfil &#& \hfil#\hfil &#& \hfil#\hfil &#& \hfil#\hfil &#&
\hfil#\hfil &#& \hfil#\hfil &#& \hfil#\hfil &#\tabskip=0pt\cr
\noalign{\hrule}\noalign{\smallskip}\noalign{\hrule}
\noalign{\medskip}
&& \hfil Class && Examples && Q1 && Q2 && Q3 && Q4 &\cr
&& && && ($a_\ssw,a_\ssz$) && ($y_f, z_f$) && ($c_\gamma$)
&& ($c_g$) &\cr
\noalign{\medskip}\noalign{\hrule}\noalign{\medskip}
&& I &&  SM, 2HDM (+)  && Y && Y && Y && Y &\cr
&& &&  LRSM (+), SUSY (+)  &&  &&  &&  &&  &\cr
&& II &&  Triplet ($\nu$,+)  && Y && Y && Y && N &\cr
&& III &&  && Y && N && Y && Y &\cr
&& IV && TechniPGBs, LRSM ($-$) && N && Y && Y && Y &\cr
&& && 2HDM ($-$), SUSY ($-$)  &&  &&  &&  &&  &\cr
&& V && Higher Representation (+) && Y && N && Y && N &\cr
&& VI && Triplet ($\nu,-$)  && N && Y && Y && N &\cr
&& VII &&    && N && Y && N && Y &\cr
&& VIII &&  && N && N && Y && Y &\cr
&& IX && Singlet with RH neutrino ($\nu$) && N && Y && N && N &\cr
&& && Conserved Q.No. ($\nu$) &&  &&  &&  &&  &\cr
&& X && && N && N && Y && N &\cr
&& XI &&    && N && N && N && Y &\cr
&& XII &&  Higher Representation ($-$)   && N && N && N && N &\cr
\noalign{\medskip}\noalign{\hrule}\noalign{\smallskip}
\noalign{\hrule}
}}$$
\medskip
\centerline{\bf Table (2):}
{\eightrm The twelve categories of models,
based on the size of their effective couplings. The positions
of some representative models are indicated, where CP conserving
scalar couplings are assumed for simplicity. ($\ss \pm$) denotes
the CP quantum number of the observed light scalar state. A $\ss \nu$
in brackets indicates that the large Yukawa coupling may
be restricted to neutrinos only. Categories XIII through XVI
are not listed because models having $\ss O(e)$ couplings to
the $\ss W$ and $\ss $ generally also have $\ss O(\alpha/2 \pi)$
effective couplings to photons. Triplet indicates a doublet-triplet
model for which the observed light scalar is dominantly from the
triplet component. }
\endinsert

Four of the possible sixteen categories which are obtained
in this way are not of much interest, however, because
models having $O(e)$ couplings to the $W$ and $Z$ bosons
generally also have $O(\alpha/2\pi)$ couplings to the photon.
The twelve remaining categories, together with some of the
models which represent them, are listed in Table (2).

Several points about this table bear emphasis.
\item{1.}
The most basic point is that the very first discovery
experiments can do little more than identify which category
of model is supported by the properties of the newly-discovered
scalar.
\item{2.}
The next most basic point is to observe that most of the models
listed are distributed over a number of different categories,
so the first experiments will immediately give nontrivial
information, although they are unlikely to permit an immediate
discrimination between theoretically popular
alternatives --- like multi-doublet models and
supersymmetric models, for instance.
\item{3.}
It should be remarked that the assignment of sample
models to various categories can vary, depending on
what other field content the underlying model is assumed
to have. For instance, the entries for triplet models assume
two features about the model. First, that the observed light neutral scalar
is dominantly triplet, with any admixture with the usual SM-style
doublet suppressed. Second, that the only new nonstandard
electroweak multiplets in the model are the colour-singlet
triplet field. If other fields were permitted, such as new
exotic fermions, then loops of these new particles might generate
a one-loop gluon coupling, forcing the recategorization of
the model from either II or VI (depending on the CP quantum
number of the light scalar) to either I or IV.

To make more refined decisions requires more precise measurements
of the effective couplings (or the discovery of more new particles!).
This is possible because most of the above categories, including
the most interesting ones -- like I and IV -- may be further
subdivided according to the relative size of the various couplings
within any given class. For example, inspection of eqs.~\dimthreeform\
shows what information about the scalar quantum numbers can be
gleaned from a comparison of the $WWh$ to the $ZZh$ coupling
strengths. As was argued in section 5.2, a comparison of these
to the $W$ and $Z$ boson masses can, in some circumstances,
rule out the possibility that the new scalars are doublets or
singlets.

More information is potentially available once the relative stengths
of the Yukawa couplings become measured. We next summarize some of
the conclusions which can follow from such more precise information.

\subsubsection{Tree Level Relations Amongst Yukawa Couplings:}
An obvious first test to perform is for the proportionality of
the Yukawa coupling strength to the mass of the fermion involved.
That is, tests of the prediction
\eq
\label\yukmasstest
{y_f \over y_{f^\prime}} = {m_f \over m_{f^\prime}}.
\eeq
Although the SM predicts this relation to hold for all fermions,
as we have seen it typically holds with modifications for other
models. Some of the main subcategories of potential Yukawa coupling
relations are:
\topic{1}
If the new scalar is a Goldstone boson, then its couplings must
vanish in the limit of vanishing momentum, and so its Yukawa
couplings must strictly all vanish, although derivative
scalar-fermion couplings need not do so. This makes a measurement
of the energy dependence of the scalar-fermion coupling instructive,
if performed over a large enough energy range.

\topic{2}
Pseudo Goldstone bosons, and elementary scalar multiplets which
do not acquire vacuum expectation values, need not couple to
fermions with strength proportional to mass at all. Among the most striking
experimental signatures would be a strong coupling to light
fermions, like electrons or $u,d$ quarks, since these are
observable through their implications for the differential
production rates, for example, yet the direct experimental
limits on their existence are quite weak.

\topic{3}
Models of the Type II class of multi-doublet theories --- including
supersymmetric models --- share the SM prediction (at tree
level) so long as the ratio is made only between up-type fermions,
or only among down-type fermions. In this case the ratio of ratios,
$(y_u/y_{d})/(m_u/m_{d})$ is given by scalar mixing angles (\eg\ by
eq.~\ttwoyuk), and so this kind of model may be tested by
comparing this ratio with predictions (like eq.~\thdmzzh) for
the scalar - gauge boson couplings, which also depends only
on these mixing angles.

\topic{4}
Multidoublet models of the Type I category share the SM prediction,
eq.~\yukmasstest\ for all fermions, but differ in their prediction
for the constant of proportionality, $y_f/m_f$. The same is true for
higher representation models which acquire their Yukawa couplings
by mixing with the ordinary SM doublet. As eqs.~\ykmstrip\ and
\toneyuk\ show, these two categories may be differentiated by
comparing the mixing angles inferred from $y_f/m_f$ with those
inferred from the scalar - $W,Z$ couplings of eqs.~\awztrip\
and \thdmzzh.

\subsubsection{Loop Corrections to Yukawa Coupling Relations:}

Because the tree-level Yukawa couplings and masses are often
very small, it becomes important to consider also radiative
corrections to these predictions. In particular large radiative
corrections have recently been proposed as a means to distinguish
supersymmetric from generic Type II 2HDMs \cmw,\bak.
We here describe the logic of these tests within a more general
framework.

Loop corrections to the tree-level predictions relating the
effective couplings of dimension-4 or less interactions
are particularly important in this regard because they do
not decouple, in the sense that the contributions of heavy
particles in loops need not be suppressed on dimensional
grounds by inverse powers of the heavy particle mass.

As was discussed in Section 2, tree-level predictions of small
Yukawa couplings can be prevented from being significantly changed
by radiative corrections if the Yukawa coupling of interest is
protected by a chiral symmetry. This is the case for the SM,
for which the proportionality of $y_f$ to $m_f$ is not
significantly modified by loops. It is instructive to ask
in more detail why this is true for the ratio $y_b/m_b$, since
the $b$ quark couples to the $t$ quark \via\ the charged-current
weak interactions, and so any chiral symmetry involving a
rephasing of the $b$ quark must also require the $t$ quark to
rotate. But the large $t$ quark mass then should strongly
break the invariance of the SM lagrangian under
chiral rotations of the top quark. Superficially one might
therefore expect loop corrections to $y_b$ to be proportional
to $y_t$ instead of $y_b$, and so to be significantly large.

The flaw in this reasoning is this: in order for the SM
charged-current weak interactions to be invariant under
$\delta b = i \epsilon \, \gamma_5 \; b$, provided it is
sufficient to rotate the top quark by a {\it vectorlike} rotation:
$\delta t = i \epsilon \; t$, which is also a symmetry of
the top-quark mass term. A vectorlike $t$ rotation suffices
because the charged-current weak interactions are purely
left-handed. And so SM loop corrections to $y_b$ must indeed
therefore be proportional to $y_b$.

We therefore see what is required in order to ruin the chiral
symmetry protection of small Yukawa couplings in a model.
One of two things is required:
\item{1.}
Right-handed effective couplings of fermions to the
ordinary $W$ boson, together with nonzero Kobayashi
Maskawa mixing with a heavy fermion like the top quark;
\item{2.}
Direct chiral-symmetry breaking interactions, either
through new kinds of Yukawa interactions whose couplings
are not small, or through new flavour-changing gauge
interactions with heavy fermions, having both left-
and right-handed couplings.

It is now easy to see which of our models satisfy either
of these two criteria. For instance, neither is satisfied by a
Type I or Type II 2HDM, since the only chiral symmetry breaking
couplings in these models are the small Yukawa couplings
themselves, and the only flavour-changing couplings to the only
significantly massive fermion -- the top quark -- are the
purely left-handed charged-current interactions. Loop corrections
therefore cannot significantly modify the tree-level predictions
these models make relating Yukawa couplings to fermion masses.
Of course, this is particularly clear in models for which
the Type I or Type II couplings are enforced by a discrete
symmetry of some sort.

Supersymmetric models, on the other hand, are of Type II form
because of supersymmetry. However, in practice supersymmetry
is broken and loops involving heavy superpartners of light particles
might be expected to generate couplings of the `wrong' Higgs doublet
to fermions. These new terms have certain nondecoupling
properties, i.e. they do not vanish if the masses of quarks and gauginos,
trilinear soft-breaking terms,
and $\mu$ parameter are simultaneously made heavy.
Furthermore, as  was discussed earlier, the supersymmetric models
can satisfy criterion 2, depending on whether the
same chiral symmetry is preserved by the supersymmetry-breaking couplings
between superpartners of left- and right-handed squarks.
Of course, the SUSY-breaking couplings can themselves be set up
to preserve the chiral symmetry in the small-fermion-mass limit,
and this is often the choice made within some of the MSSM's
found in the literature. In this case, it can also happen that the
MSSM predictions can approach those of the 2HDM (and so,
for some parameter values, also those of the SM) in the limit
that all superpartners are taken to be arbitrarily heavy.

Clearly LR models satisfy criterion 1, and so top-quark
loops may be expected to significantly modify the $b$-quark
yukawa coupling. In order of magnitude one expects $\delta y_b
\sim (\alpha_w/4 \pi) \epsilon y_t$ where $\epsilon$
is the relative strength of the
right- and left-handed couplings of the light $W$ boson, which
is constrained by experiment to be $\epsilon \lsim 10^{-3}$.
 Since LR models are neither Type I nor
Type II in the couplings of the bidoublet scalar to fermions,
the resulting correction is more difficult to test experimentally.

\subsubsection{Loop Contributions to $c_\gamma$ and $c_g$:}

Another class of loop corrections which can be important are
the contributions to the effective couplings $c_\gamma$ and
$c_g$ and the CP-odd $\tw c_{\gamma}$, $\tw c_g$.  Comparing
the measured size of these couplings with the measured
Yukawa and gauge-boson couplings of the new scalar provides
independent information about the nature of the underlying
physics.

We have seen that measurements of  these couplings with sufficient precision
can usefully distinguish models. For example, the relative size of $c_g$
compared to $c_\gamma$ reflects differences in the kinds of coloured
and electrically-charged heavy states in the model. Similarly, 
comparing non vanishing CP-odd
($\tw c_{g,\gamma}$) with non vanishing CP-even $c_{g,\gamma}$ 
couplings
gives information about
the P and T breaking scalar interactions, \etc.

Unfortunately, the utility of this comparison is limited
until these couplings are known with relatively good precision. This is
because the $c_{g,\gamma}$'s are comparatively difficult to disentangle from the
$\tw c_{g,\gamma}$'s experimentally, and because new-physics
contributions to $c_g$ and $c_\gamma$ must be extracted from
relatively large, competing SM effects.

\subsection{Outlook}

In conclusion, we have shown how the properties of a new scalar
particle may be encapsulated into a low-energy effective lagrangian,
so long as other new particles are not also light enough to
be close to discovery. (We should be so lucky!) These effective scalar
couplings can be related to observables once-and-for-all (we give explicit
formulae), and provide a complete summary of the information
which experiments can provide about the scalar's properties.
Finally, we have compared how these couplings depend on more
microscopic parameters within a reasonably wide selection of models.

We conclude that the first round of experiments are likely to distinguish
only amongst broad classes of models, as summarized in Table (2). Later,
more precise, experiments can sharpen this process and in the event that
no other new particles are discovered,  much
can be  learned in principle about the underlying model if all effective
couplings become measured to levels of a few percent.

\bigskip

\centerline{\bf Acknowledgments}

\bigskip

We thank G. Azuelos and F. Corriveau for helpful conversations.
This research was partially funded by N.S.E.R.C.\ of
Canada, les   Fonds F.C.A.R.\ du Qu\'ebec, and was supported in part by
the Department of Energy under Grant No. DE-FG02-94ER40823.
C.B. thanks the University of Barcelona for its
generous support, and congenial hospitality,
while part of this research was being carried out.
J.M acknowledges financial support from a Marie Curie EC grant (TMR-ERBFMBICT
972147) and thanks the Physics Department of McGill University for the
kind hospitality and nice atmosphere when completing this work.

\vfill\eject

\listrefs

\bye